\documentclass[twocolumn]{aastex631}

\usepackage[T1]{fontenc}
\usepackage{ae,aecompl}

\usepackage{graphicx}	
\usepackage{amsmath}	
\usepackage{amssymb}	

\usepackage{natbib}

\bibpunct{(}{)}{;}{a}{}{,}

\usepackage{url}

\usepackage{longtable}
\usepackage{booktabs}

\usepackage[flushleft, referable]{threeparttablex}

\usepackage{subfloat}

\DeclareGraphicsExtensions{.jpg,.pdf,.png}

\usepackage{parskip}

\usepackage{lipsum}

\usepackage{dsfont}

\usepackage{comment}

\usepackage{tablefootnote}

\newcommand\degrees{^{\circ}}

\newcommand\lenstool{\textsc{lenstool}}

\newcommand\hii{\rm H {\sc ii}}

\defcitealias{Frye:2024aa}{F24}

\defcitealias{Abdurrouf:2018aa}{A18}

\defcitealias{Harrington:2016aa}{H16}

\graphicspath{{./}{figs/}}

\shorttitle{Stellar mass assembly in a pair of strongly lensed $z=2$ DSFGs}
\shortauthors{Kamieneski et al.}

\begin{document}

\title{Birds of a Feather: 
Resolving Stellar Mass Assembly With JWST/NIRCam
in a Pair of Kindred $\boldsymbol{z\sim 2}$ Dusty Star-forming Galaxies Lensed by the PLCK G165.7+67.0 Cluster}

\correspondingauthor{Patrick S. Kamieneski}
\email{pkamiene@asu.edu}

\author[0000-0001-9394-6732]{Patrick S. Kamieneski}
\affiliation{School of Earth and Space Exploration, Arizona State University, 
PO Box 871404,
Tempe, AZ 85287-1404, USA}

\author[0000-0003-1625-8009]{Brenda L. Frye}
\affiliation{Steward Observatory, University of Arizona, 933 N Cherry Ave, Tucson, AZ, 85721-0009}

\author[0000-0001-8156-6281]{Rogier A. Windhorst}
\affiliation{School of Earth and Space Exploration, Arizona State University, 
PO Box 871404,
Tempe, AZ 85287-1404, USA}

\author[0000-0001-5429-5762]{Kevin C. Harrington}
\affiliation{European Southern Observatory, Alonso de C{\'o}rdova 3107, Vitacura, Casilla 19001, Santiago de Chile, Chile}

\author[0000-0001-7095-7543]{Min S. Yun}
\affiliation{Department of Astronomy, University of Massachusetts, Amherst, MA 01003, USA}

\author[0000-0003-1832-4137]{Allison Noble}
\affiliation{School of Earth and Space Exploration, Arizona State University, 
PO Box 871404,
Tempe, AZ 85287-1404, USA}

\author[0000-0002-2282-8795]{Massimo Pascale}
\affiliation{Department of Astronomy, University of California, 501 Campbell Hall \#3411, Berkeley, CA 94720, USA}

\author[0000-0002-7460-8460]{Nicholas Foo}
\affiliation{School of Earth and Space Exploration, Arizona State University, 
PO Box 871404,
Tempe, AZ 85287-1404, USA}

\author[0000-0003-3329-1337]{Seth H. Cohen}
\affiliation{School of Earth and Space Exploration, Arizona State University, 
PO Box 871404,
Tempe, AZ 85287-1404, USA}

\author[0000-0003-1268-5230]{Rolf A. Jansen}
\affiliation{School of Earth and Space Exploration, Arizona State University, 
PO Box 871404,
Tempe, AZ 85287-1404, USA}

\author[0000-0001-6650-2853]{Timothy Carleton}
\affiliation{School of Earth and Space Exploration, Arizona State University, 
PO Box 871404,
Tempe, AZ 85287-1404, USA}

\author[0000-0002-6610-2048]{Anton M. Koekemoer}
\affiliation{Space Telescope Science Institute, 3700 San Martin Dr., Baltimore, MD 21218, USA}

\author[0000-0001-9262-9997]{Christopher N. A. Willmer}
\affiliation{Steward Observatory, University of Arizona, 933 N Cherry Ave, Tucson, AZ, 85721-0009}

\author[0000-0002-7265-7920]{Jake S. Summers}
\affiliation{School of Earth and Space Exploration, Arizona State University, 
PO Box 871404,
Tempe, AZ 85287-1404, USA}

\author[0000-0003-3418-2482]{Nikhil Garuda}
\affiliation{Steward Observatory, University of Arizona, 933 N Cherry Ave, Tucson, AZ, 85721-0009}

\author[0009-0001-7446-2350]{Reagen Leimbach}
\affiliation{Steward Observatory, University of Arizona, 933 N Cherry Ave, Tucson, AZ, 85721-0009}

\author[0000-0002-4884-6756]{Benne W. Holwerda} 
\affiliation{Department of Physics and Astronomy, University of Louisville, Louisville KY 40292, USA}

\author[0000-0002-2361-7201]{Justin D. R. Pierel}
\affiliation{Space Telescope Science Institute, 3700 San Martin Dr., Baltimore, MD 21218, USA}

\author[0000-0002-2640-5917]{Eric F. Jim\'{e}nez-Andrade}
\affiliation{Instituto de Radioastronom\'{i}a y Astrof\'{i}sica, Universidad Nacional Aut\'{o}noma de M\'{e}xico, Antigua Carretera a P\'{a}tzcuaro \# 8701, Ex-Hda. San Jos\'{e} de la Huerta, Morelia, Michoac\'{a}n, C.P. 58089, M\'{e}xico}

\author[0000-0002-9895-5758]{S. P. Willner}
\affiliation{Center for Astrophysics | Harvard \& Smithsonian, 60 Garden Street, Cambridge, MA, 02138, USA}

\author[0000-0002-4140-0428]{Bel\'{e}n Alcalde Pampliega}
\affiliation{European Southern Observatory, Alonso de C{\'o}rdova 3107, Vitacura, Casilla 19001, Santiago de Chile, Chile}

\author[0000-0002-4444-8929]{Amit Vishwas}
\affiliation{Cornell Center for Astrophysics and Planetary Science, Cornell University, Space Sciences Building, Ithaca, NY 14853, USA}

\author[0000-0002-6131-9539]{William C. Keel} 
\affiliation{Department of Physics and Astronomy, University of Alabama, Box 870324, Tuscaloosa, AL 35404, USA}

\author[0000-0002-9279-4041]{Q. Daniel Wang}
\affiliation{Department of Astronomy, University of Massachusetts, Amherst, MA 01003, USA}

\author[0000-0003-0202-0534]{Cheng Cheng}
\affiliation{Chinese Academy of Sciences South America Center for Astronomy, National Astronomical Observatories, CAS, Beijing, 100101, China}

\author[0000-0001-7410-7669]{Dan Coe} 
\affiliation{Space Telescope Science Institute, 3700 San Martin Dr., Baltimore, MD 21218, USA}
\affiliation{Association of Universities for Research in Astronomy (AURA) for the European Space Agency (ESA), STScI, Baltimore, MD 21218, USA}
\affiliation{Center for Astrophysical Sciences, Department of Physics and Astronomy, The Johns Hopkins University, 3400 N Charles St. Baltimore, MD 21218, USA}

\author[0000-0003-1949-7638]{Christopher J. Conselice} 
\affiliation{Jodrell Bank Centre for Astrophysics, Alan Turing Building,
University of Manchester, Oxford Road, Manchester M13 9PL, UK}

\author[0000-0002-9816-1931]{Jordan C. J. D'Silva} 
\affiliation{International Centre for Radio Astronomy Research (ICRAR) and the
International Space Centre (ISC), The University of Western Australia, M468,
35 Stirling Highway, Crawley, WA 6009, Australia}
\affiliation{ARC Centre of Excellence for All Sky Astrophysics in 3 Dimensions
(ASTRO 3D), Australia}

\author[0000-0001-9491-7327]{Simon P. Driver} 
\affiliation{International Centre for Radio Astronomy Research (ICRAR) and the
International Space Centre (ISC), The University of Western Australia, M468,
35 Stirling Highway, Crawley, WA 6009, Australia}

\author[0000-0001-9440-8872]{Norman A. Grogin}
\affiliation{Space Telescope Science Institute, 3700 San Martin Dr., Baltimore, MD 21218, USA}

\author[0009-0008-0376-3771]{Tyler Hinrichs}
\affiliation{School of Earth and Space Exploration, Arizona State University, 
PO Box 871404,
Tempe, AZ 85287-1404, USA}

\author[0000-0001-9969-3115]{James D. Lowenthal}
\affiliation{Smith College, Northampton, MA 01063, USA}

\author[0000-0001-6434-7845]{Madeline A. Marshall}
\affiliation{National Research Council of Canada, Herzberg Astronomy \& Astrophysics Research Centre, 5071 West Saanich Road, Victoria, BC V9E 2E7, Canada}
\affiliation{ARC Centre of Excellence for All Sky Astrophysics in 3 Dimensions (ASTRO 3D), Australia}

\author[0000-0001-6342-9662]{Mario Nonino}
\affiliation{INAF-Osservatorio Astronomico di Trieste, Via Bazzoni 2, I-34124 Trieste, Italy}

\author[0000-0002-6150-833X]{Rafael Ortiz III}
\affiliation{School of Earth and Space Exploration, Arizona State University, 
PO Box 871404,
Tempe, AZ 85287-1404, USA}

\author[0000-0001-9369-6921]{Alex Pigarelli}
\affiliation{School of Earth and Space Exploration, Arizona State University, 
PO Box 871404,
Tempe, AZ 85287-1404, USA}

\author[0000-0003-3382-5941]{Nor Pirzkal}
\affiliation{Space Telescope Science Institute, 3700 San Martin Dr., Baltimore, MD 21218, USA}

\author[0000-0001-7411-5386]{Maria del Carmen Polletta}
\affiliation{INAF – Istituto di Astrofisica Spaziale e Fisica cosmica (IASF) Milano, Via A. Corti 12, 20133 Milan, Italy}

\author[0000-0003-0429-3579]{Aaron S. G. Robotham}
\affiliation{International Centre for Radio Astronomy Research (ICRAR) and the International Space Centre (ISC), The University of Western Australia, M468, 35 Stirling Highway, Crawley, WA 6009, Australia}

\author[0000-0003-0894-1588]{Russell E. Ryan, Jr.}
\affiliation{Space Telescope Science Institute, 3700 San Martin Dr., Baltimore, MD 21218, USA}

\author[0000-0001-7592-7714]{Haojing Yan}
\affiliation{Department of Physics and Astronomy, University of Missouri, Columbia, MO 65211, USA}

\begin{abstract}
We present a new parametric lens model for the G165.7+67.0 galaxy cluster,
which was discovered with {\it Planck} through its bright submillimeter flux, originating from a pair of extraordinary dusty star-forming galaxies (DSFGs) at $z\approx 2.2$.
Using JWST and interferometric mm/radio observations, we characterize the intrinsic physical properties of 
the DSFGs, which are separated by only $\sim 1\arcsec$ (8 kpc) and a velocity difference $\Delta V \lesssim 600~{\rm km}~{\rm s}^{-1}$ in the source plane,
and thus likely 
undergoing
a major merger.
Boasting intrinsic star formation rates ${\rm SFR}_{\rm IR} = 320 \pm 70$ and $400 \pm 80~ M_\odot~{\rm yr}^{-1}$, stellar masses ${\rm log}[M_\star/M_\odot] = 10.2 \pm 0.1$ and $10.3 \pm 0.1$, and dust attenuations $A_V = 1.5 \pm 0.3$ and $1.2 \pm 0.3$, they are remarkably similar objects.
We perform spatially-resolved pixel-by-pixel SED fitting using rest-frame near-UV to near-IR imaging from JWST/NIRCam,
%
resolving some stellar structures down to 100 pc scales. 
Based on their resolved specific SFRs and $UVJ$ colors, both DSFGs 
are 
experiencing significant galaxy-scale star formation events.
If they are indeed interacting gravitationally, this strong starburst could be the hallmark of gas that has been disrupted by an initial close passage.
In contrast, the 
host galaxy of the recently discovered triply-imaged SN H0pe has a much lower SFR than the DSFGs,
and we present evidence for the onset of 
inside-out quenching and large column densities of dust even in regions of low specific SFR. 
Based on the intrinsic 
SFRs of the DSFGs inferred from UV through FIR SED modeling, 
this pair of objects alone is predicted to yield an observable
$1.1 \pm 0.2~{\rm CCSNe~yr}^{-1}$,
making this cluster field ripe for continued monitoring.
\end{abstract}



\section{Introduction} 
\label{sec:intro}

Out to at least $z \sim 3$, the fraction of obscured star formation in galaxies appears to increase nearly monotonically with increasing stellar mass \citep{Bourne:2017aa, Whitaker:2017aa, McLure:2018aa}, a relation that has also been recovered by simulations at $z\lesssim 2$ (e.g., \citealt{Zimmerman:2024aa}). 
Recent results from the Atacama Large Millimeter/submillimeter Array (ALMA) have suggested even the presence of a population of galaxies that are both massive and dust-rich as early as $z \approx 7$ \citep{Algera:2024aa}.
The dust-obscured mode of star formation has been found to dominate the cosmic star formation rate density (SFRD) for 12 billion years, to at least $z\sim 4$ (e.g., \citealt{Bouwens:2020aa, Zavala:2021aa}).
\citet{Long:2023ab} predicted that, at the peak of the Universe's star formation history at $z\sim 2$ (``Cosmic Noon"), $25 - 60 \%$ of the stellar mass density of the Universe may be contributed by dusty star-forming galaxies\footnote{There is no widely-adopted strict definition for DSFGs, but they are generally considered to have large SFRs $\gtrsim 100~M_\odot~{\rm yr}^{-1}$ and significant quantities of dust capable of reprocessing a sizable fraction of the ultraviolet (UV) emission from young stars into the infrared.
A subset of DSFGs, hyperluminous infrared galaxies (HyLIRGs) are more clearly defined, such that they exceed $L_{\rm IR} > 10^{13}~L_\odot$.} (DSFGs) and their quiescent descendants. 
Together, this implies that our understanding of the growth and evolution of stellar mass in the Universe from rest-frame UV/optical studies alone would be horribly incomplete.

Conversely, our knowledge of the typical stellar masses and properties of older, extant stellar populations of DSFGs is still woefully lacking, due in large part to intense dust attenuation at crucial rest-frame UV/optical wavelengths. Now, the sensitivity and resolution of JWST in this regime is beginning to facilitate novel interpretations of this frequently perplexing population\textemdash often also referred to as submillimeter galaxies (SMGs; \citealt{Blain:2002aa}), although these may more accurately be said to comprise a subset of DSFGs \citep{Casey:2014aa}.
They are widely believed to be the progenitors of massive quiescent, early-type galaxies seen at lower redshifts \citep{Lilly:1999aa, Farrah:2006aa, Toft:2014aa}, and given their association with galaxy overdensities and protoclusters \citep{Geach:2006aa, Chapman:2009aa, Clements:2014aa, Dannerbauer:2014aa, Casey:2015aa, Umehata:2015aa, Casey:2016aa, Clements:2016aa, Hung:2016aa, Miller:2018ab, Long:2020aa, Caputi:2021aa, Calvi:2023ab}, it is presumed that many will evolve into the ultra-massive population of brightest cluster galaxies (BCGs).

DSFGs at Cosmic Noon are typically considered to be dominated by compact nuclear starburst events, occurring within a thick shroud of dust. These have been postulated to signal the inside-out growth of galactic bulges that will persist in the massive descendents (e.g., \citealt{Gullberg:2019aa, Nelson:2019aa}). This activity may be driven by major mergers \citep{Hopkins:2008aa}, or more secularly through {\it in situ} star formation in gravitationally-unstable, gas-rich disks, followed by inward clump migration (e.g. \citealt{Bournaud:2007aa, Bournaud:2008aa, Elmegreen:2008aa, Ceverino:2010aa, Elmegreen:2011aa, Wuyts:2012aa}). 
However, for DSFGs, this picture often lacks the context of how the extant stellar mass (or even ongoing unobscured star formation) is distributed.
Many studies have found a factor of $>2$ larger effective radii in the rest-frame UV/optical vs. FIR continuum for DSFGs (including \citealt{Ikarashi:2015aa, Simpson:2015ab, Barro:2016aa, Hodge:2016aa, Pantoni:2021ab}).
However, these half-light sizes are subject to gradients in the mass-to-light (M/L) ratios, which may be the result of spatial variation in dust attenuation (e.g., \citealt{Byun:1994aa, Jansen:1994aa, Tuffs:2004aa, Mollenhoff:2006aa, Pastrav:2013aa, Popping:2022ac}), stellar population ages (e.g., \citealt{Carrasco:2010aa, van-Dokkum:2010aa, Tacchella:2015aa}), and metallicities \citep{Franx:1990aa, Molla:1997aa, Carrera:2008aa, Cresci:2010aa, Kewley:2010aa, Pilkington:2012aa}\textemdash or realistically, some combination of all three (e.g., \citealt{Keel:2001aa, Miller:2022ab}).

\begin{figure*}[ht!]
\centering
\includegraphics[width=0.99\textwidth]{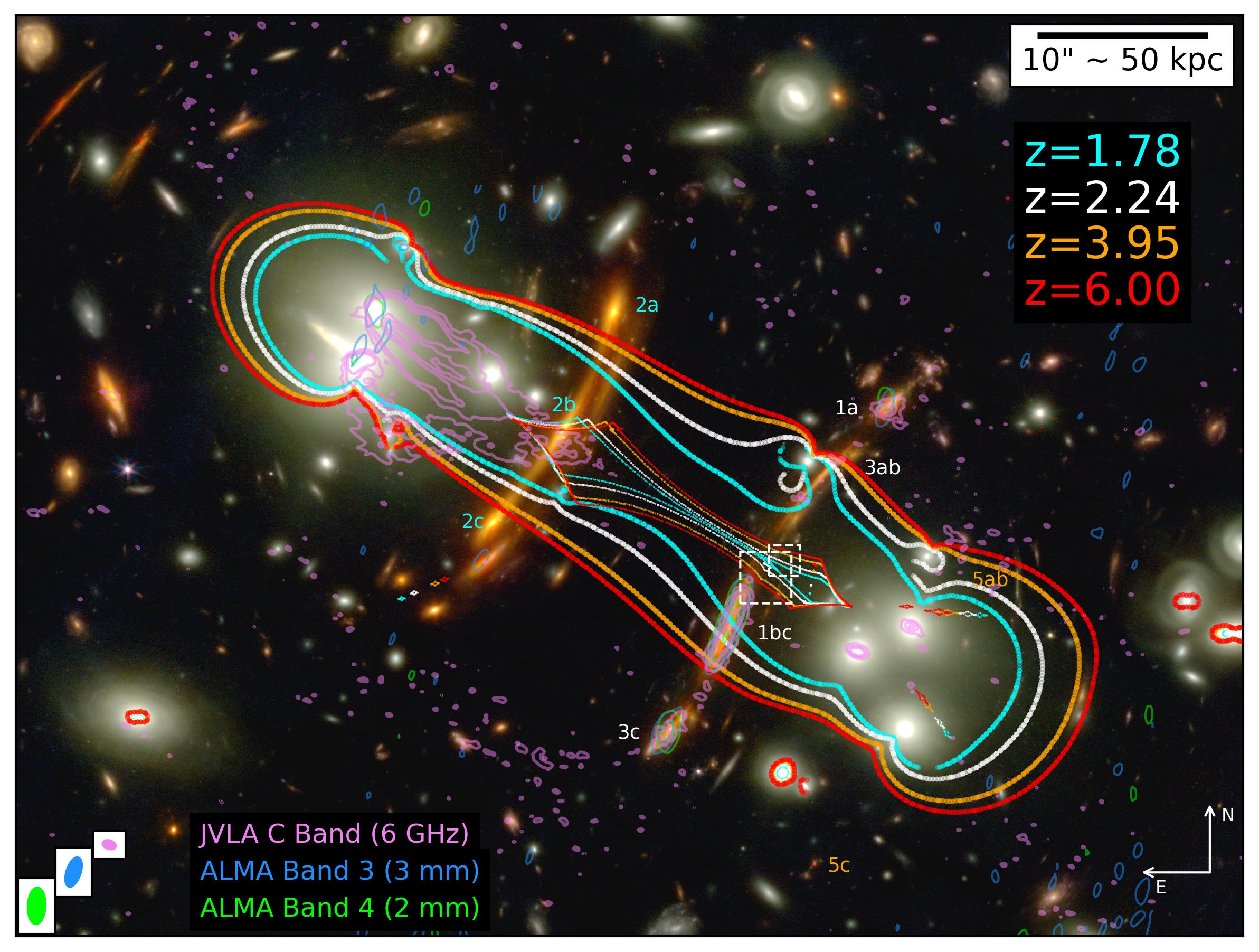}
\caption{
North-aligned RGB image of the G165 cluster, with (R, G, B) = (F444W + F356W, F277W + F200W, F150W + F090W). Colored curves show the best-fit model-derived lensing caustics at the source-plane redshifts of Arc system 2abc (the host of SN H0pe; $z=1.78$), Arc 1abc ($z=2.24$), 
Arc 5abc ($z=3.95$), 
and at $z=6$; see Fig. 2 of \citetalias{Frye:2024aa} for the full set of arcs.
Violet contours show JVLA 6 GHz continuum, blue contours show ALMA 3 mm continuum, and green contours show ALMA 2 mm continuum. Their respective synthesized beams are shown in the lower left. 
Spurious features in the ALMA contours towards the edge of the field of view are due to lower sensitivity near the edge of the primary beam. Arcs 1a, 1bc, and 3c are strongly detected at 2 mm, 3 mm, and 6 GHz. Arc 3ab yields only a marginal detection at 6 GHz, which is due to the lack of dust/radio continuum in the region of the galaxy that is lensed into this merging pair of arcs (see Fig.~\ref{fig:SP_DSFG}, for example).
Dashed white boxes show the source-plane locations of DSFG-1 and DSFG-3 (smaller northwest and larger southeast boxes, respectively), and match the fields of view shown in Fig.~\ref{fig:SP_1a_3c}. 
    \label{fig:lens_model}
}
\end{figure*}

In the era of JWST, great progress is being made in directly capturing the rest-frame near-IR structure of galaxies at Cosmic Noon and earlier, offering a significantly less biased view of the actual distribution of stellar mass (\citealt{Lang:2019aa, Chen:2022ae, Cheng:2022ab, Suess:2022aa, Abdurrouf:2023aa, Cheng:2023aa, Colina:2023aa, Kamieneski:2023ab, Kokorev:2023aa, Le-Bail:2023aa, Liu:2023aj, Rujopakarn:2023aa, Smail:2023aa, Amvrosiadis:2024aa, Kalita:2024aa, Sun:2024aa}), including their clumpiness, compactness, and lopsidedness.
This pixel-by-pixel comparison of stellar mass and star formation provides a key snapshot of the structural history of a galaxy, especially when supplemented with mapping of the molecular gas kinematics and dust continuum at comparable resolutions with facilities like ALMA.
For example, an examination of how the star-forming main sequence of galaxies translates on resolved scales at $z>1$ has now been made feasible (e.g., \citealt{Abdurrouf:2018aa}; hereafter \citetalias{Abdurrouf:2018aa}), along with its connection to fundamental scaling relations between the surface density of gas mass vs. star formation rate and between gas mass surface density vs. stellar mass surface density. While a full systematic study has not yet been carried out, results for a small number of objects can offer preliminary insights.

\begin{figure*}[ht!]
\centering
\includegraphics[width=0.97\textwidth]{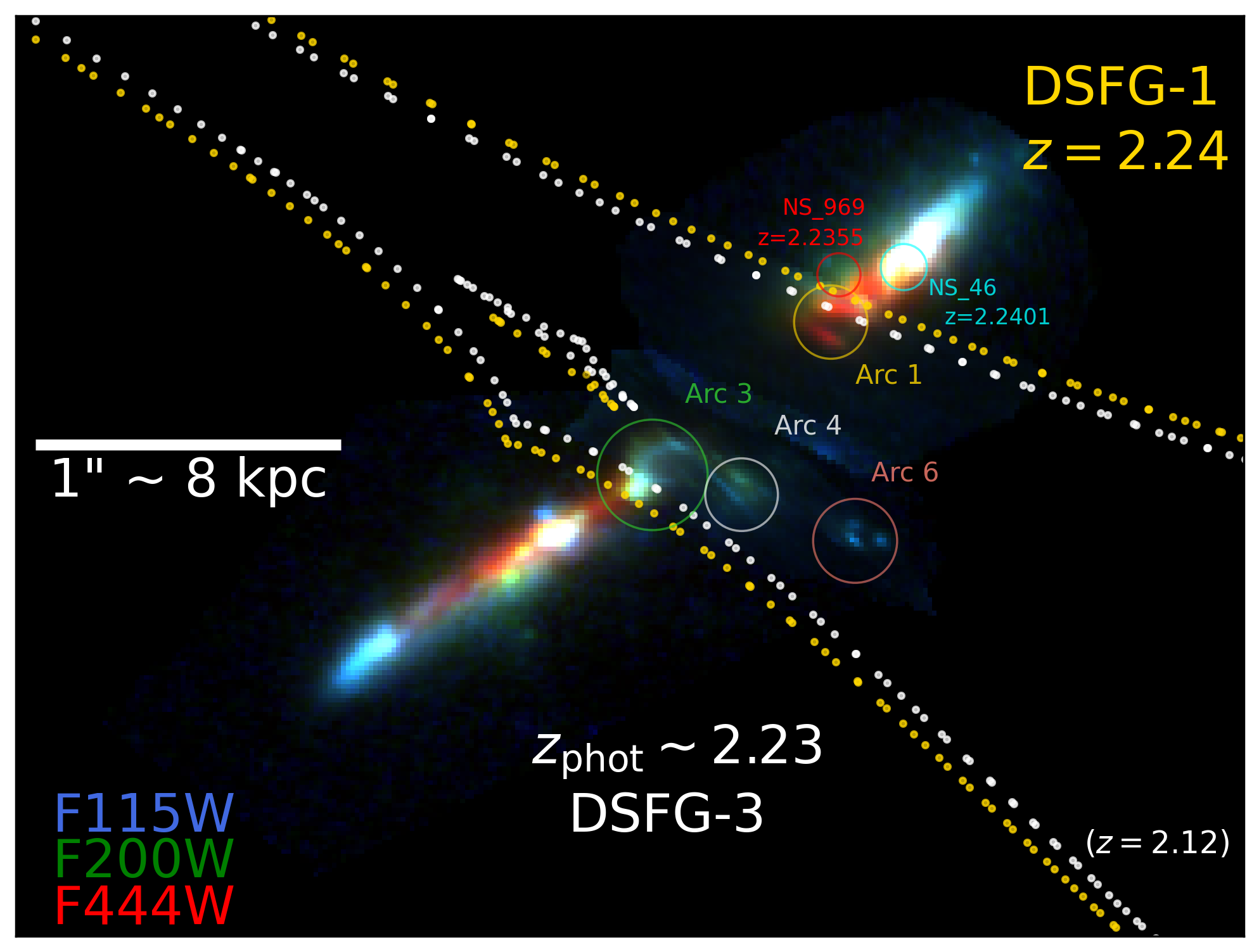}
\caption{
    Source-plane reconstruction of the DSFG system at $z=2.24$. The background RGB image shows JWST filters F444W, F200W, and F115W (respectively). The gold/white dotted line shows the lens model-derived caustic curve at $z=2.24 / z=2.12$ (the latter of which is a secondary redshift solution for DSFG-3, as discussed in Appendix \ref{sec:appendix1}). The portions of the source-plane interior to the caustics are triply-imaged; structures above and below the caustics in this figure are only singly-imaged (but still magnified by $\mu \gtrsim 5$, primarily in the direction perpendicular to the caustic).
    The constituent arcs (Arcs 1, 3, 4, and 6), as labeled by \citet{Frye:2024aa}, are indicated with circles, in addition to the NS\_969 and NS\_46 components that are associated with DSFG-1 (but at slightly different redshifts, indicating perhaps the presence of two orbiting clumps seen in projection). 
    \label{fig:SP_DSFG}
}
\end{figure*}

In this work, we focus on 
two DSFGs and a massive red galaxy at Cosmic Noon that have been strongly lensed by a massive galaxy cluster at $z=0.35$.
PLCK G165.7+67.0 (hereafter G165) was first identified by \citet{Canameras:2015aa} and \citet{Harrington:2016aa}\footnote{The cluster has also been denoted in literature as PJ112714.5.}, hereafter \citetalias{Harrington:2016aa}, as an exceptionally IR-bright source in {\it Planck} and {\it Herschel} surveys, but was first examined in-depth by \citet{Frye:2019aa} and \citet{Pascale:2022aa}. 
Multi-wavelength follow-up imaging at higher resolution revealed that the substantial submillimeter flux is driven by a highly-magnified DSFG (Table~\ref{tab:magnifications}) at $z=2.236$ \citepalias{Harrington:2016aa}, lensed into a giant arc by a galaxy cluster that appears to be undergoing a merger of two primary mass components (Fig.~\ref{fig:lens_model}, and \citealt{Pascale:2022aa}).
Hereafter, we refer to this known DSFG,
PLCK\_G165.6295+67.0026\footnote{Galactic coordinates are used to remain consistent with the name of the cluster, and the location of Arc 1a is chosen as it contains the entire object, in contrast with the fold image 1bc.},
as G165-DSFG-1 (or just DSFG-1),
given its appearance as Arcs 1a and 1bc, per the labeling introduced by \citet{Frye:2019aa}. With this work, we identify another strongly-lensed submillimeter source, which is believed to lie at the same redshift. Our current photometric redshift and lens model-predicted geometric redshift 
are consistent with $z\approx 2.23$, making it likely that the two are gravitationally interacting (see Fig.~\ref{fig:SP_DSFG}). For this second object, G165.6354+67.0087, we use the nickname G165-DSFG-3 (or DSFG-3), in accordance with its appearance in Arcs 3c and 3ab.
Through its inferred properties (including $L_{\rm IR} > 10^{12.6}$), it is clearly also safely classified as a DSFG.
\citet{Frye:2024aa}, hereafter \citetalias{Frye:2024aa}, 
revealed substantial candidate galaxy overdensities at $z\approx 1.8$ and at $z\approx 2.2$ using
a combination of spectroscopic redshift identifications with JWST NIRSpec and photometric redshifts from 8-filter NIRCam photometry.
The number of candidates identified for these structures ($N>150$) rivaled even the number of objects with photometric redshifts consistent with the G165 cluster itself.

As first reported by \citet{Frye:2023ab} and \citet{Polletta:2023aa}, initial inspection of the NIRCam observations immediately revealed a conspicuous multiply-imaged supernova in a set of arcs on the opposite side of the cluster from the DSFGs. Named SN H0pe, this serendipitous discovery was quickly followed-up with additional NIRCam imaging and NIRSpec spectroscopy to track the transient's light curve and classify its type. At $z=1.78$, it was confirmed to be one of the most distant known Type Ia supernovae. 
This discovery may not be purely coincidence: the fortuitous alignment of a rich collection of star-forming $z\sim 2$ objects with a massive and highly elliptical foreground cluster ($M_{\rm 600 kpc} = (2.6 \pm 0.3) \times 10^{14}~M_\odot$; \citetalias{Frye:2024aa}) along the line of sight provides ideal conditions for the detection of Type Ia and core-collapse supernovae. As the area of 
time-delay cosmography has 
become increasingly fruitful 
in recent years \citep{Treu:2022ac}, the G165 cluster is likely to endure as a subject of intense interest and legacy value for years to come.

This paper is organized as follows: in Section \ref{sec:data}, we outline the near-IR, radio, and mm-wave imaging used in this study. In Section \ref{sec:methodology}, we discuss our approach to lens modeling
and performing pixel-by-pixel SED fitting. The results of these analyses are outlined in Section \ref{sec:all_results}, and their implications are discussed in Section \ref{sec:discussion} and summarized in Section \ref{sec:summary}.

In this work, we adopt a $\Lambda$CDM cosmological model of $\Omega_m=0.3$, $\Omega_\Lambda = 0.7$, and $H_0 = 70~{\rm km}~{\rm s}^{-1}~{\rm Mpc}^{-1}$, and a \citet{Kroupa:2001aa} initial mass function (IMF). At the redshift of the cluster ($z=0.348$), 
the angular-to-physical size conversion is $1 \arcsec = 4.921$ kpc, and at the redshift of the DSFGs ($z=2.236$),
$1 \arcsec = 8.243$ kpc
\citep{Wright:2006aa}.
To comment on our notation:
throughout, we frequently use DSFG-1 and DSFG-3 as shorthand for 
PLCK\_G165.6295+67.0026 
and 
G165.6354+67.0087,
respectively.
These names themselves are introduced to avoid confusion when referring to the actual objects in the background of the cluster. To clarify, DSFG-1 is visible in its entirety in Arc 1a, while only a fraction of the intrinsic source is visible in the merging pair Arc 1bc. Likewise, DSFG-3 is visible entirely in Arc 3c, with a small portion of the source visible in the merging pair Arc 3ab.

\begin{figure*}[htb]
\centering
\includegraphics[width=0.49\textwidth]{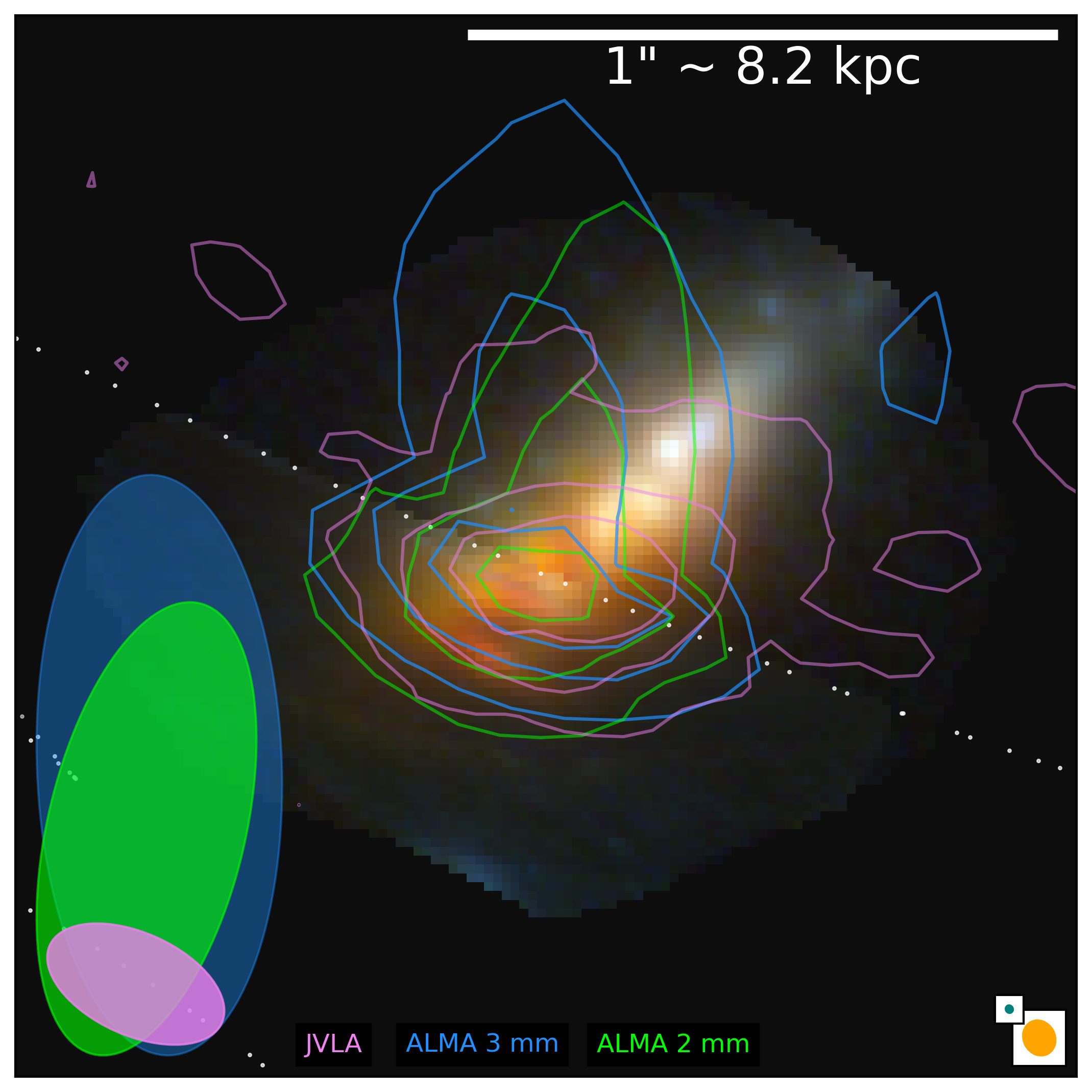}
\includegraphics[width=0.49\textwidth]{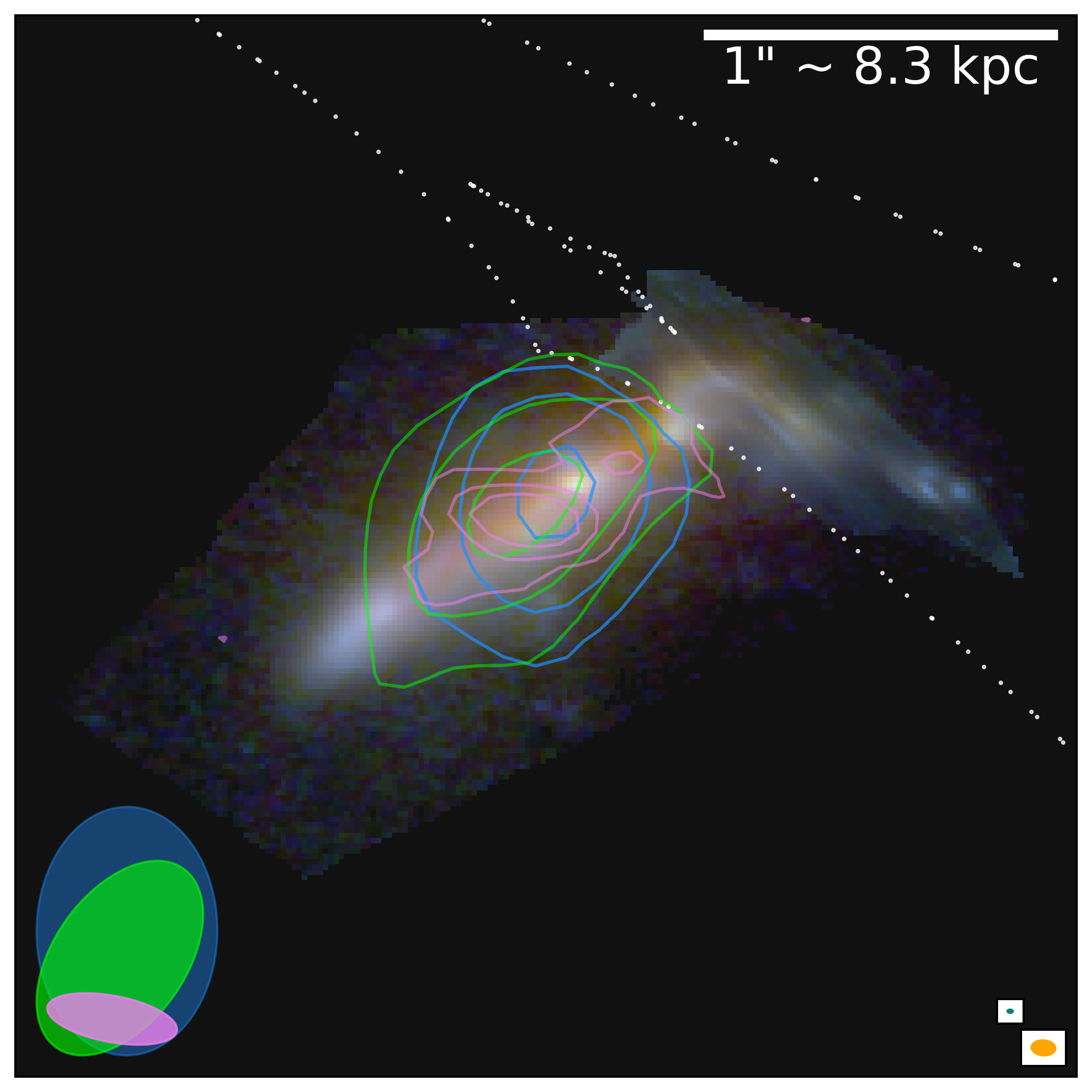}
\caption{
    Source-plane reconstruction of Arcs 1abc at $z=2.236$ ({\it left}) and 3abc at the lens model-preferred redshift of $z=2.12$ ({\it right}), with 8-filter RGB images scaled according to the Trilogy prescription. 
    The locations of these reconstructions are indicated in Fig.~\ref{fig:lens_model}.
    The filters are assigned to the red/green/blue channels as (R,G,B) = (F356W + F410M + F444W, F200W+F277W, F090W + F115W + F150W).
    Caustics for their respective redshifts are shown as white dotted curves, while contours show the reconstruction of JVLA 6 GHz (violet), ALMA 3 mm (light blue), and ALMA 2 mm (lime green), 
    as in Fig.~\ref{fig:lens_model}. 
    Representative source-plane elliptical PSF beams (\S \ref{sec:SP_reconstruction}) for the radio/mm images are shown in the lower left, and for the finest/coarsest NIRCam filters in the lower right (teal = F090W, orange = F444W). 
    Only the regions of the source inside the caustic curves (southeastern end of G165-DSFG1, northwestern end of G165-DSFG3) are included in their respective highly-magnified fold images, 1bc and 3ab (Fig.~\ref{fig:SP_high_mag}). However, by reconstructing together with the much lower-magnification images 1a and 3c, they are effectively convolved to lower resolution. 
    \label{fig:SP_1a_3c}
}
\end{figure*}

\section{Data} 
\label{sec:data}

\subsection{JWST NIRCam} 
\label{sec:jwst}

A more in-depth description of the JWST NIRCam data is given in \citet{Windhorst:2023aa} and \citet{Frye:2024aa}, but we provide relevant details here.
G165 was observed with NIRCam in three epochs: first through the Prime Extragalactic Areas for Reionization and Lensing Science (PEARLS) program (PID 1176; PI: Windhorst) on 2023 March 30 (Epoch 1), followed by two later epochs on 2023 April 22 (Epoch 2) and 2023 May 9 (Epoch 3) through a DDT program (PID 4446, PI: Frye), which were designed to track the light curve of SN H0pe following its discovery in Epoch 1.
As the exposure times of Epoch 1 were longer (see Table 1 of \citetalias{Frye:2024aa}), and 8 filters were used instead of 6, for simplicity we performed most of our analysis in this work on data from Epoch 1 only, in part to minimize variance in the PSF which could impact the spatially-resolved SED analysis. Moreover, the primary targets here are detected with very high significance, and do not benefit enormously from stacking additional exposures. However, 
the image families used in constraining the lens model (which are often faint) benefit greatly from this additional effective exposure time, and so the stacked images are employed for that purpose. 
Limiting magnitudes are $m_{\rm AB} \sim 28.4 - 28.5$ for the 4 short wavelength (SW) filters (F090W / F115W / F150W / F200W) and $m_{\rm AB} \sim 28.6 - 28.9$ for the 4 long wavelength (LW) filters (F277W / F356W / F410M / F444W); see Figure 6 of \citetalias{Frye:2024aa}.
Data were reduced in line with \citet{Windhorst:2023aa}, using the STScI JWST Pipeline\footnote{\url{https://github.com/spacetelescope/jwst}}  version 1.11.2 \citep{Bushouse:2023aa} and context file pmap\_1100, but with additional improvements in ``wisp" removal as described by \citet{Robotham:2023aa}. Images in this work were drizzled on to 30 milliarcsecond square pixels.

\subsection{ALMA Bands 3 and 4}
\label{sec:ALMA}

G165 was observed with ALMA band 3 (representative wavelength 2.7 mm, 111 GHz; hereafter referred to as 3 mm) and band 4 (2.2 mm, 138 GHz; hereafter referred to as 2 mm) as part of program 2021.1.00607.S (PI: R. Ca\~{n}ameras). These data were retrieved from the public ALMA archive\footnote{\url{https://almascience.nrao.edu/aq/}}.

Continuum multi-frequency synthesis images and spectral cubes were created using the Common Astronomy Software Applications ({\sc casa}) version {6.2.1} \citep{McMullin:2007aa}, with \texttt{robust}$=$0.5 Briggs weighting \citep{Briggs:1995aa}.
The unweighted baselines ranged from $0.03 - 0.3$ km (5th to 80th percentile).
This results in synthesized beams of $2 \farcs 0 \times 0 \farcs 9$ at PA$=-2.7\degrees$ for 3 mm (maximum recoverable scale of $18 \farcs 4$) and  $1 \farcs 6 \times 0 \farcs 7$ at PA$=-18.4\degrees$ for 2 mm (maximum recoverable scale of $15 \farcs 1$)\footnote{Imaging with $\texttt{robust}=-1$ (i.e., closer to uniform weighting) was tested in an attempt to improve angular resolution, but S/N was insufficient to yield useful results.}.
The continuum noise rms levels achieved are 22 $\mu$Jy/beam (3 mm) and 20 $\mu$Jy/beam (2 mm), both over a 3.3 GHz bandwidth.
Weather conditions were favorable, with mean precipitable water vapor PWV $=0.9$ mm for the 3 mm observations (executed on 2022 May 26) and PWV$=1.4$ mm for the 2mm observations (executed on 2022 May 28).

As the lensed objects of interest are spatially resolved, photometry was performed using the flood-filling algorithm {\sc blobcat} \citep{Hales:2012aa}.
In Table~\ref{tab:magnifications}, we compare the measured 2 mm and 3 mm fluxes against the 1 mm flux from the Large Millimeter Telescope (LMT) AzTEC, reported by \citetalias{Harrington:2016aa}, under assumption of a modified blackbody SED, and find relative consistency.

\subsection{VLA C Band} 
\label{sec:JVLA}

G165 was observed with the {\it Karl G. Jansky} Very Large Array in C band (4 - 8 GHz) in full polarization during the most extended A configuration through our program 18A-399 (PI: P. Kamieneski), which targeted 26 members of the {\it Planck} All-Sky Survey to Analyze Gravitationally-lensed Extreme Starbursts (PASSAGES) sample; see \citet{Kamieneski:2024aa} for additional details.
The target was observed on 2018 April 6 during a shared track (switching with another nearby object) lasting a total of 3.0 hrs, with an effective total on-source integration time of 1.1 hrs.
Weather conditions were favorable during the observations: the phase RMS measured by the Atmospheric Phase Interferometer at the VLA site was $\approx 2\degrees$.
Antenna baselines ranged from $0.8 - 36.6$ km, yielding a maximum recoverable scale of $9\arcsec$.
The sensitivity reached was $2.7~ \mu{\rm Jy}~{\rm beam}^{-1}$.
Natural weighting achieved a synthesized beam of $0 \farcs 65 \times 0 \farcs 37$ at PA$=72.2\degrees$.

The data were reduced using {\sc casa} v6.1.0. Initial flagging and calibration was carried out through the VLA Calibration pipeline v2018.1. The \citet{Hogbom:1974aa} CLEAN algorithm was used to deconvolve the image down to a $2\sigma$ threshold.
As with the ALMA images, photometry was performed using {\sc blobcat} \citep{Hales:2012aa}, also reported in Table~\ref{tab:SFR_IR}.
As discussed by \citet{Pascale:2022aa}, these data also reveal a number of point sources and narrow angle tails (e.g., \citealt{Rudnick:1977aa, Venkatesan:1994aa}), which offer some insight into the intracluster medium and dynamical state of the galaxy cluster.

\section{Methodology} 
\label{sec:methodology}

\subsection{Lens modeling} 
\label{sec:lens_modeling}

We constructed a parametric lens model for the G165 cluster using \textsc{lenstool}\footnote{\url{https://projets.lam.fr/projects/lenstool/wiki}} \citep{Kneib:1993aa, Kneib:1996aa, Jullo:2007aa, Jullo:2009aa}, which has been tested extensively for galaxy clusters (e.g., \citealt{Johnson:2014aa, Meneghetti:2017aa}).
The model
is constrained by {41} families of multiply-imaged features\textemdash of which {14} 
have spectroscopic redshift information \citep{Frye:2019aa, Pascale:2022aa, Frye:2024aa}, covering 4 distinct source planes. Some details of this model and the cluster members are presented in the appendix of \citet{Pascale:2024aa}, but we include additional information here. For this iteration of the model, only the image-plane positions were considered, which are reported in \citetalias{Frye:2024aa}. To estimate the astrometric uncertainty in these positions, 
multiple team members independently marked the positions for the known image family members, and found a median offset of $0\farcs 03$, which is consistent with the pixel scale used for identifying images. For the mass profiles, we used two cluster halo-scale profiles for the merging cluster components, in addition to perturbations to this large-scale potential in the form of the {165} identified cluster member galaxies at $z\approx 0.35$.
Each of these profiles was chosen to be a pseudo-isothermal elliptical mass distribution, or PIEMD \citep{Kassiola:1993aa}. 
To reduce the number of free parameters, the position of each cluster member was held fixed to the centroids of their luminous components. Likewise, the ellipticities and orientations were held fixed to match their morphologies in the F200W filter. The masses of the cluster members were scaled all together in a power law, according to their F200W flux, per \citet{Limousin:2005aa}. The power law was normalized such that a characteristic $L_\star$ galaxy at the redshift of the cluster ($m_{\rm F200W} = 17.0$ AB mag) has a velocity dispersion of $\sigma_0^\star = 120~{\rm km~s}^{-1}$, a core radius of $r_{\rm core}^\star = 0.15$ kpc, and a cut radius of $r_{\rm cut}^\star = 30$ kpc. All galaxies were scaled in mass based on their flux relative to $L_\star$. We settled on these choices after some trial and error, as it was determined that the current set of lensing evidence does not sufficiently constrain the masses of perturbing cluster galaxies, so they were instead held fixed. The choices of $\sigma_0^\star$, $r_{\rm core}^\star$, and $r_{\rm cut}^\star$ match those used in several other works that use \textsc{lenstool}, including \citet{Limousin:2008aa} and \citet{Johnson:2014aa}. However, one foreground galaxy near Arc 3ab, located 
at 
$(\alpha, \delta) =$ (11$^h$27$^m$14\fs 304, $+42^d$28$^m$32\fs 32),
was optimized independently from the other cluster members, given the large effect it has on the arc.
Its velocity dispersion $\sigma$ was kept as a free parameter, assuming a uniform prior of 0 to 250 km s$^{-1}$.

As for the two cluster-scale halos, their free parameters include centroid position, ellipticity, orientation, velocity dispersion, and core radius.
As their cut/truncation radii are poorly constrained, these were fixed to values of 1000 kpc. 
Uniform priors were used for these parameters, including a wide range of $\pm 5\arcsec$ in $x$ and $y$ position from the approximate center of both cluster cores, a very wide range in ellipticity from 0 to 0.85, a maximum velocity dispersion of 800 km s$^{-1}$, and a maximum core radius of $20\arcsec \approx 100$ kpc.
We also left as free parameters the redshifts for 16 background lensed image families for which no spectroscopic information is currently available.
To simplify the parameter space, we placed relatively tight priors on these, however, based on their photometric redshifts. 
This was done iteratively, where we widened or adjusted the redshift range if prior iterations of the model optimization resulted in any parameters clearly being overly limited by the prior boundaries. 
Still, the uncertainties in these ``geometric" redshifts (often as small as $\Delta z = \pm 0.01$) are likely drastic underestimates. 
We present the full set of best-fit parameter values, and the corresponding convergence map, in Appendix \ref{sec:appendix}. 
In this work, we adopt the median of each parameter's posterior distribution, which is only appropriate if the posteriors are well-behaved and approximately normally distributed (and especially only if the distributions are not multi-modal). 
We also provide the minimum-$\chi^2$ solutions, which are nearly always in close agreement with the median solution.

To determine magnifications, we used slightly different approaches for JWST vs. mm/radio images.
For the ALMA and VLA images, which are not contaminated by nearby foreground sources\footnote{While several of the cluster members are detected in 6 GHz continuum, they are sufficiently far from the lensed arcs of interest to have negligible impact.}, we measured magnifications simply by taking the ratio of image-plane to source-plane area.
For the JWST images, we measured representative magnifications using the F200W filter, and compute an average of the magnification map weighted by source-plane flux. 
In both cases, to include the relevant statistical uncertainties, we resampled 300 iterations uniformly from the MCMC-derived posterior distribution, resulting in an approximately Gaussian distribution of magnifications. The results for the Arc 1abc and Arc 3abc systems are summarized in Table~\ref{tab:magnifications}.
Any slight discrepancies beyond statistical uncertainty between the magnifications for rest-frame near-IR, far-IR, and radio continuum are likely due in part to very different observing parameters (e.g. PSFs), but also due to differences in source-plane distribution (as seen in Fig.~\ref{fig:SP_1a_3c}).

\subsubsection{Redshift of DSFG-3} 
\label{sec:z_Arc3}

Since DSFG-3 lacks a secure spectroscopic redshift, we left this as a free parameter in our lens model, and allowed the redshift to vary from $z=2.0 - 2.2$ in the final model. This range was chosen based on previous iterations of the model with wider redshift priors. This is also in line with the photometric redshift of 
$z_{\rm phot} = 2.23 \pm 0.09$
and with previous iterations of the lens model using a wider prior range, which suggested a geometric redshift of 
$z_{\rm geom} \approx 2.1$. 
The MCMC posterior distribution yields a quite precise value of $z_{\rm geom} = 2.11 \pm 0.01$ (although the uncertainty is likely underestimated). This narrow range is probably a result of the constraint offered by DSFG-1, which is lensed by a similar portion of the foreground cluster and has a known redshift from the LMT Redshift Search Receiver (RSR) and other single-dish sub-mm observations (\citetalias{Harrington:2016aa}; \citealt{Harrington:2021aa})
and from NIRSpec \citepalias{Frye:2024aa}.

There is no clear evidence of a second line detected in the LMT/RSR spectrum \citepalias{Harrington:2016aa}, which means either that the CO(3--2) line from DSFG-3 is blended with the line from DSFG-1 at $z=2.236$, or is undetected.
Indeed, this target has the widest linewidth ($\Delta V = 650 \pm 10$ km s$^{-1}$) of the sample
presented by \citetalias{Harrington:2016aa}, and a possible secondary component at $\sim -400$ km s$^{-1}$, giving weight to the hypothesis that the lines are blended.
Since DSFG-3 is inferred to have a similar stellar mass, size, and star formation rate as DSFG-1 (as we discuss in \S \ref{sec:all_results}), it is reasonable to expect comparable CO flux (or only $\sim 4$ times greater flux for DSFG-1 given its larger magnification, Table~\ref{tab:magnifications}). A non-detection by virtue of a faint line flux is thus unlikely.
However, the pointing center of the LMT observations coincides with the peak of the sub-mm flux, which is dominated by the high-magnification Arc 1bc. 
At an angular separation of $<11\arcsec$, Arc 3c would be near the edge of the primary beam ($\sim 13 \farcs 1$ at 106.86 GHz), so it is also somewhat possible that any lines were simply missed by the observations. 

We adopt the redshift of DSFG-3 to be $z=2.23$ in this work ($\Delta V \sim 600~{\rm km~s}^{-1}$ from DSFG-1), but we briefly discuss the possibility and implications of the secondary redshift solution, $z\approx 2.12$, in Appendix \ref{sec:appendix1}. Also, we do not hold fixed the redshift of DSFG-3 in our lens model given this lingering uncertainty. For the source-plane reconstructions shown in Fig.~\ref{fig:SP_DSFG} and \ref{fig:SP_1a_3c}, we ray trace back to $z=2.12$ for the highest-fidelity view of the source, but this has no impact on our interpretation of results.

\startlongtable
\begin{deluxetable*}{c|cccccccc}
\tablecaption{Model-derived magnifications and observed mm/radio fluxes. \label{tab:magnifications}}
\tablehead{
\colhead{Arc} & \colhead{$\mu_{2\mu{\rm m, JWST}}$} & \colhead{$\mu_{2{\rm mm}}$} & \colhead{$\mu_{3{\rm mm}}$} & \colhead{$\mu_{6{\rm GHz}}$} & \colhead{$S_{\rm 2 mm}$} & \colhead{$S_{\rm 3 mm}$} & \colhead{$S_{\rm 6 GHz}$} \\
\colhead{} & \colhead{} & \colhead{} & \colhead{} & \colhead{} & \colhead{[$\mu$Jy]} & \colhead{[$\mu$Jy]} &\colhead{[$\mu$Jy]} 
}
\startdata
Arc 1a  & $5.5 \pm 0.1$ & $7.6 \pm 0.2$    &  $8.1 \pm 0.2$ 	& $6.6 \pm 0.2$	& $260 \pm 30$	& $110 \pm 30$ & $190 \pm 10$	  	\\
Arc 1bc  & $42 \pm 1$ &  $44 \pm 3$   &   $41 \pm 4$	& $46 \pm 5$	& $1620 \pm 90$	& $860 \pm 60$ & $960 \pm 50$	  	\\
Arc 1abc & \textemdash & $28.4 \pm 2.4$    & $29.1 \pm 1.7$  	& $24^{+6}_{-3}$ 	& $1900 \pm 100^\dagger$	& $980 \pm 60^\dagger$ & $1150 \pm 50$	  	\\
\hline
Arc 3ab  &  $40 \pm 2$ &  \textemdash   &  \textemdash 	& $29 \pm 5$	& $<80$ & $<90$	& $39 \pm 5$	  	\\
Arc 3c  & $5.9 \pm 0.1$ & $6.8 \pm 0.2$    &  $7.5 \pm 0.3$ 	& $5.9 \pm 0.2$	&	$620 \pm 40$ & $310 \pm 30$ & $132 \pm 8$	  	\\
Arc 3abc & \textemdash & $7.3 \pm 0.7$     &  $7.4 \pm 0.6$ 	& $6.8^{+1.1}_{-0.4}$ 	& $620 \pm 40^\dagger$	& $310 \pm 30^\dagger$ & $171 \pm 9$	  	\\
\enddata
\tablenotetext{^\dagger}{The combined flux of of Arcs 1abc and 3abc at 2 mm ($S\approx2.5$ mJy) is consistent with that expected from a modified blackbody (dust emissivity index $\beta = 1.8$) with $S_{\rm 1mm} = 24 \pm 2.0$ mJy \citepalias{Harrington:2016aa}, which is $S_{\rm 2mm}\approx 2.5$ mJy. The observed flux at 3 mm (1.3 mJy) exceeds the modified blackbody prediction (0.5 mJy), possibly owing to contamination from synchrotron emission.}
\tablecomments{Magnifications are measured for the F200W filter (as representative for all of the NIRCam filters) by taking the source-plane flux-weighed average over the magnification map. For the mm and radio images, magnifications are measured by simply taking the ratio of image-plane to source-plane areas.
}
\end{deluxetable*}

\subsubsection{Source-plane reconstruction} 
\label{sec:SP_reconstruction}

The reconstruction of all images in this work in the source plane does not impose regularization or parameterization of the intrinsic structure of the source. Instead, the observed images are ray-traced to the source plane pixel-by-pixel through the {\tt cleanlens} function of {\sc lenstool}, with the image-plane oversampled by a factor of 4 and the source-plane by a factor of 2. Further details of our specific procedure may be found in Section 3.3 of \citet{Kamieneski:2024aa}.
Fig.~\ref{fig:SP_1a_3c} reconstructs the full structure of G165-DSFG-1 and G165-DSFG-3 by including both low- (1a and 3c) and high-magnification (1bc and 3ab) arcs. 
Effectively, this simple approach convolves the high-magnification arcs with the PSF of the low-magnification counterparts (e.g., \citealt{Sharma:2018aa}).
We use this version to estimate the source-plane size of the dust continuum, where we use the {\sc casa} {\sc imfit} task to fit the emission with 2-d Gaussian profiles (and adopt a minimum 10\% fractional uncertainty). 
Despite the smaller beam size of the 3 mm observations,
these dust sizes are measured for the 2 mm images given their higher S/N, as reported in 
Table~\ref{tab:SFR_IR}.
Additionally, the possible contamination from synchrotron at 3 mm is expected to result in slightly larger sizes than from thermal dust alone. 
In 
Appendix \ref{sec:appendix2}, 
we show the reconstructions from including only the high-magnification images, which are at essentially the finest achievable resolution.
There is general agreement with the structure seen in Fig.~\ref{fig:SP_1a_3c}, adding confidence to this reconstruction.

RGB images of the 8 NIRCam filters are constructed according to the Trilogy\footnote{\href{https://www.stsci.edu/~dcoe/trilogy}{https://www.stsci.edu/\(\sim \)dcoe/trilogy}} prescription \citep{Coe:2012aa}. 
In all cases, we also show representative source-plane PSFs to illustrate the approximate resolution of the reconstructions. 
These are created by placing a simulated PSF or beam at a location in the image plane and ray-tracing it to the source-plane in the same manner as the image itself (then refitting with a 2-dimensional Gaussian).
The distorted source-plane PSFs arise because the image-plane PSFs are relatively circularly symmetric to first order, but this symmetry is lost through the ray-tracing process.
However, as there is a gradient in magnification over each of the arcs (which is much more extreme for 1bc and 3ab), the source-plane PSF is finer in regions closer to the caustic curve than those further away. 
While the source-plane reconstructions assist in our interpretation of these objects, most of this work computes properties in the image plane and applies a magnification correction (or relies on magnification-invariant properties). This approach is less subject to any pitfalls of the intricacies of source-plane PSFs.

The approximate source-plane major axes of the ALMA 3 mm and 2 mm beams (roughly oriented along the major axis of the galaxies) are $1 \farcs 0 \approx 8.2$ kpc and $0 \farcs 8 \approx 6.6$ kpc for G165-DSFG-1, and $0 \farcs 7 \approx 5.8$ kpc and $0 \farcs 6 \approx 5.0$ kpc for G165-DSFG-3.
For VLA 6 GHz, the source-plane beam minor axes are approximately oriented along the galaxies' major axes, which are $0 \farcs 2 \approx 1.6$ kpc for G165-DSFG-1, and $0 \farcs 1 \approx 0.83$ kpc for G165-DSFG-3.
These are indicated in Fig.~\ref{fig:SP_1a_3c}.

\subsection{Pixel-by-pixel SED fitting with {\sc piXedfit} and {\sc bagpipes}}
\label{sec:pixelbypixel}

As the cluster lensing provides an areal magnification and a flux amplification for the DSFGs, it becomes easier to ascertain the spatially-resolved properties of their stellar populations through so-called pixel-by-pixel SED fitting.
To accomplish this, we used the publicly-available package {\sc piXedfit}\footnote{\url{https://pixedfit.readthedocs.io/en/latest/}} (\citealt{Abdurrouf:2021ab}; also \citealt{Abdurrouf:2022aa, Abdurrouf:2022ab, Abdurrouf:2023aa}) to bin pixels such that each satisfies a S/N condition in all filters, while preserving similarity in SED shape between pixels within each bin. In brief, starting with the brightest unbinned pixel, each bin is required to be larger than the F444W PSF FWHM ($0 \farcs 14$). If a S/N threshold for each filter is met (a conservative S/N$=20$, in our case), the bin is established; otherwise, the radius of the bin is increased. Before adding the new pixels from the larger radius to the bin, their SED shape is confirmed to be consistent through a $\chi^2$ calculation.
This process then repeats, starting with the brightest pixel not already in a bin, and terminates once no additional bins can be constructed. Full details are provided in Section 3.3 of \citealt{Abdurrouf:2021ab}.

With the bin-wise photometry collected for the 8 NIRCam bands ($0.9 - 4.4~\mu$m), all convolved to the resolution of F444W, we fitted the SED of each bin with {\sc bagpipes} (Bayesian Analysis of Galaxies for Physical Inference and Parameter EStimation; \citealt{Carnall:2018aa}). The star formation history (SFH) was parameterized as an exponentially declining or ``tau" model, where star formation promptly begins at time $T_0$ and decays relative to a timescale $\tau$. The stellar population age for each bin could vary from 0.1 Gyr to the age of the Universe at the given redshift, and $\tau$ was allowed to vary from $0.3 - 10$ Gyr. 
We consider this simple smooth (i.e. not bursty) parameterization of the SFH to be appropriate given the small physical scale of each bin. In other words, given the high spatial resolution, these sub-kpc regions are 
perhaps 
best described with low time resolution (to avoid overfitting), and thus a less stochastic star formation history than a galaxy in aggregate.
The total mass formed was allowed to vary widely, from ${\rm log}_{10}(M / M_\odot) = 1 - 15$, and metallicity could vary from 0 to 3.5 $Z_\odot$.
The standard nebular component in the {\sc bagpipes} fits was accounted for through the ionization parameter, log($\cal{U}$), which was allowed to vary from $-4$ to $-2$, pre-computed from {\sc cloudy} photoionization models \citep{Ferland:2017aa}. While higher values (up to $-1$) can be found for individually resolved \hii\ regions (e.g., \citealt{Snijders:2007aa}), an upper limit of $-2$ is typically sufficient for starbursts resolved on kpc scales (e.g., for M82's inner 500 pc; \citealt{Thornley:2000aa, Forster-Schreiber:2001aa}).
A \citet{Calzetti:2000aa} dust attenuation model was imposed, with attenuation allowed to vary from $A_V = 0 - 3$, which we found to be a sufficient range\footnote{To test this, we examined the 84th percentile of the derived $A_V$ posteriors for the DSFG-1 and DSFG-3 bins. Only 12\% and 8\% (respectively) exceeded $A_V > 2.5$.}.

\begin{figure*}[ht!]
\centering
\includegraphics[width=0.9705\textwidth]{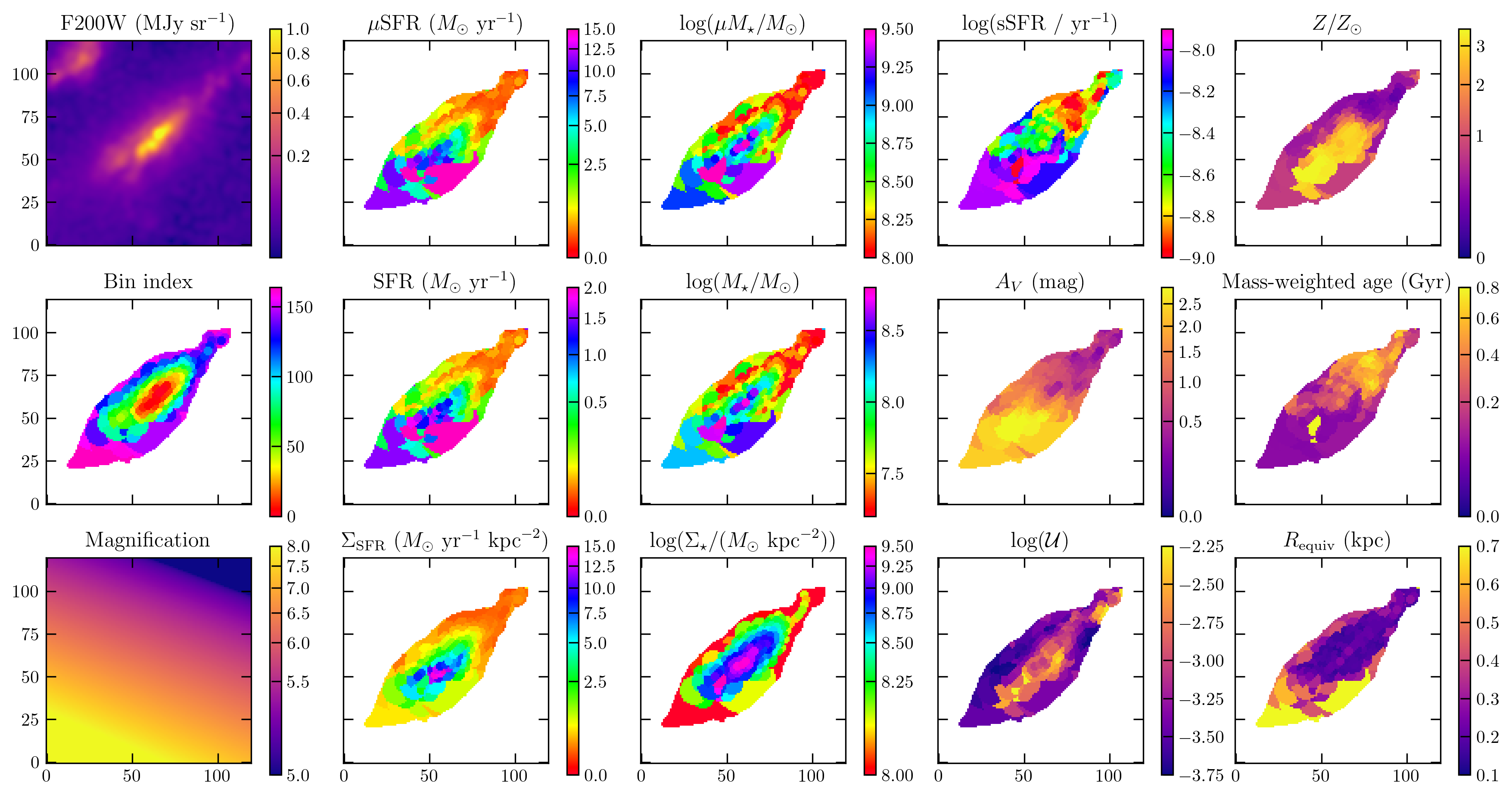}
\caption{
    Spatially resolved properties of DSFG-1 (Arc 1a), including image-plane properties $\mu {\rm SFR}$ and $\mu M_\star$, and their magnification-corrected intrinsic values, SFR and $M_\star$. As bin size varies, these properties are best assessed through their respective surface densities, $\Sigma_{\rm SFR}$ and $\Sigma_\star$, which are also unaffected by magnification, as are the specific star formation rates, sSFR. Dust attenuation $A_V$, ionization parameter log($\cal{U}$), stellar metallicity $Z$, and mass-weighted ages are likewise assumed to not be affected by lensing magnification. These properties are inferred through SED fitting with {\sc bagpipes} at a fixed redshift, $z=2.236$.
    The bin index map shows the order in which the {\sc piXedfit} bins are allocated.
    The $R_{\rm equiv}$ panel shows the radius of a circle with equivalent area to each bin corrected for areal magnification factor $\mu$, such that $R_{\rm equiv} \equiv (A \mu^{-1} / \pi)^{0.5}$.
   The field of view of each panel is $120\times 120$ pix, or $3 \farcs 6 \times 3 \farcs 6$. 
    \label{fig:pixbin_1a}
}
\end{figure*}

\begin{figure*}[ht!]
\centering
\includegraphics[width=0.9705\textwidth]{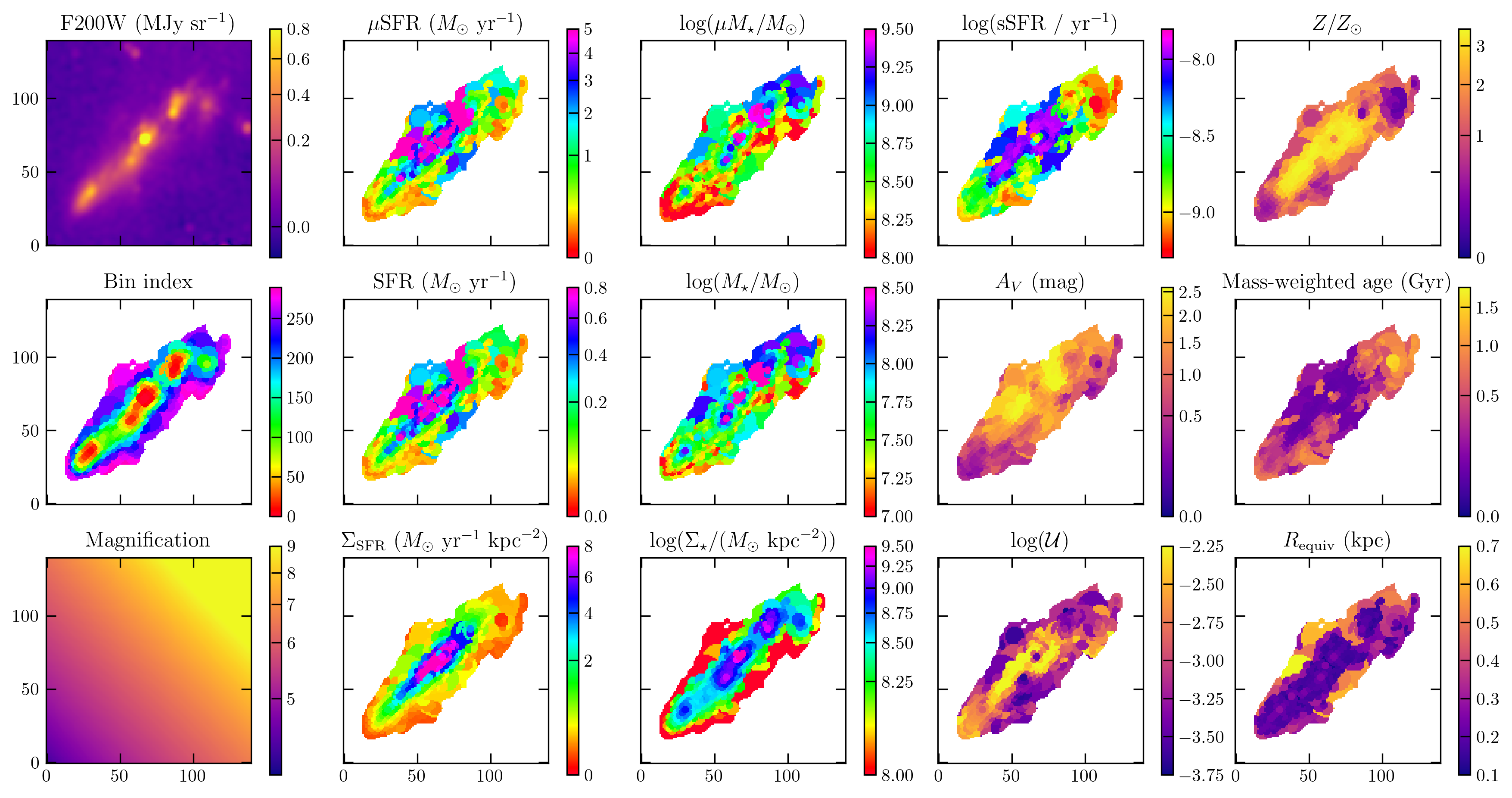}
\caption{
    Spatially resolved image-plane properties for DSFG-3 (Arc 3c), 
    as with Fig.~\ref{fig:pixbin_1a}. 
    Redshift is held fixed at an assumed $z=2.23$.
    The field of view is $140\times 140$ pix, or $4 \farcs 2\times 4 \farcs 2$.
    \label{fig:pixbin_3c}
}
\end{figure*}

\section{Results} 
\label{sec:all_results}

\subsection{The complex, galaxy-wide assembly of stellar mass in two DSFGs at Cosmic Noon}
\label{sec:resolved_SED_maps}

One of the fundamental questions we have sought to address through our resolved pixel-by-pixel SED modeling of G165-DSFG-1 and G165-DSFG-3 is how their significant ongoing star formation is spatially distributed.
If DSFGs are primarily powered by violent major mergers of massive galaxies, as for local ultra-luminous infrared galaxies (ULIRGs; \citealt{Armus:1987aa, Soifer:1987aa, Sanders:1988aa, Sanders:1996aa, Lonsdale:2006aa}), is it reasonable to expect that their burst of star formation might be confined only to nuclear regions, with shocks and tidal forces driving gas inwards (e.g., \citealt{Hernquist:1989aa, Stockton:1990aa, Barnes:1991aa,Barnes:1992aa})?
Or if other modes of extreme star formation are prevalent, such as gravitational instabilities within massive gas-rich
disks\textemdash fueled continuously by cold gas streams (e.g., \citealt{Keres:2005aa, Dekel:2009ab, Dave:2010aa})\textemdash would we expect to find a greater number of global star formation events, reaching well beyond the nucleus?

Numerous recent works have demonstrated that the effective radii of DSFGs at rest-frame far-IR are compact ($\sim 1 - 2$ kpc) and smaller than at rest-frame UV/optical (e.g., \citealt{Simpson:2015ab, Ikarashi:2015aa, Barro:2016aa, Hodge:2016aa, Tadaki:2017ab}).
On the other hand, some of the most luminous DSFGs\textemdash so-called hyperluminous infrared galaxies (HyLIRGs), which exceed $L_{\rm IR} > 10^{13}~L_\odot$\textemdash may be more extended, in part explaining their ability to sustain SFRs above $1000~M_\odot~{\rm yr}^{-1}$ without significantly breaching the Eddington limit (e.g., \citealt{Bussmann:2013aa, Kamieneski:2024aa}). 
In a large sample of 
{\it Herschel}-selected candidate lensed DSFGs, \citet{Borsato:2024aa} found 
rest-optical sizes 
to be $\approx 3$ times 
smaller
than far-IR sizes, determined by \citealt{Enia:2018aa} to be $\sim 2 - 3$ kpc.
However, these isophotal rest-optical sizes are SNR-dependent, and the far-IR sizes are derived from lower-resolution Submillimeter Array imaging, both of which may bias results towards smaller optical-to-FIR size ratios.

In this work, we find 2-mm dust continuum 
effective 
radii of $R_{e, 2{\rm mm}} = 1.6 \pm 0.2$ kpc and $2.1 \pm 0.2$ kpc for DSFG-1 and DSFG-3, respectively (Table~\ref{tab:SFR_IR}).\footnote{At 3 mm, we find $\sim 1.5\times$ larger sizes of $R_{e, 3{\rm mm}} = 2.6 \pm 0.3$ for both DSFGs, consistent with contribution from synchrotron \citep{Thomson:2019aa}.}
These values are quite consistent with other members of PASSAGES \citep{Kamieneski:2024aa} and other lensed DSFGs (e.g., \citealt{Bussmann:2013aa}), 
and in line with what
might be expected from the size-luminosity relation derived by \citet{Fujimoto:2017aa} and their intrinsic luminosities of ${\rm log}[L_{\rm IR} / L_\odot] \approx 12.5$.
Given their apparent elongated morphology suggestive of high-inclination disks (Fig.~\ref{fig:SP_DSFG}), this seems to suggest a large region of ongoing obscured-mode star formation. However, higher-resolution dust continuum imaging is required to comment on this more robustly, as Arcs 1a and 3c are only marginally resolved at 2 mm.

The UV-NIR photometry captures primarily the unobscured mode of star formation, which is likely to be a small fraction of the total (obscured plus unobscured) SFR$_{\rm UV + IR}$ for such massive, highly star-forming galaxies \citep{Whitaker:2017aa}.
Regardless, it is still informative to investigate the spatial scales over which unobscured star formation is occurring, as an insight into the driving mechanisms and present evolutionary stages of the galaxies.
For example, galaxies with a higher central specific SFR (sSFR$\equiv {\rm SFR} / M_\star$) are likely undergoing inside-out growth, by which a spheroidal bulge component is built up {\it in situ} through gas compaction (e.g., \citealt{Dekel:2014aa, Zolotov:2015aa, Tacchella:2016aa, Tacchella:2018aa}).
In later stages, sSFR may be elevated in outer regions of the galaxy, pointing to inside-out quenching. This may be driven by morphological quenching \citep{Martig:2009aa, Genzel:2014aa, Tacchella:2015aa}, where the central spheroidal component renders the disk stable against gravitational fragmentation of the molecular gas, even without requiring energetic feedback from a central starburst or active galactic nucleus (AGN).

\begin{figure*}[ht!]
\centering
\raisebox{30pt}{
\includegraphics[width=0.36\textwidth]{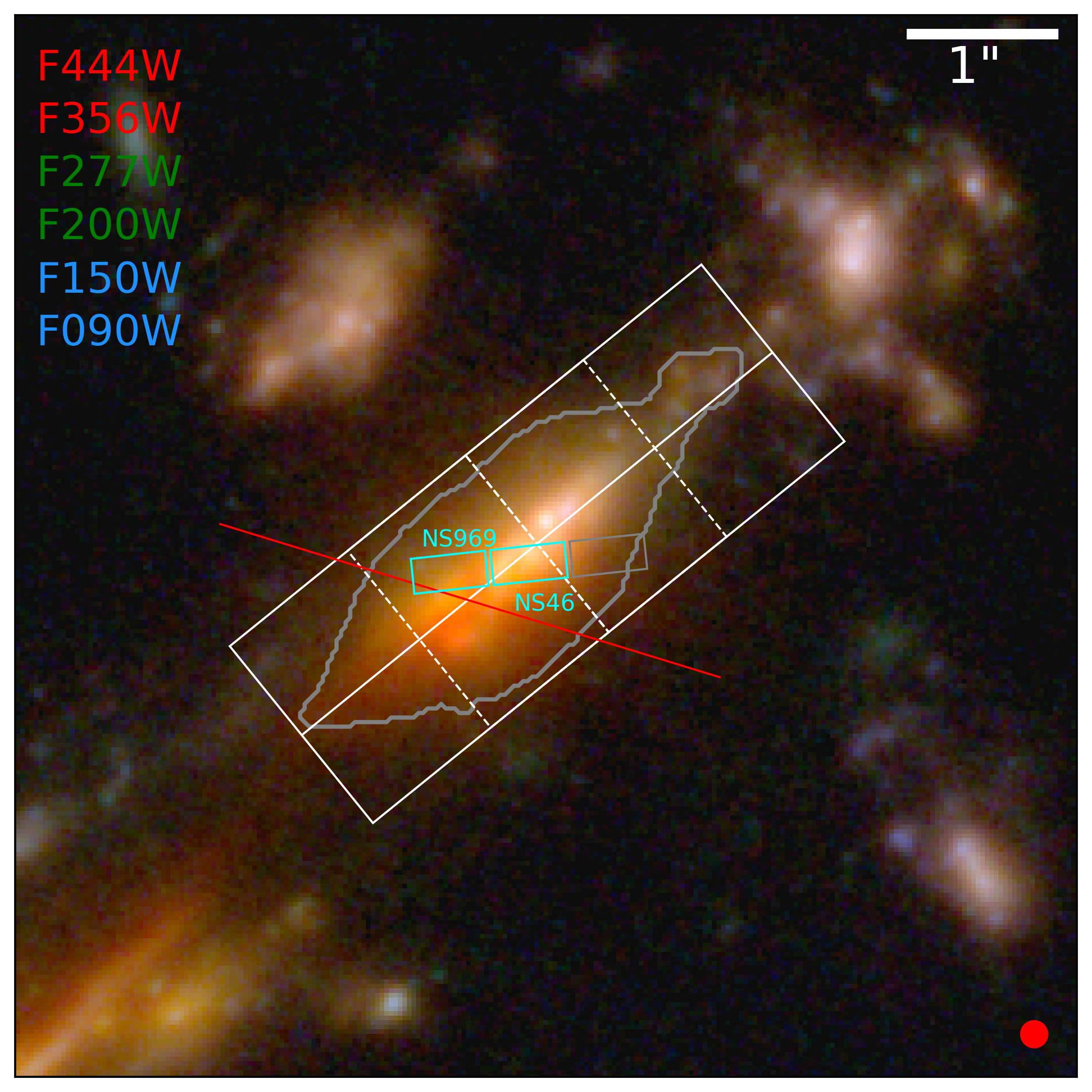}}
\includegraphics[width=0.62\textwidth]{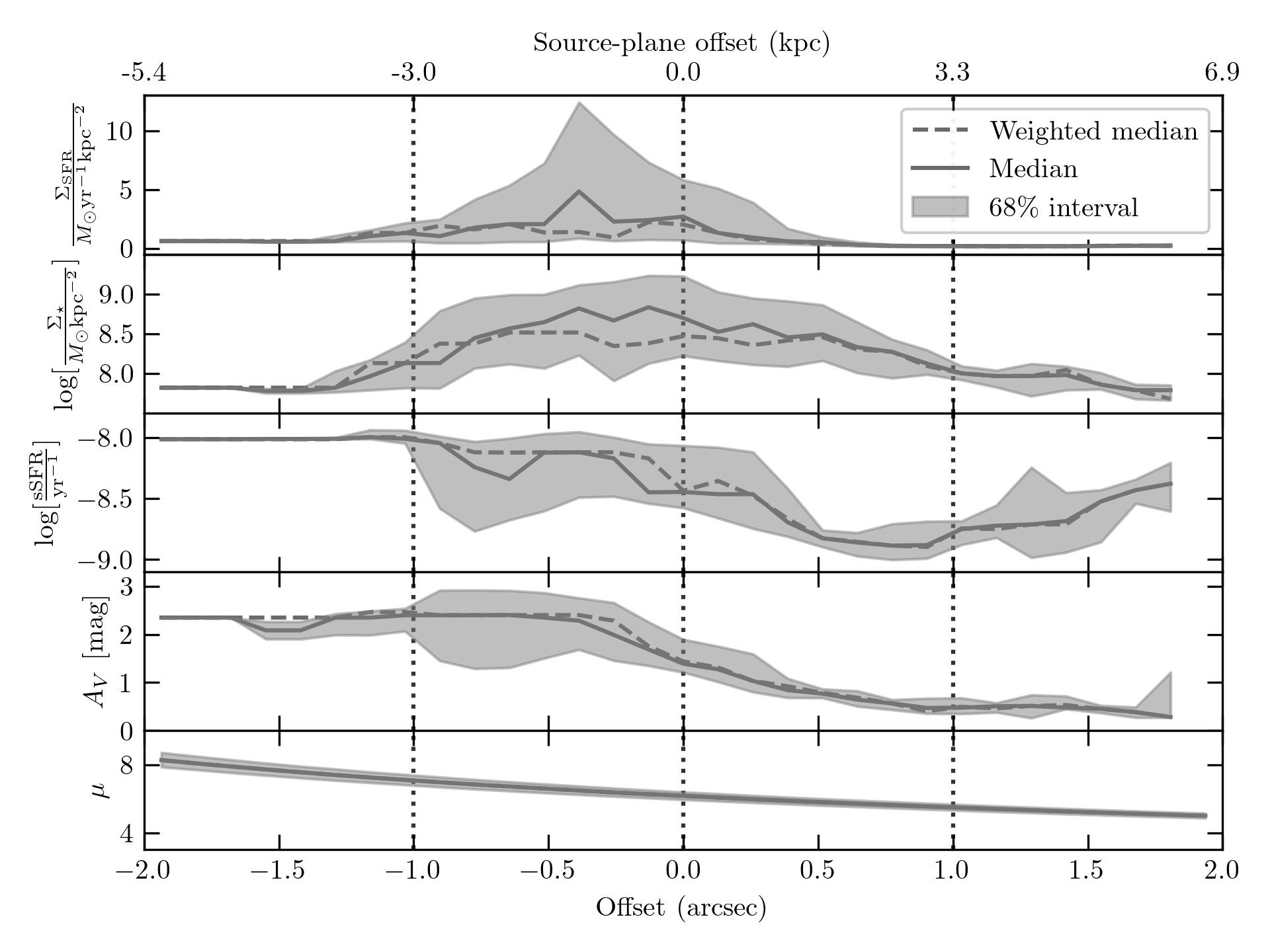}
\caption{
Linear 
profiles ({\it right}) of $\Sigma_{\rm SFR}$, log$(\Sigma_\star)$, log(sSFR), $A_V$, and magnification $\mu$ created by taking bins along the 
major axis, shown over the RGB image-plane image of Arc 1a ({\it left}). Dashed white lines perpendicular to the major axis correspond to those in the right panel (for the center and at an offset of $\pm 1\arcsec$, where positive is northwest).  
The relative source-plane offsets in physical units (kpc) corresponding to the image-plane offsets of $\pm 1\arcsec$ and $\pm 2\arcsec$ are shown at the top, which are asymmetric due to the gradient in magnification.
The F444W PSF is shown as a red beam in the left panel. The gray contour in the left panel shows the border of the bins used to make the profiles (as seen also in Fig.~\ref{fig:pixbin_1a}). 
In each bin along the major axis, we show the weighted median (dashed curve) for each property, which does not account for overcounting pixels within each bin, so they are effectively weighted by the area of each SED fit bin within the major axis bin. The median (solid curve) counts each SED bin only once. The inner 68\% interval is shown as shaded curves. 
A red line denotes where the source crosses the caustic curve according to our lens model; in this case, regions south of the line is triply-imaged, and north are singly-imaged. 
The locations of the NIRSpec MSA slits from \citetalias{Frye:2024aa} are shown in cyan and labeled for reference.
As a caveat, properties at an offset $> +1 \farcs 0$ are less reliable, given possible blending with a nearby object of unknown redshift.
    \label{fig:slit_1a}
}
\end{figure*}

\begin{figure*}[ht!]
\centering
\raisebox{30pt}{
\includegraphics[width=0.36\textwidth]{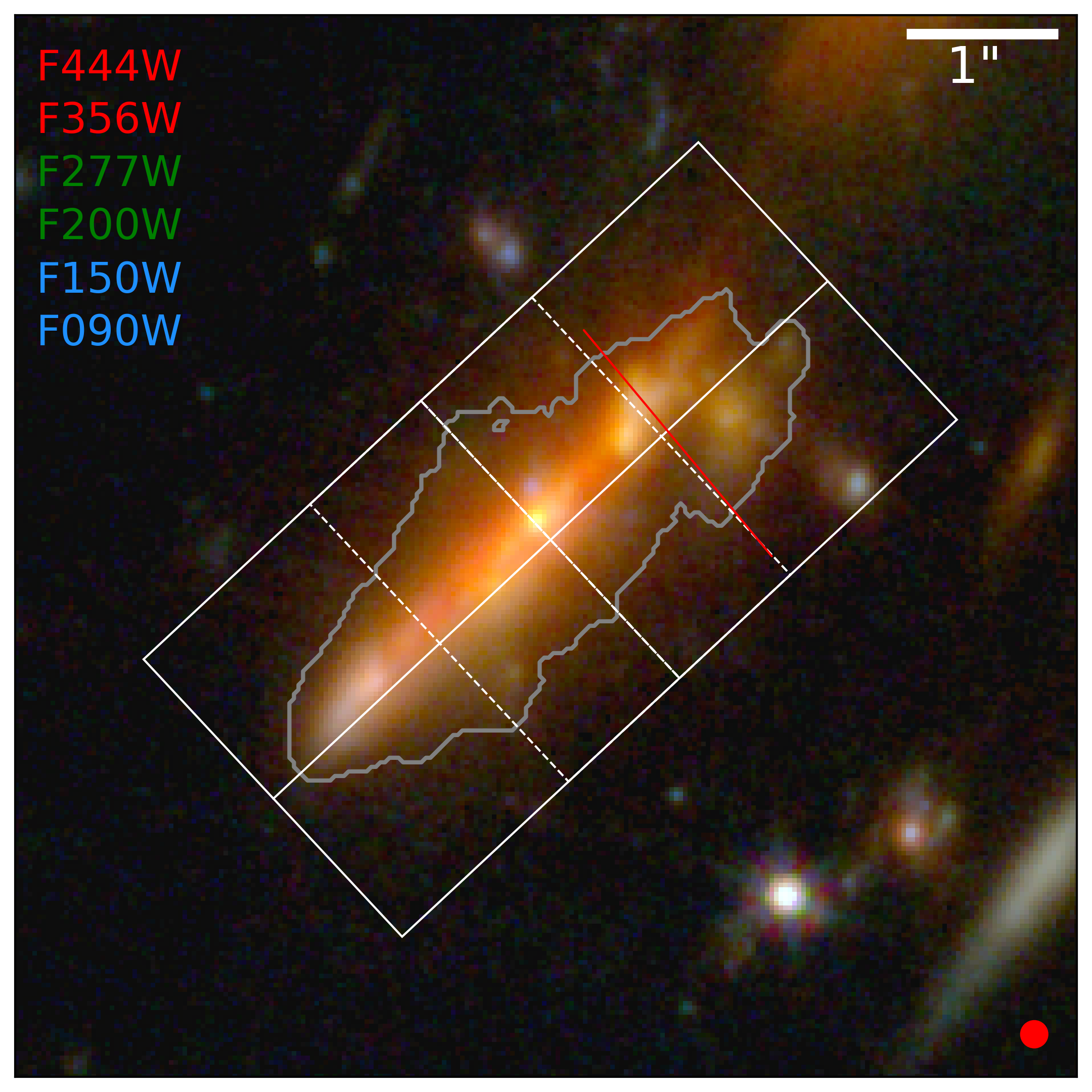}}
\includegraphics[width=0.62\textwidth]{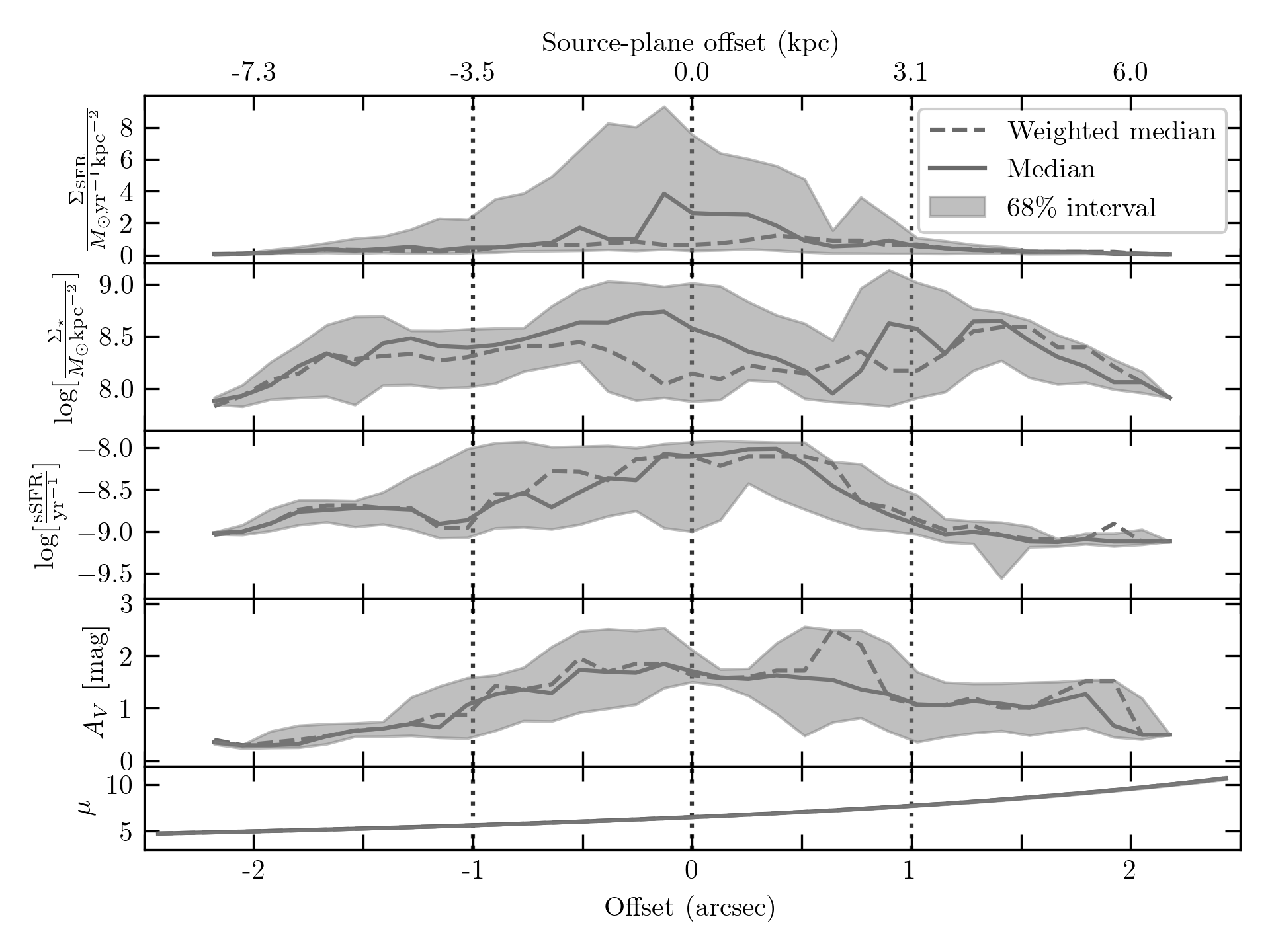}
\caption{
As with Fig.~\ref{fig:slit_1a}, but for Arc 3c. 
In this case, the portion of the object northwest of the red line is triply-imaged, and the rest is singly-imaged.
    \label{fig:slit_3c}
}
\end{figure*}

Figs.~\ref{fig:pixbin_1a} and \ref{fig:pixbin_3c} show the spatially-resolved bin-wise properties derived through SED fitting with {\sc piXedfit} and {\sc bagpipes}, including corrections for magnification from our lens model.
Given the variation in bin size, we primarily examine the surface densities of SFR and $M_\star$, or $\Sigma_{\rm SFR}$ and $\Sigma_\star$. 
As gravitational lensing conserves surface brightness, these properties do not need to be corrected for magnification.
We also perform the same analysis for the high-magnification Arcs 1bc and 3ab as a consistency check in 
Appendix \ref{sec:appendix2}.
These arcs arise only from a portion of the galaxy that crosses inside the caustics (see Fig.~\ref{fig:SP_DSFG} and the red demarcation line in left panels of Figs.~\ref{fig:slit_1a} and \ref{fig:slit_3c}). 
For that reason, we focus here primarily on Arcs 1a and 3c.

For G165-DSFG-1 (Arc 1a), there appears to be a single central peak in $\Sigma_{\rm SFR}\approx 15~M_\odot~{\rm yr}^{-1}~{\rm kpc}^{-2}$, and a roughly axially symmetric radial decrease (except for perhaps lower $\Sigma_{\rm SFR}$ in the northwest region). 
The stellar mass surface density $\Sigma_\star$ also peaks in the center of the arc\textemdash log$[\Sigma_\star / (M_\odot~{\rm kpc}^{-2})]\approx 9.5$\textemdash although the peak is more consistent with an elongated ellipse parallel to the major axis of the arc. In fact, there appear to be two peaks in $\Sigma_\star$, with a slight dip between them corresponding to the peak in $\Sigma_{\rm SFR}$. 
Of possible relevance is the discovery by \citetalias{Frye:2024aa} that Arc 1a is actually composed of components at slightly different redshifts seen in projection, as indicated in the source-plane reconstruction in Fig.~\ref{fig:SP_DSFG}.
The redder, southeasternmost portion (NIRSpec aperture NS\_969) is found to be at $z=2.2355 \pm 0.0003$, while the bluer region $0 \farcs 2$ (1.6 kpc in projection) more to the northwest (NS\_46) is at $z=2.2401 \pm 0.0002$ (i.e., a rest-frame redshift difference of $\approx$ 400 km s$^{-1}$ relative to the southeast clump).
Despite this, we still refer to the system as G165-DSFG-1, as it is not possible to segregate them with the current data, and as it seems likely that they will ultimately merge into one object. 

There is 
uncertainty in 
the spectroscopic redshift of G165-DSFG-3, 
although 
it shares a similar color gradient as DSFG-1 (Fig.~\ref{fig:SP_DSFG}), and an even clumpier distribution of $\Sigma_\star$ (peaking at $\Sigma_\star \approx 10^{9.5}~M_\odot~{\rm kpc}^{-2}$), with as many as five distinct peaks along the major axis.
This may be suggestive even of spiral arm structures within the plane of a disk, a history of recent minor mergers, or the result of gravitational fragmentation of molecular clouds throughout the galaxy. Without kinematic information, our interpretation is necessarily limited.
The distribution of $\Sigma_{\rm SFR}$ (peaking at $\approx 8~M_\odot~{\rm yr}^{-1}~{\rm kpc}^{-2}$) is elongated along the major axis and decreasing radially, not dissimilar to G165-DSFG-1.

Combining the SFR and $M_\star$ information yields a map of sSFR (or alternatively the mass doubling time, sSFR$^{-1}$) on sub-kpc scales. 
This provides a direct measure of how and where the galaxies are actively growing (in tandem with the distribution of mass-weighted ages, also shown in Figs.~\ref{fig:pixbin_1a} 
and \ref{fig:pixbin_3c}).
In \S \ref{sec:rSFMS}, we discuss the relation of these sSFR values to the resolved star-forming main sequence, but first we examine the relative values of sSFR as a function of projected galaxy radius. 
For G165-DSFG-1, there is a clear bimodality in sSFR, with more starbursting regions in the southeast vs. the northwest (by nearly 1 dex). 
This may also support the picture of DSFG-1 comprising multiple orbiting clumps seen in projection, with one both forming stars more actively and containing greater dust content, based on the $A_V$ map. The map of mass-weighted stellar ages also supports this picture, with the southeast component more or less entirely younger than $\sim100$ Myr.
Alternatively, if the orbiting clumps are not responsible for this effect, then there must be some other mechanism by which gas is driven preferentially to one side of the galaxy to form stars. 
While it remains to be confirmed that DSFG-3 lies at the same redshift, one might speculate that the asymmetry is driven by a recent gravitational interaction between the two DSFGs, with molecular gas driven towards the center of mass of the system to form stars.

If a similar morphology were seen in DSFG-3, 
this might support an interpretation of tidal interaction between DSFG-1 and DSFG-3.
There is no such obvious asymmetry in DSFG-3, however.
In fact, the scenario for DSFG-3 is rather different: here there is an extensive central starbursting region with sSFR$ > 10~{\rm Gyr}^{-1}$, with patchy regions of more mild star formation in the outskirts of the disk. 
However, the very central region (where $\Sigma_\star$ peaks) has a slight depression in sSFR, which we speculate might reveal the very early stages of inside-out quenching.
In a similar resolved SED-fitting analysis, \citet{Gimenez-Arteaga:2023aa} suggested that dramatic spatial variations in sSFR could signal the important role of bursty star formation on small (kpc-sized) sub-galactic scales.
This stochastic star formation at fine spatial and temporal resolutions also makes it difficult to determine if lower sSFR is due to quenching or merely a period of inactivity between bursts.

To condense the overwhelming amount of information contained in Figs.~\ref{fig:pixbin_1a} and \ref{fig:pixbin_3c}, we also extract profiles of $\Sigma_{\rm SFR}$, $\Sigma_\star$, sSFR, and $A_V$ by taking 
linear profiles along the major axes of Arcs 1a and 3c, which are shown in Figs.~\ref{fig:slit_1a} and \ref{fig:slit_3c}. 
This approach is largely independent of inclination \citep{Devour:2017aa, Devour:2019aa}.
In these plots, we take the {\sc piXedfit} bins from Figs.~\ref{fig:pixbin_1a} and \ref{fig:pixbin_3c} and compile these into binned slices of various properties along the major axis.
We do this in two ways: for the ``weighted median" curves, we do not account for overcounting pixels. That is, the {\sc piXedfit}-binned properties consist of regions of multiple pixels that are all assigned the same sSFR, $A_V$, etc. By not removing duplicate pixels, the result is effectively weighted by the area of each {\sc piXedfit} bin contained within the major axis slices. For the ``median" curve, these duplicates are removed, which gives all {\sc piXedfit} bins within the major axis slice equal weight. 
The inner 68\% confidence intervals of pixel values are also shown as shaded curves. 

Here, the skew in sSFR and $A_V$ is even more readily observable for DSFG-1, with both transitioning between higher and lower values at a similar offset of $\approx +0 \farcs 5$. 
It is also clear that the $\Sigma_{\rm SFR}$ peak is skewed relative to the broad $\Sigma_\star$ plateau\footnote{The eastern NIRSpec slit from \citetalias{Frye:2024aa} was consistent with a redshift of $z=2.2355 \pm 0.0003$, while the western slit yields $z=2.2401 \pm 0.0002$. This redshift difference is insufficient to affect the derived properties, but it is worth recalling that both components are contained within the offset range of $-1 \farcs 0$ to $0 \farcs 0$, and it is not clear yet how to visually separate them in these images.}.
For DSFG-3, the sSFR and $A_V$ profiles are less pronounced, but are ultimately greater in the center than in the outskirts. They appear largely symmetric, thus more consistent with a simpler picture of a more centrally-concentrated dusty starburst.

The ionization parameter is found to generally peak at ${\rm log}(\mathcal{U}) \approx -2.3$, consistent with the inner 500 pc core of the starburst M82 (\citealt{Forster-Schreiber:2001aa}; see also \citealt{Engelbracht:1998aa, Thornley:2000aa}).
This is also slightly below the upper limit of ${\rm log}(\mathcal{U}) = -2$ used for our SED fitting.
For DSFG-1, the inner 68\% confidence interval ranges from ${\rm log}(\mathcal{U}) = -3.6$ to $-2.7$, while for DSFG-3 the range is ${\rm log}(\mathcal{U}) = -3.3$ to $-2.3$ (median uncertainty $0.3 - 0.4$ dex), but these are likely poor estimates from SED fitting alone without spectroscopy.
This is likewise the case for the estimation of stellar metallicity. While robust super-solar gas-phase metallicity has been found in the nucleus of DSFGs even at $z\sim 4$ (e.g., \citealt{Birkin:2023ac, Peng:2023aa}), we do not place much confidence in the peak of $Z/Z_\odot \approx 3$ for G165-DSFG-1 and G165-DSFG-3 (apart from perhaps a believable radial decrease in both cases). However, we also find similar results of $Z/Z_\odot = 3.1 \pm 0.3$ for DSFG-1 and $Z/Z_\odot = 3.2 \pm 0.3$ from the global galaxy-wide SED fits described in \S \ref{sec:integrated}, but we consider the unaccounted-for systematic uncertainties too great to further interpret these results.

\subsection{A favorable dust geometry for detection of supernovae?}
\label{sec:dust_geomtry_SN}

We also remark that the widespread star formation and starbursting regions\textemdash along with some regions of decreased dust attenuation\textemdash have implications for the observable rate of supernovae, which we discuss in detail in \S \ref{sec:SN_rates}. 
If the formation of massive stars (i.e. progenitors of core-collapse supernovae, or CCSNe) is not limited just to highly dust-obscured nuclear starbursts \citep{Mattila:2001aa}, but is apportioned more evenly throughout the disk, then the turbulence in the ISM may result in a patchy, porous, and generally complex star vs. dust geometry that would facilitate easier detection of SNe than a scenario of a uniform dust screen \citep{Casey:2014ab, Narayanan:2018ab}.
Likewise, star formation in the outskirts of the galaxy might also result in a ``frosting" of unobscured UV-bright massive stars (e.g., \citealt{Safarzadeh:2017ab}).
This has not yet been tested in practice, but such DSFGs with evidence for galaxy-wide star formation events and inhomogeneous distributions of dust may thus be ideal candidates for continued monitoring for supernovae.

\begin{figure}[ht!]
\centering
\includegraphics[width=0.95\columnwidth]{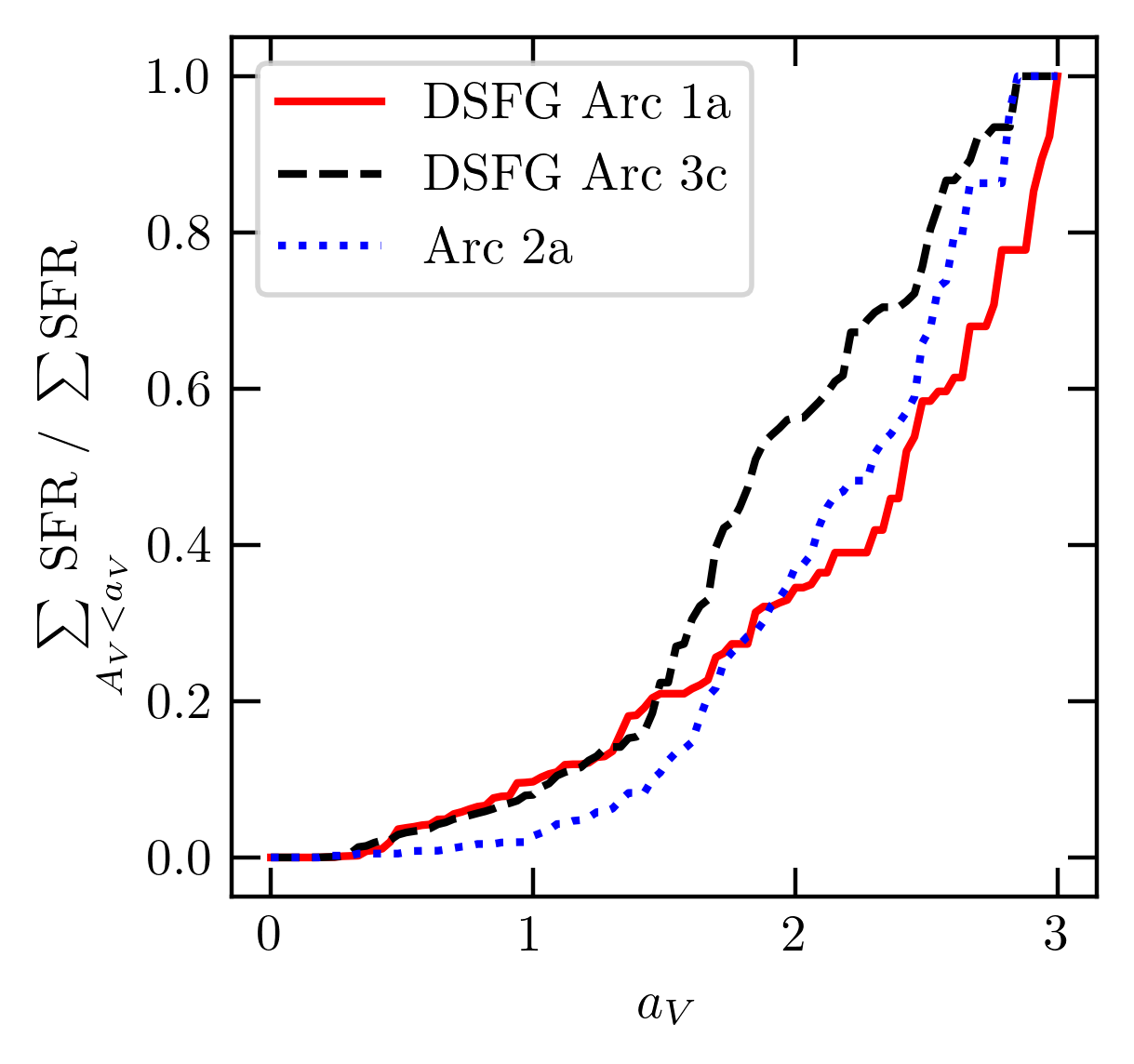}
\caption{
    Cumulative distribution function of SFR$_{\rm UV-NIR}$ below a given threshold of dust attenuation, $A_V < a_V$, as determined through {\sc piXedfit} binning. 
    For both the DSFGs and Arc 2a, 50\% of the star formation is confined to regions with $A_V < 2$. 
    \label{fig:CDF}
}
\end{figure}

To get a sense of what supernova rates would be observed given different observing depths, we sum the resolved SFRs for only bins with attenuations below a certain value, shown in Fig.~\ref{fig:CDF}. 
For example, all bins of DSFG-1 are consistent with $A_V < 3$, so the summed (magnification-uncorrected) $\mu {\rm SFR}_{A_V < 3} \approx 380~M_\odot~{\rm yr}^{-1}$ (see Table \ref{tab:SFR_SED}), while
below $A_V < 2$ yields $\mu {\rm SFR}_{A_V < 2} \approx 130~M_\odot~{\rm yr}^{-1}$, or 34\% of the total. 
Likewise, for DSFG-3, the total SFR satisfies $\mu {\rm SFR}_{A_V < 3} \approx 340~M_\odot~{\rm yr}^{-1}$ (Table \ref{tab:SFR_SED}), while
$\mu {\rm SFR}_{A_V < 2} \approx 190~M_\odot~{\rm yr}^{-1}$ (56\% of total).
Essentially, this implies that a shallower point-source sensitivity accounting for only up to 2 mag dust attenuation for supernovae would still detect them at a rate of $30 - 60$\% the intrinsic rate.
This contrasts 
with the estimated $\lesssim 20$\% recovered rate of supernovae at optical wavelengths for some local ULIRGs (e.g., Arp 220 and Arp 299; \citealt{Leaman:2011aa, Mattila:2012aa}; also \citealt{Richmond:1998aa}).
\citet{Mattila:2012aa} and \citet{Dahlen:2012aa} recognized that the increasing fraction of the cosmic star formation rate occurring within luminous infrared galaxies (LIRGs; $10^{11} < L_{\rm IR} / L_\odot < 10^{12}$) and ULIRGs \citep{Magnelli:2011aa} might result in dire, prohibitively large fractions of missed CCSNe at $z > 1$. 
However, the concrete differences in the distribution and driving mechanisms of star formation in distant vs. local ULIRGs that have become apparent in recent years may work immensely in our favor, as we contend here.

Moreover, a more dispersed distribution of CCSNe is favorable for lensed objects like G165-DSFG-1 and G165-DSFG-3, as it improves the odds that any given supernova will occur within a multiply-imaged region of the galaxy.
If all SNe were confined to the galaxy nucleus, this would effectively require that this region falls by chance within the caustics in order to ever observe a multiply-imaged supernova.

\begin{figure*}[ht!]
\centering
\includegraphics[height=0.3\textheight]{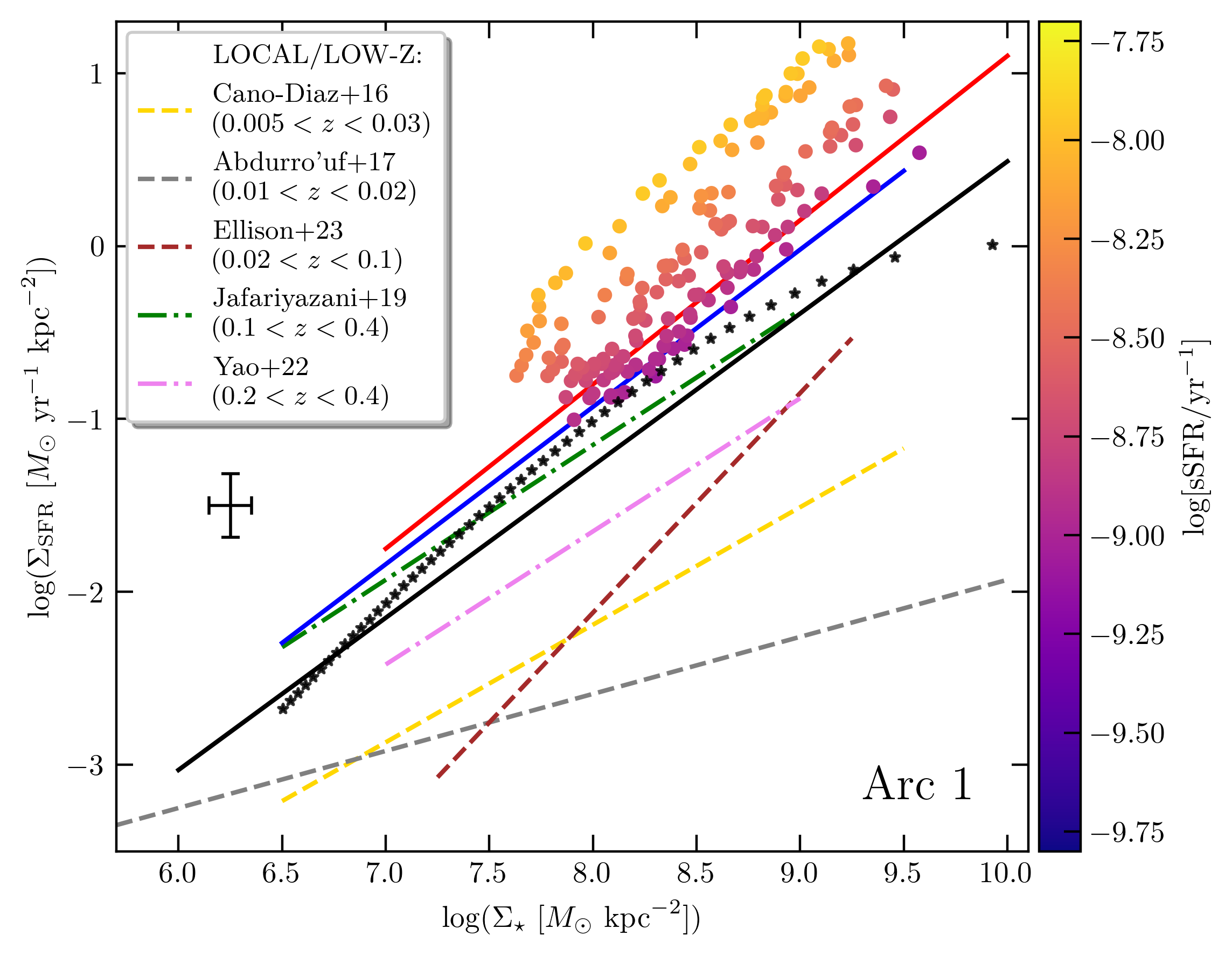}
\includegraphics[height=0.3\textheight]{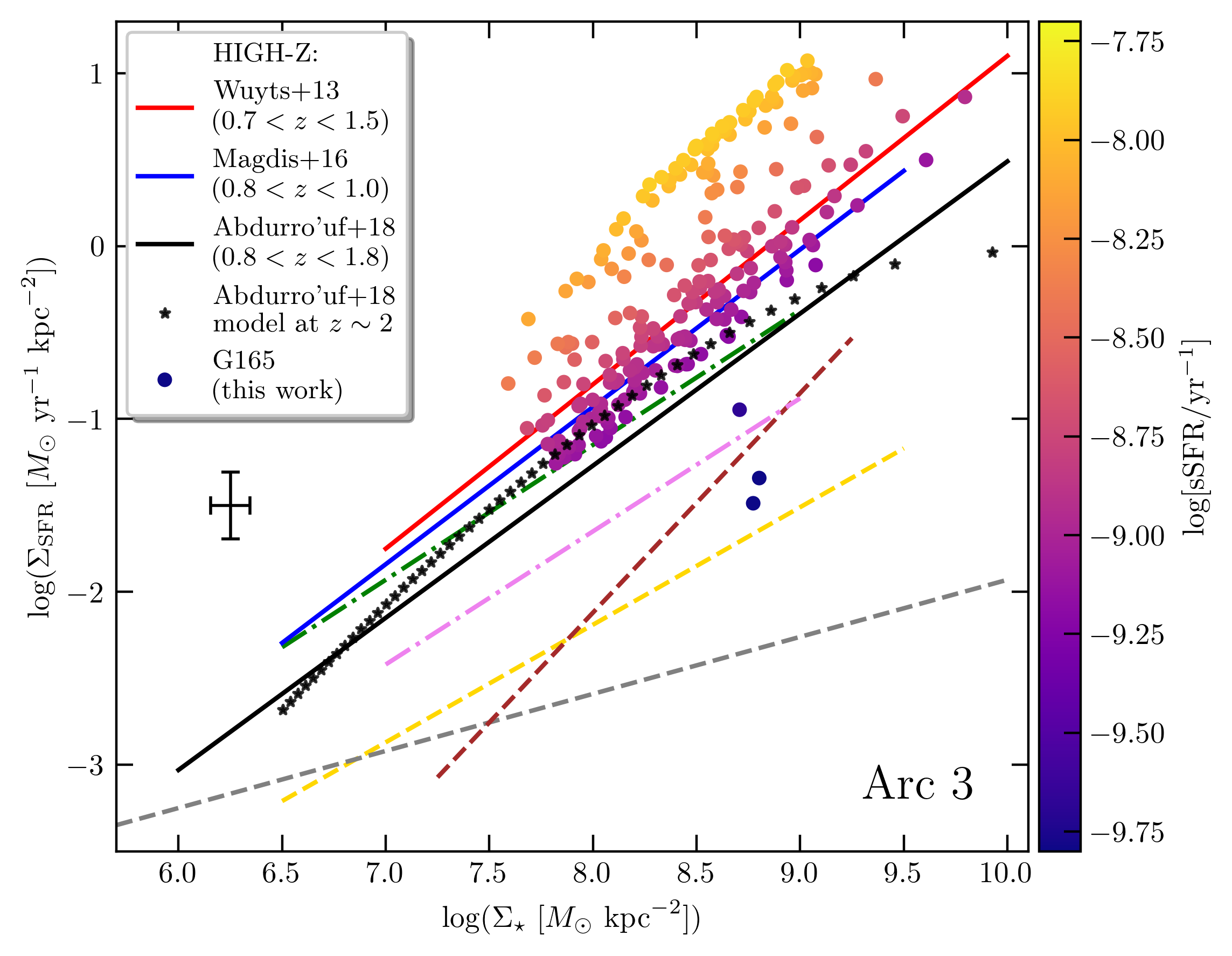}
\caption{
    Individual {\sc piXedfit} bins of G165 DSFGs (Arc 1, left panel; Arc 3, right panel) placed on the resolved star formation main sequence, or $\Sigma_{\rm SFR}$ vs. $\Sigma_\star$.
    The color scale indicates sSFR. Median uncertainties are indicated in the center left. These properties are not corrected for lensing, as they are related to surface brightness and thus independent of magnification. However, we filter bins to only include those with areas equivalent to $0.15 - 1.5$ kpc in radius, where we assume a fiducial magnification factor of $\mu=6$ for Arc 1 and $\mu=7$ for Arc 3 (i.e., image-plane areas are divided by this factor before filtering). A number of measurements of the rSFMS in the local Universe \citep{Cano-Diaz:2016aa, Abdurrouf:2017aa, Ellison:2023aa}, at low-$z$ \citep{Jafariyazani:2019aa,Yao:2022aa}, and at $z\sim 1$ (\citealt{Wuyts:2013aa, Magdis:2016aa}; \citetalias{Abdurrouf:2018aa}) are shown as dashed, dash-dotted, and solid lines, respectively. Legends in the two panels indicate the redshift ranges included for each. To make an approximate comparison at $z\sim 2$, we extrapolate the redshift evolution model of $\Sigma_{\rm SFR}(r)$ and $\Sigma_\star(r)$ from \citetalias{Abdurrouf:2018aa} backwards to $z=2.24$,
    which is shown as black stars. The vast majority of pixels in each galaxy appear to be on or above the rSFMS, as 99\% (89\%) of the bins in Arc 1 (Arc 3) lie above this backward-evolved curve. In terms of areal fractions, these translate to $\sim 100\%$ (Arc 1) and $\sim 85\%$ (Arc 3).
    \label{fig:rSFMS}
}
\end{figure*}

\subsection{Resolved star-forming main sequence in DSFG-1 and DSFG-3}
\label{sec:rSFMS}

Remarkably, most star-forming galaxies (SFGs) fall on a tight, nearly linear scaling relation (that evolves with redshift) between SFR and $M_\star$, known commonly as the star-forming main sequence, as has been borne out extensively
(including
\citealt{Brinchmann:2004aa, Daddi:2007aa, Elbaz:2007aa, Noeske:2007aa, Noeske:2007ab, Salim:2007aa, Whitaker:2012aa, Speagle:2014aa, Renzini:2015aa}).
While the meaning behind this result is still somewhat subject to debate, it is commonly interpreted as an indication that most SFGs grow steadily when considered over $>100$ Myr timescales (although individual objects are likely to oscillate around the main sequence; e.g., \citealt{Tacchella:2016ab}).
What remains highly uncertain is if and how this relation holds for $\sim$kpc-scale regions within galaxies.
In parallel with the integrated properties of SFR and $M_\star$, their respective surface densities $\Sigma_{\rm SFR}$ and $\Sigma_\star$ are correlated through what is referred to as the spatially resolved star formation main sequence (rSFMS),
first explored by \citet{Sanchez:2013aa} and \citet{Cano-Diaz:2016aa} locally and at higher redshift by \citet{Wuyts:2013aa}.
A number of works since have supported the existence of the correlation, suggesting that it is physically motivated at sub-galactic scales
\citep{Hemmati:2014aa, Gonzalez-Delgado:2016aa, Magdis:2016aa, Abdurrouf:2017aa,  Hsieh:2017aa, Maragkoudakis:2017aa, Abdurrouf:2018aa, Ellison:2018aa, Hall:2018aa, Liu:2018aa, Medling:2018aa, Cano-Diaz:2019aa, Vulcani:2019aa, Bluck:2020aa, Bluck:2020ab, Enia:2020aa, Morselli:2020aa, Pessa:2021aa, Pessa:2022aa, Ellison:2023aa}.

In particular, \citet{Erroz-Ferrer:2019aa} demonstrated that 
%
the relation between $\Sigma_{\rm SFR}$ and $\Sigma_\star$ may be more fundamental; the global SFR$-M_\star$ relation could be in part a consequence of the generally small variation for $\Sigma_{\rm SFR}$ and $\Sigma_\star$ throughout a galaxy, combined with the size-mass relation \citep{Shen:2003aa, van-der-Wel:2014aa} that scales both up and yields a linear relation between the integrated properties SFR and $M_\star$ \citep{Vulcani:2019aa}.
The rSFMS has also been observed through simulations (e.g. \citealt{Trayford:2019aa, Hani:2020ab, McDonough:2023aa}).
However, we caution that the relation is likely non-universal, with non-trivial variations in the rSFMS from galaxy to galaxy (e.g., \citealt{Vulcani:2019aa, Ellison:2021aa}), and the slope of the sequence may be influenced by the minimum size scales that are probed (e.g., \citealt{Hani:2020ab}).

Fig.~\ref{fig:rSFMS} compares the distribution of individual {\sc piXedfit} bins in Arcs 1 and 3 in the $\Sigma_{\rm SFR}-\Sigma_\star$ plane. 
To facilitate comparison with work at lower-$z$, we include only the bins with an area approximately equivalent to a radius of $150 - 1500$ pc. 
Additionally, we expect that SED fitting at $\lesssim 100$ pc physical scales (comparable to individual giant molecular clouds) becomes much less physically meaningful.
As \citet{Vulcani:2019aa} point out, it is highly possible for star clusters to migrate somewhat from the natal molecular cloud where they formed, which perturbs the one-to-one relationship between $\Sigma_{\rm SFR}$ and $\Sigma_{\rm gas}$ (and by extension, between $\Sigma_{\rm SFR}$ and $\Sigma_\star$).
Since our method is applied to the magnified image-plane arcs, we first divide the bin areas by a fiducial magnification ($\mu=6$ for Arcs 1 and 3, and $\mu = 8$ for Arc 2) before filtering to include only intrinsic sizes from $\approx 150 - 1500$ pc.

So far, there are rather few studies that have examined the rSFMS beyond the local Universe.
\citet{Jafariyazani:2019aa} used integral-field spectroscopy to place a sample covering $0.1 < z < 0.4$ on the rSFMS, while 
\citet{Yao:2022aa} likewise covered $0.2 < z < 0.4$. 
Further afield, our comparison is primarily limited to \citet{Wuyts:2013aa} at $0.7 < z < 1.5$, \citet{Magdis:2016aa} at $0.8 < z < 1.0$, \citetalias{Abdurrouf:2018aa} at $0.8 < z < 1.8$,
and 
\citet{Nelson:2021aa} at $0.8 < z < 1.5$.
Fig.~\ref{fig:rSFMS} shows the best-fit rSFMS relations from these works (over approximately the pertinent range in $\Sigma_\star$ that each study dealt with), in addition to local relations from \citet{Cano-Diaz:2016aa}, \citet{Abdurrouf:2017aa}, and \citet{Ellison:2023aa}.
\citet{Abdurrouf:2017aa} and \citetalias{Abdurrouf:2018aa} first examined how the radial profiles of $\Sigma_{\rm SFR}(r)$, $\Sigma_\star(r)$, and ${\rm sSFR}(r)$ generally evolve with redshift, and by extension, how the rSFMS evolves. We can use this to predict where the pixels of the G165 DSFGs lie relative to the rSFMS\textemdash 
with the caveat that the rSFMS is not universal from one galaxy to another.

For now,  we employ the redshift evolutionary model of \citetalias{Abdurrouf:2018aa}, which tracked progenitors from $z\approx 1.06$ ($t(z) \approx 5.5$ Gyr in our assumed cosmology) to $z\approx 0.02$ ($t(z) \approx 13.2$ Gyr), and extend it backwards in time to $z=2.24$
($t(z) \approx 2.9$ Gyr).
The result is shown as a dotted curve of black stars in Fig.~\ref{fig:rSFMS} (and see also Fig. 15 of \citetalias{Abdurrouf:2018aa} for $z=0$ to 1.1 in steps of $\Delta z = 0.11$). Relative to this curve, we find that the overwhelming majority ($\approx 99\%$) of the bins in Arc 1 lie above the model rSFMS at $z \approx 2.2$ (accounting for nearly $\approx 100\%$ of the total area), whereas $\approx 89\%$ of the bins in Arc 3 lie above (here accounting for $85\%$ of the area)\footnote{While the main sequence has large intrinsic scatter, this binary calculation of the areal fraction ``above" vs. ``below" the main sequence treats it as an exact discrete curve, in part to draw clear contrast with Arc 2 in \S \ref{sec:SN_host_maps}. In this instance, ``above the main sequence" does not equate to ``starbursting."}.
This suggests that a clear majority of the bins shown in Figs.~\ref{fig:pixbin_1a} and \ref{fig:pixbin_3c} are actively star-forming (and some of which are starbursting). For these objects, at least, it appears to be the case that star formation is not confined purely to a central starburst region, but is widely distributed (often in clumps) throughout the plane of the disk. This has been demonstrated to be the case in recent years for a number of high-$z$ DSFGs, especially in contrast with local ULIRGs (e.g., \citealt{Chapman:2004aa, Carilli:2010aa, Ivison:2010ad, Ivison:2011aa, Rujopakarn:2011aa, Hodge:2012aa, Hatsukade:2015aa, Miettinen:2015aa, Simpson:2015ab, Swinbank:2015aa, Hodge:2016aa, Sharda:2018aa, Tadaki:2018aa, Hodge:2019aa, Kamieneski:2024aa}).
Here, we have demonstrated that the unobscured mode of star formation inferred from the UV-NIR SED is similarly widespread for DSFG-1 and DSFG-3, matching the widespread obscured star formation suggested by our finding of $R_e \sim 3$ kpc for the dust continuum. 
This is a result that we revisit in \S \ref{sec:UVJ}, using $U-V$ and $V-J$ colors alone for a largely independent and non-parametric characterization.

The evolutionary model we employ for the rSFMS was only designed for $z\sim 1$ to $z=0$, and it is not clear how the evolution might extend to our objects at $z\approx 2.2$. However, 
the $z\sim 1$ sample from \citetalias{Abdurrouf:2018aa} includes objects up to $z\approx 1.8$, only $500 - 700$ Myr after the G165 DSFGs. 
Their sample of high-mass objects ($M_\star > 10^{10.5}~M_\odot$, SFR $\lesssim 300~M_\odot~{\rm yr}^{-1}$) is also selected to focus on face-on (low-ellipticity) disk-like objects, so it is subject to different systematic effects than our work.
Nevertheless, properly quantifying the spatial extent of starbursting regions in G165 and other $z\sim 2$ galaxies will not be possible until a more detailed study of the rSFMS at higher redshifts has been undertaken. With JWST covering an ever-increasing number of strongly lensed objects, this is becoming more feasible, but beyond the scope of our current study.

\subsubsection{A mode of elevated star formation efficiency in the resolved star-forming main sequence}
\label{sec:SFE}

We also observe for the first time what appears to be a secondary ``starburst" sequence $\sim 1$ dex above the ``star-forming" main sequence (Fig.~\ref{fig:rSFMS}). 
This elevated mode of star formation is not dissimilar to the higher star formation efficiency (SFE) mode of the Schmidt-Kennicutt relation between molecular hydrogen mass $M_{H_2}$ and SFR on resolved scales \citep{Sharon:2019ab}. 
Recent works (e.g., \citealt{Baker:2022ab}) have demonstrated that the (integrated) SFMS is not a fundamental correlation, but rather the result of two more fundamental scaling relations with less scatter: the Schmidt-Kennicutt relation, and the Molecular Gas Main Sequence (MGMS) between $M_{H_2}$ and $M_\star$.
\citet{Lin:2019aa} found likewise (and since supported by \citealt{Ellison:2021aa} and \citealt{Baker:2022aa}) that the correlation between $\Sigma_{\rm SFR}$ and $\Sigma_\star$ is merely a byproduct of the correlations between $\Sigma_{\rm SFR}$ and $\Sigma_{H_2}$ (which is the strongest correlation) and between $\Sigma_{H_2}$ and $\Sigma_\star$.
Together, this suggests a picture where the rate of star formation is generally driven (or regulated) by the availability of molecular gas fuel (although this is not necessarily a universal slope, e.g., \citealt{Daddi:2010aa}), or neutral H\textsc{i} in the case of low-density environments at the outskirts of galaxies \citep{Leroy:2008aa}. The correlation between $\Sigma_{H_2}$ vs. $\Sigma_\star$ is suggestive of gas being preferentially accreted in regions of deeper gravitational potential, as traced by the distribution of stellar mass
\citep{Lin:2019aa}. 

The bimodalities in the rSFMS and the Schmidt-Kennicutt relation may in fact be directly related if, for example, the MGMS that is hypothesized to link the two is not itself bimodal.
In other words, regions of elevated SFE could also be detected merely through $\Sigma_{\rm SFR}$ vs. $\Sigma_\star$, under the assumption that each galaxy satisfies a unimodal $\Sigma_{H_2}$-$\Sigma_\star$ relation (without large scatter).
Without information on the molecular gas (or dust) distribution of the G165 DSFGs on similar spatial scales, or the gas excitation conditions through multiple spectral lines to accurately infer gas mass (e.g., \citealt{Harrington:2021aa}), we are missing part of the picture in characterizing what drives stellar mass assembly across the galaxy. However, the rSFMS is still helpful in characterizing the widespread distribution
of star formation and
the rapid consumption of molecular gas.

\begin{figure}[htb]
\centering
\includegraphics[width=1\columnwidth]{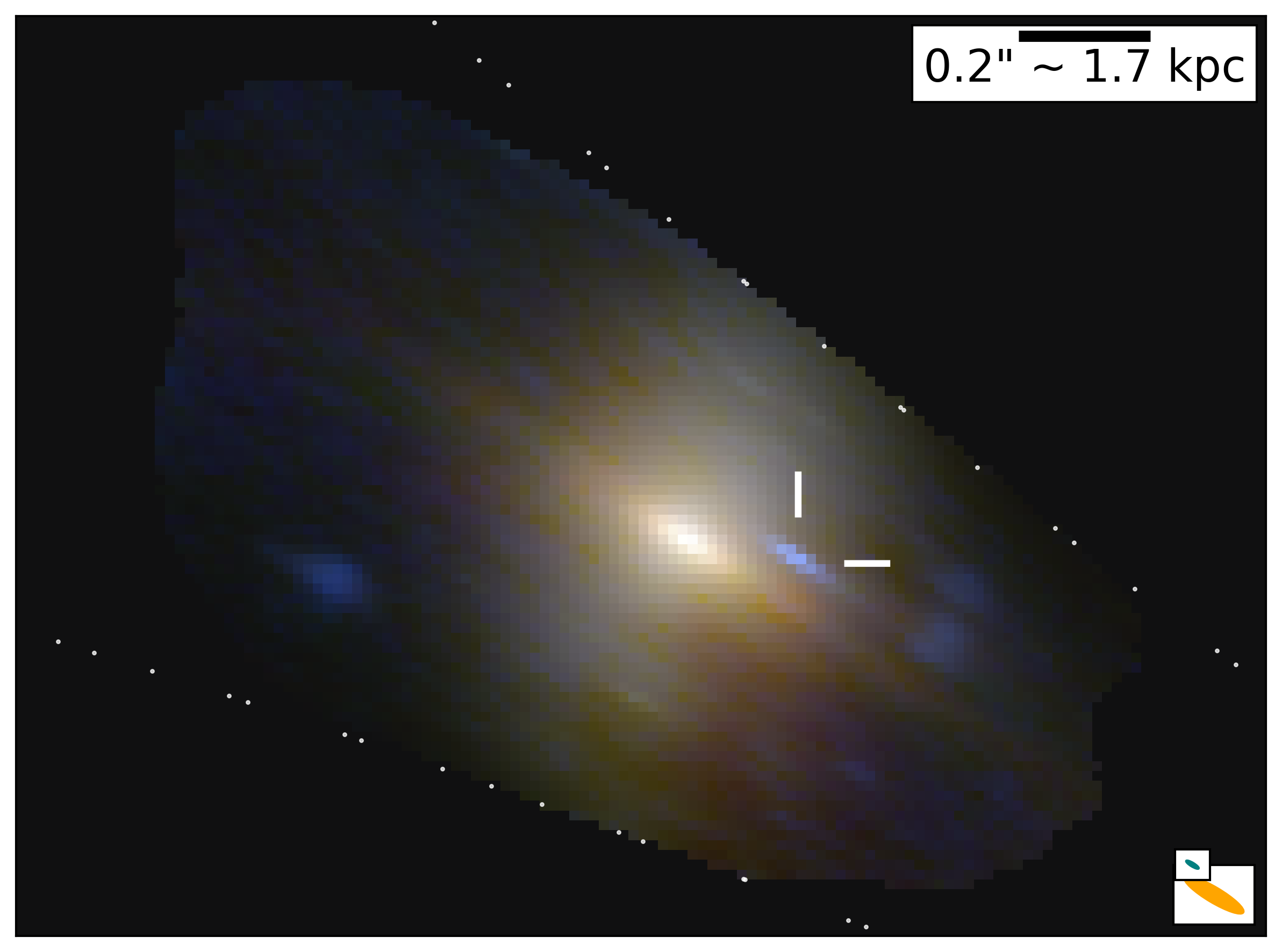}
\caption{
    RGB source-plane reconstruction of NIRCam imaging of Arc 2b at $z=1.78$, as with Fig.~\ref{fig:SP_1a_3c}. Caustics are shown as sparse white dotted curves, and F090W/F444W source-plane PSFs are shown in the lower right as teal/orange beams, respectively. Only Arc 2b is shown in this reconstruction, as small but visually detectable offsets ($\sim 0 \farcs 05$) in the source-plane location of Arcs 2a, 2b, and 2c are introduced from the lens model.  
    \label{fig:SP_2b}
}
\end{figure}

\begin{figure*}[ht!]
\centering
\includegraphics[width=0.99\textwidth]{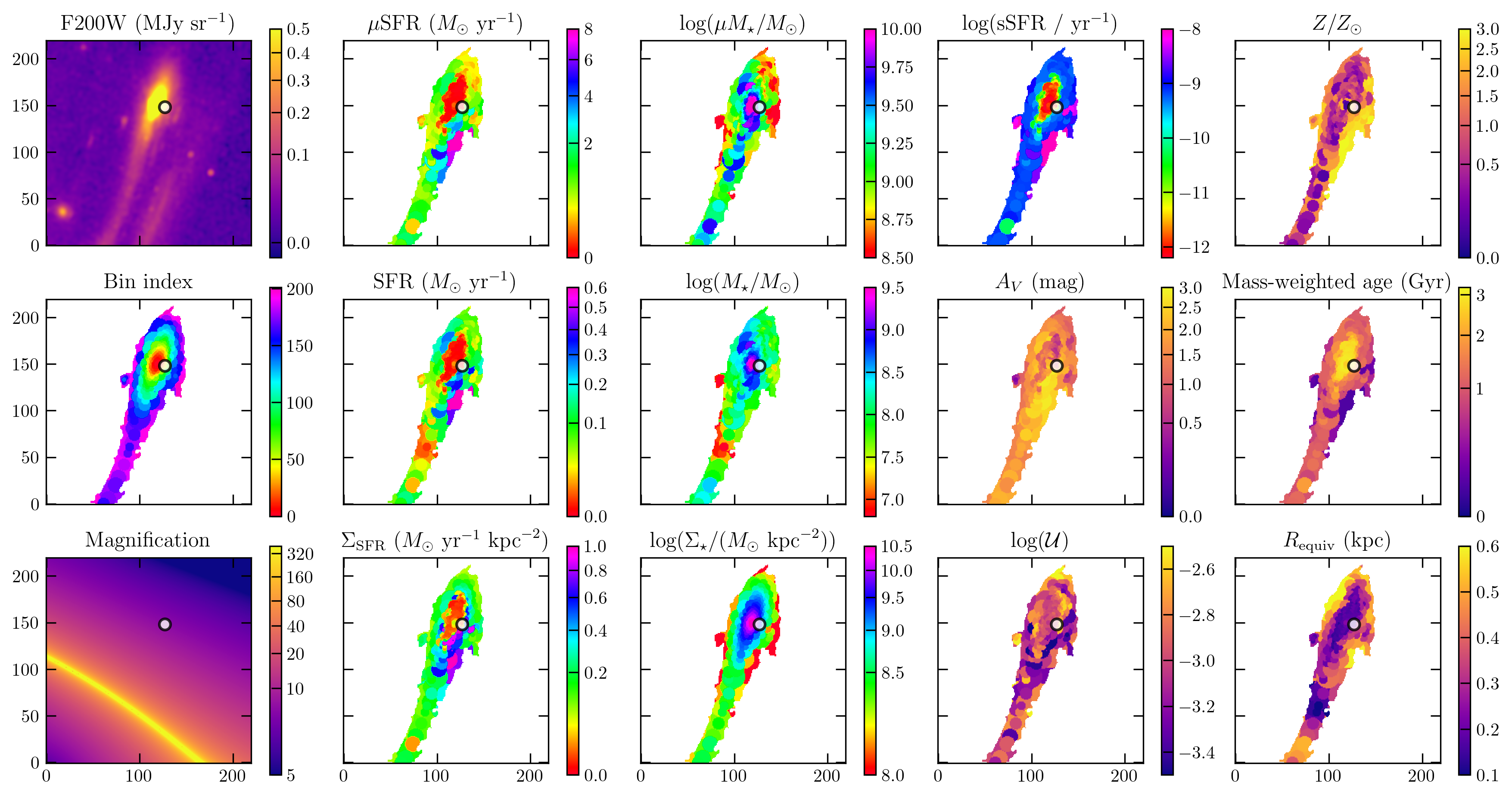}
\caption{
    As with Fig.~\ref{fig:pixbin_1a}, but for Arc 2a, which hosts SN H0pe, with a fixed redshift of $z=1.78$. A small black circle indicates the location of the supernova. We choose this image for SED fitting rather than 2b or 2c as it is the one in which the SN is faintest (and thus contaminates the least). 
    \label{fig:pixbin_2a}
}
\end{figure*}

\subsection{Resolved properties of the massive host galaxy of SN H0pe}
\label{sec:SN_host_maps}

As discussed by \citet{Polletta:2023aa} and \citetalias{Frye:2024aa}, the rest-frame optical spectrum of Arc system 2abc, host to SN H0pe, shows evidence for a mostly intermediate-age stellar population at $\sim 2$ Gyr (e.g., a strong NaD line and larger D(4000) value). The discovery of a Type Ia supernova itself
helps support this, 
as these require delay times for white dwarf progenitors ($\sim 3-8~M_\odot$) to evolve off of the main sequence and reach the end of their lifetimes. This delay time distribution appears to peak at $\sim 1 - 3$ Gyr (\citealt{Maoz:2012ab}, and references therein).
While the galaxy has evidence of some ongoing star formation, 
$10^{10.9}~M_\odot$ of stellar mass has already been assembled by the time it is observed at $z=1.78$ \citepalias{Frye:2024aa}.
We fit its UV-NIR SED with {\sc bagpipes} in Appendix \ref{sec:appendix3}, and remark on possible degeneracies with this narrow wavelength range which affect our interpretations..

Fig.~\ref{fig:pixbin_2a} shows the results of spatially-resolved SED fitting for Arc 2a, which is host to SN H0pe. As this image is understood to arrive first (followed by 2c then 2b), the supernova is faintest in this image, having already faded substantially in its light curve. We thus choose this arc for SED fitting, as it is subject to the least contamination from the SN. Moreover, it is also less contaminated by intracluster light than image 2b (which coincides with the axis of the merging cluster), and more isolated than 2c (which is nearby another very red arc, of currently-unknown redshift).
While the properties of this massive, mildly star-forming galaxy are naturally of interest in relation to SN H0pe, it also serves as a useful control for comparison with the more extreme DSFGs in this work. 

\begin{figure}[th]
\centering
\includegraphics[width=\columnwidth]{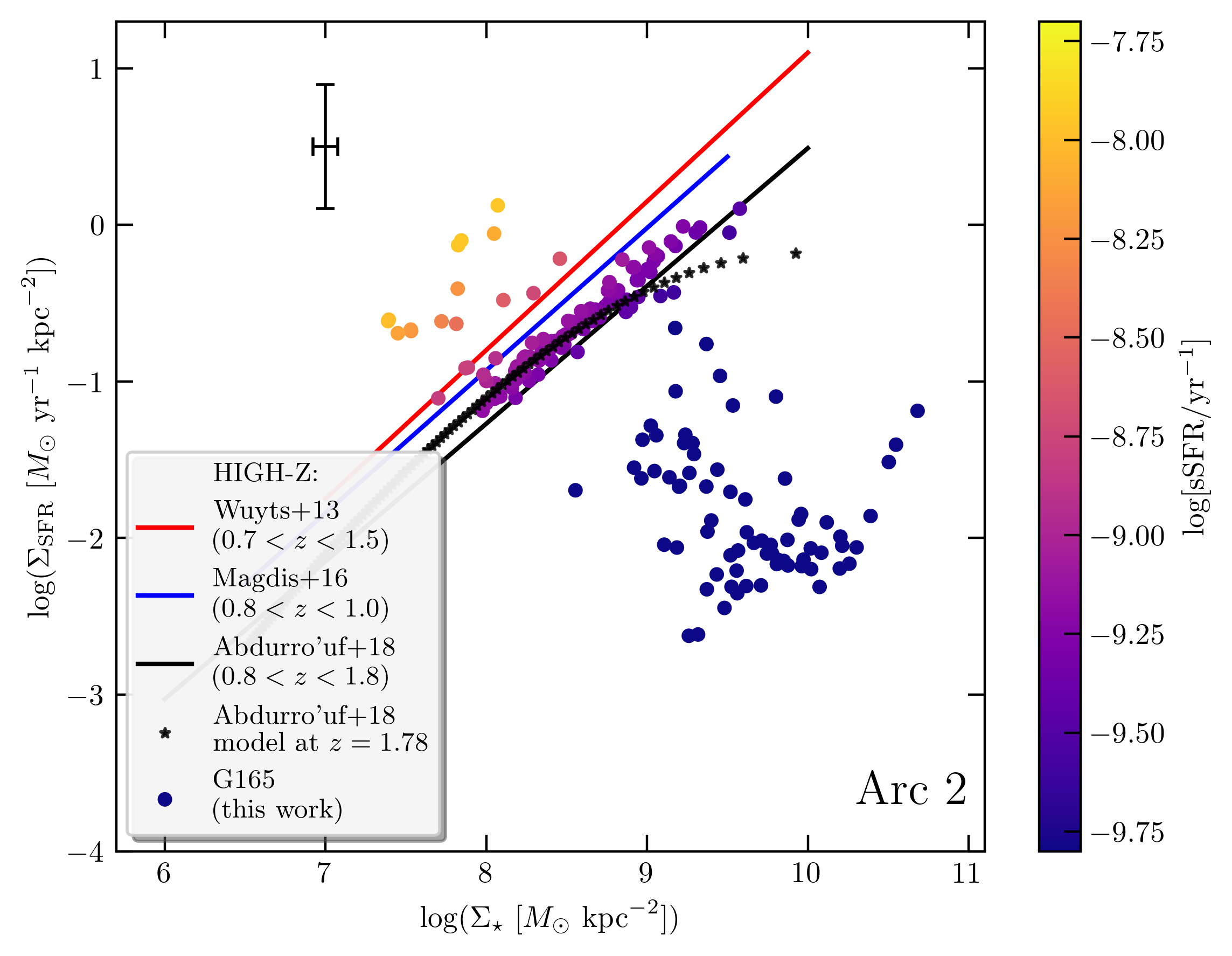}
\caption{
    The $\Sigma_{\rm SFR}-\Sigma_\star$ plane for Arc 2, as with Fig.~\ref{fig:rSFMS}. A fiducial magnification of $\mu=8$ is chosen to filter intrinsic bin sizes.
    Note that, given the larger stellar mass for this object, we shift the horizontal axis by 1 dex relative to Fig.~\ref{fig:rSFMS}. Here, for a fiducial magnification of $\mu=8$ and filtering to only include bins with intrinsic equivalent radii of $0.15 - 1.5$ kpc, we find that $\approx 44\%$ of bins lie above the redshift-evolved rSFMS to $z=1.78$ (accounting for $\approx 57\%$ of the galaxy area). 
    \label{fig:rSFMS_2a}
}
\end{figure}

Fig.~\ref{fig:rSFMS_2a} shows the relation of Arc 2a to the rSFMS, which shows clear contrasts with Fig.~\ref{fig:rSFMS}. While 99\% and 89\% of bins in Arcs 1a and 3c (respectively) lie above the center of the rSFMS evolved to their respective redshifts (or areal fractions of $\sim$100\% and 85\%), merely 44\% of bins in Arc 2a are above the rSFMS at $z=1.78$ (or $\approx 57\%$ in terms of galaxy area). This is strongly influenced by the presence of a low-$\Sigma_{\rm SFR}$ (and low-sSFR) sequence at the center of the galaxy\textemdash with high stellar mass surface density, log$(\Sigma_\star [M_\odot~{\rm yr}^{-1}])>9$. This locus is 
entirely absent for the DSFGs, although the sSFRs are still above the nominal ``green valley" transition at log(sSFR)$<-10.8$ yr$^{-1}$ \citep{Salim:2014ab}.
It also appears to be the case that many bins are clustered around the main sequence, adding some further confidence in the validity of backward-evolving the rSFMS model from \citetalias{Abdurrouf:2018aa}, although 
it is true that 
this arc actually falls within the redshift range used for the model.
There are a select number of regions that appear to be starbursting, perhaps in line with the conclusion by \citetalias{Frye:2024aa} that the galaxy is still star-forming (despite its very large stellar mass), as evidenced by the detection of H$\alpha$ emission (alongside spectral features that confirm the presence of a much older stellar population).

The large stellar mass and non-negligible ongoing star formation may in part help explain the favorable conditions leading to the discovery of SN H0pe, as the SN Ia rate is hypothesized to depend on an instantaneous term proportional to the SFR and on an extended term proportional to the stellar mass (e.g., \citealt{Scannapieco:2005aa, Smith:2012aa}).
Similarly, the Type Ia SN Requiem \citep{Rodney:2021ab} was discovered in a low-sSFR host, classified as a Massive/Magnified Red Galaxy (MRG; \citealt{Newman:2018aa, Akhshik:2020aa}). 
Since then, an additional SN Ia (``Encore") has recently been found in the same host galaxy \citep{Pierel:2024ab}.
In \S \ref{sec:SN_rates}, we discuss the prospects for discovering core-collapse SNe behind G165 given the high rate of star formation. However, given the large stellar mass of the host of SN H0pe \citep{Polletta:2023aa, Frye:2024aa}, we also consider that the G165 field may be a source for the discovery of additional Type Ia SNe, as well (e.g., \citealt{Shu:2018ab, Holwerda:2021aa}).

\begin{figure*}[ht!]
\centering
\includegraphics[height=0.31\textheight]{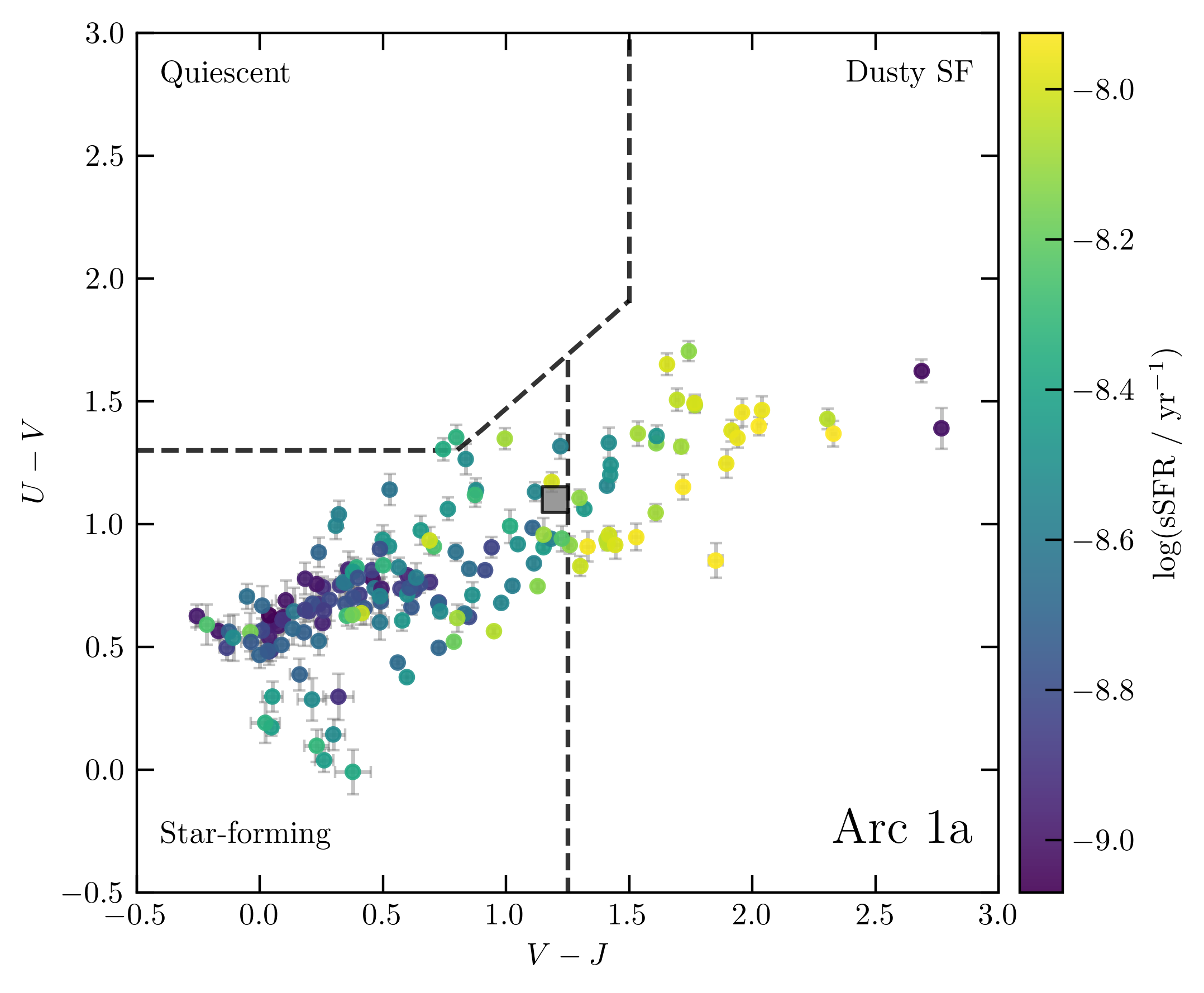}
\includegraphics[height=0.31\textheight]{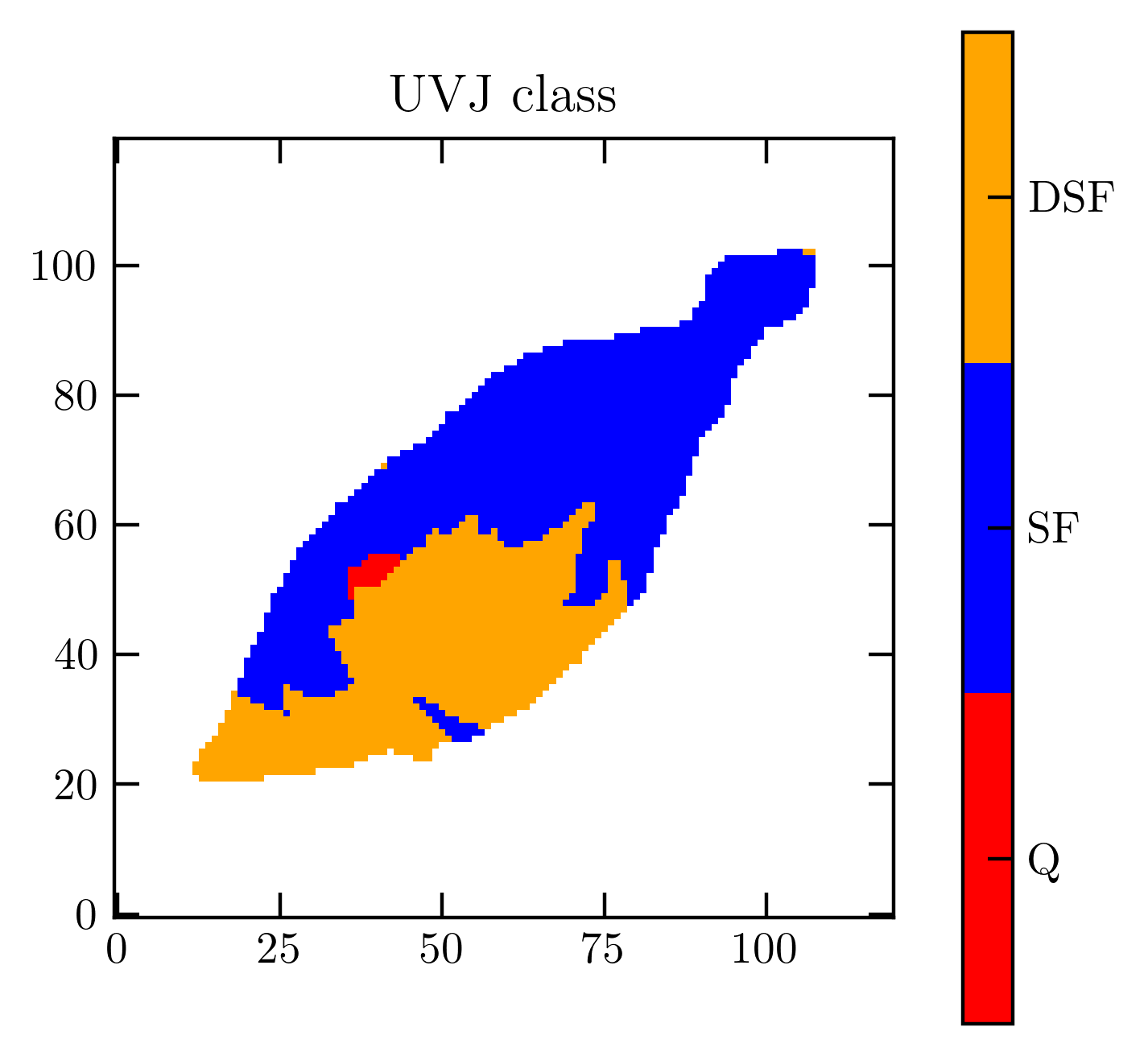}
\includegraphics[height=0.31\textheight]{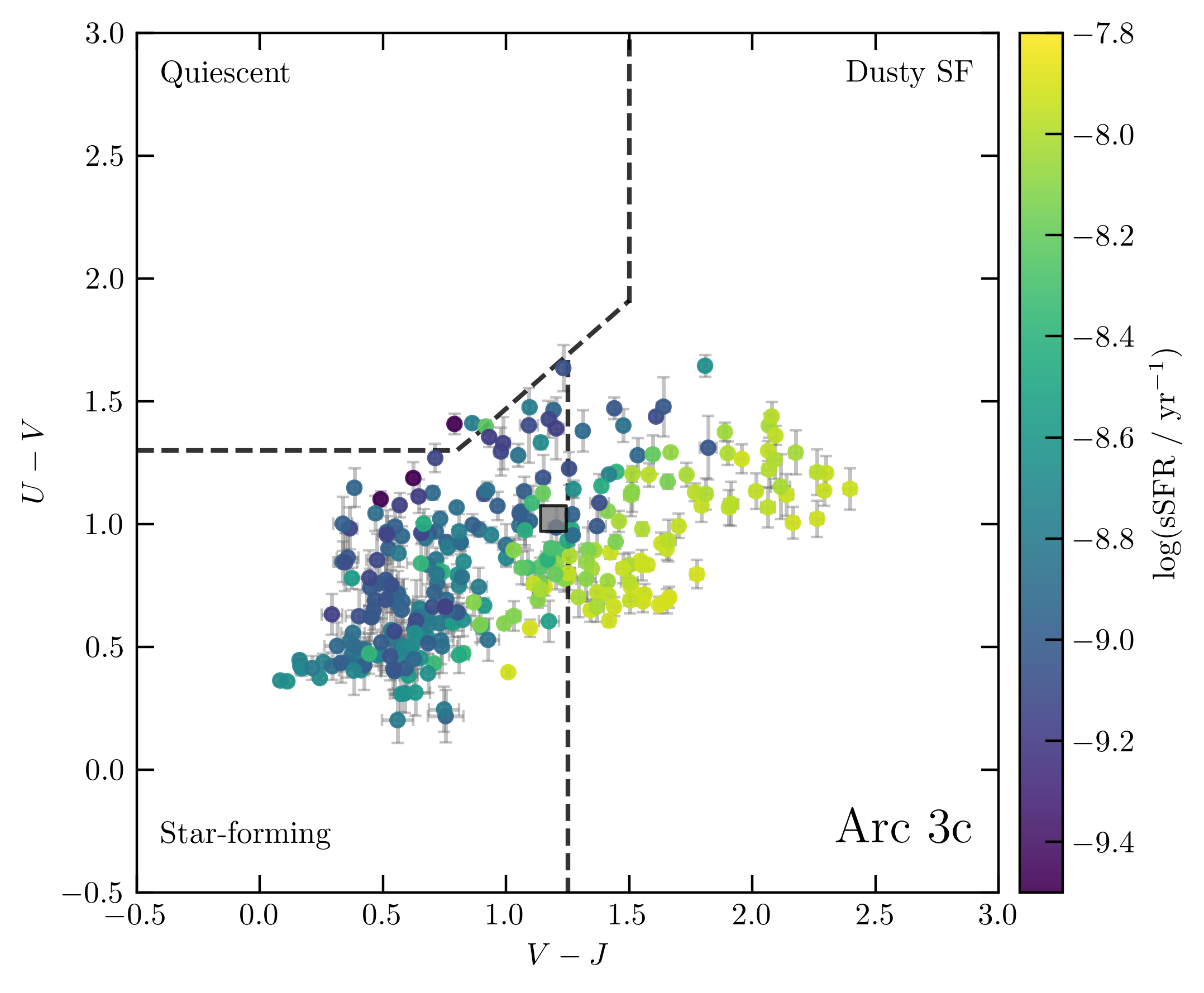}
\includegraphics[height=0.31\textheight]{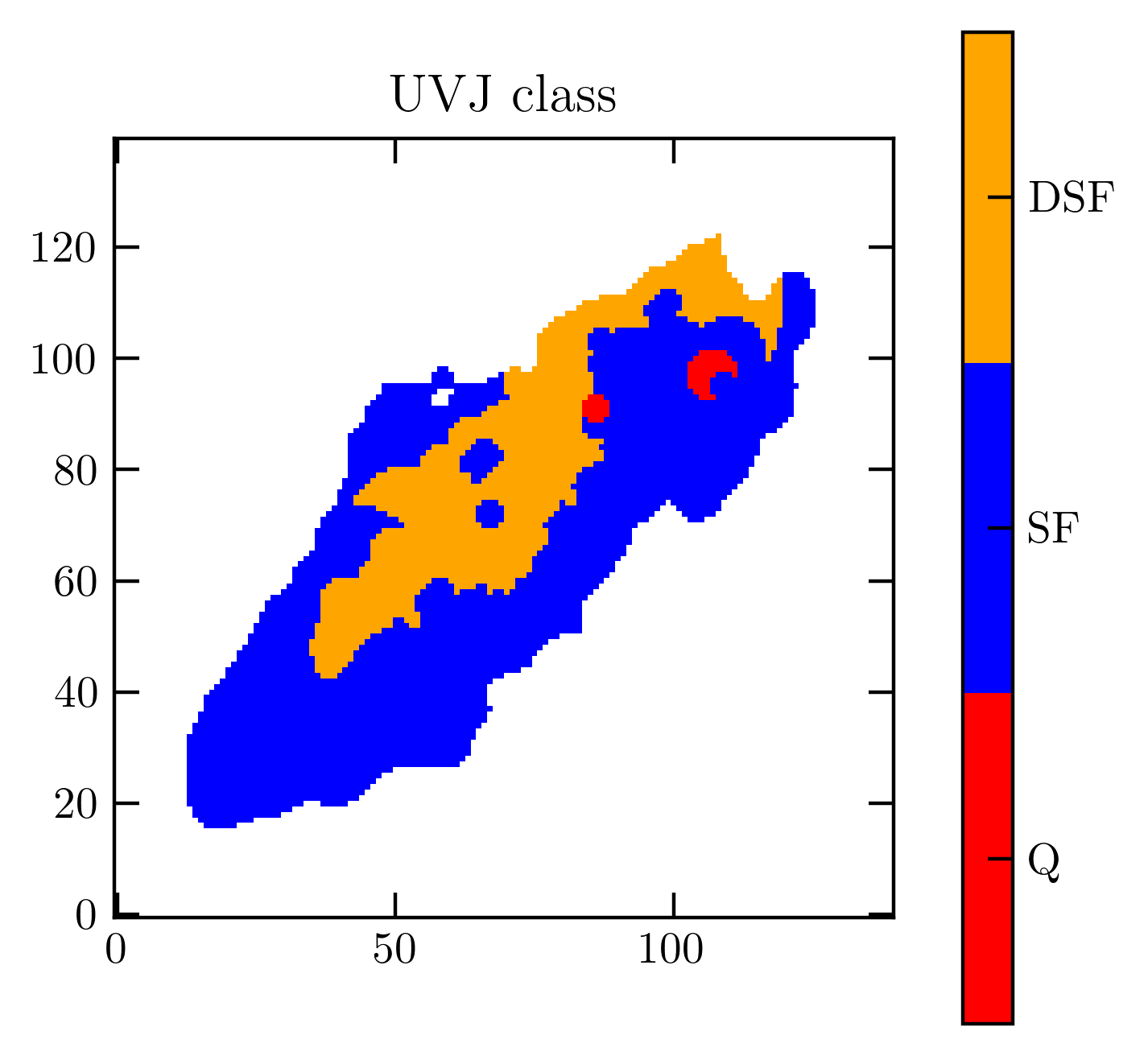}
\includegraphics[height=0.31\textheight]{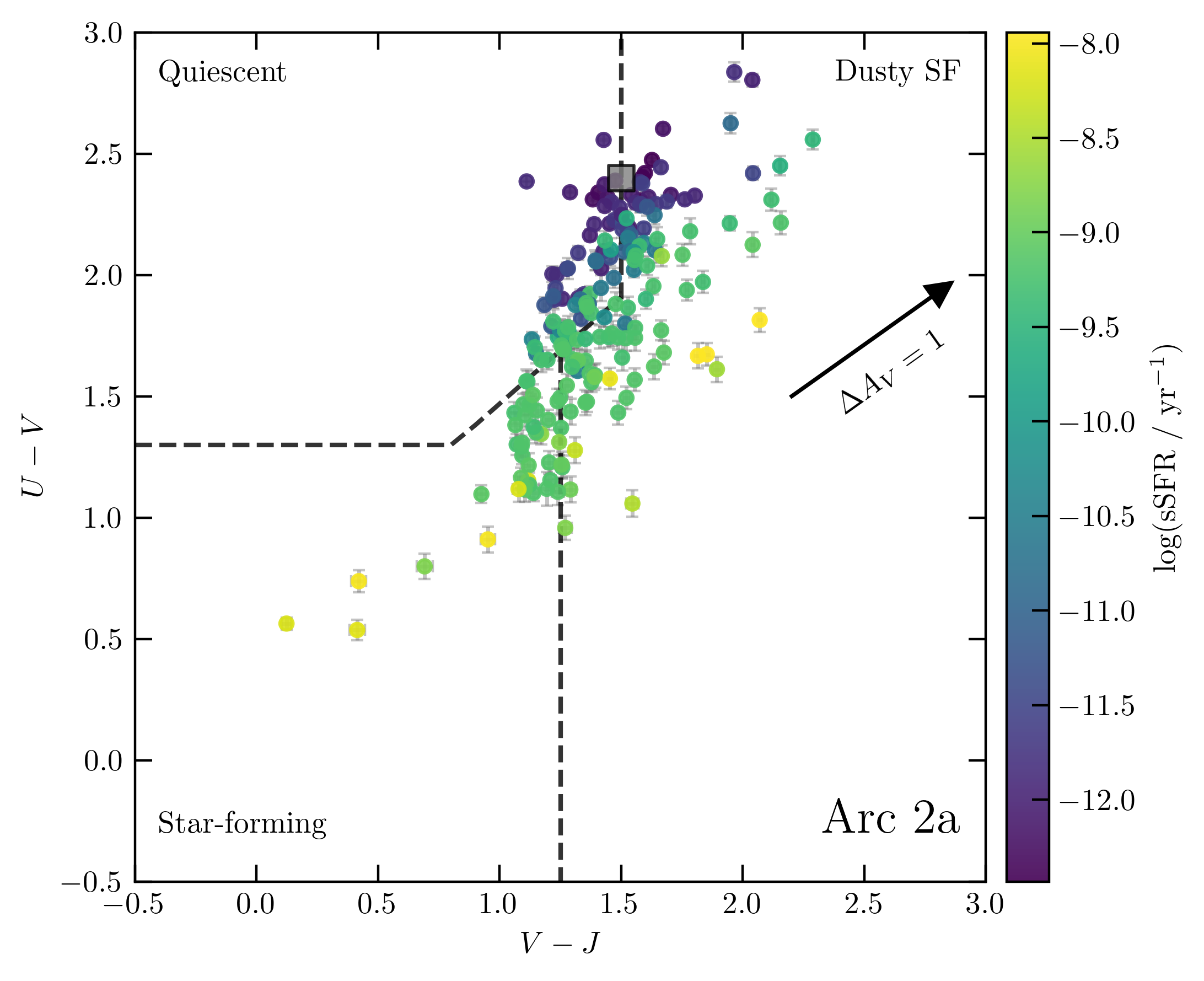}
\includegraphics[height=0.31\textheight]{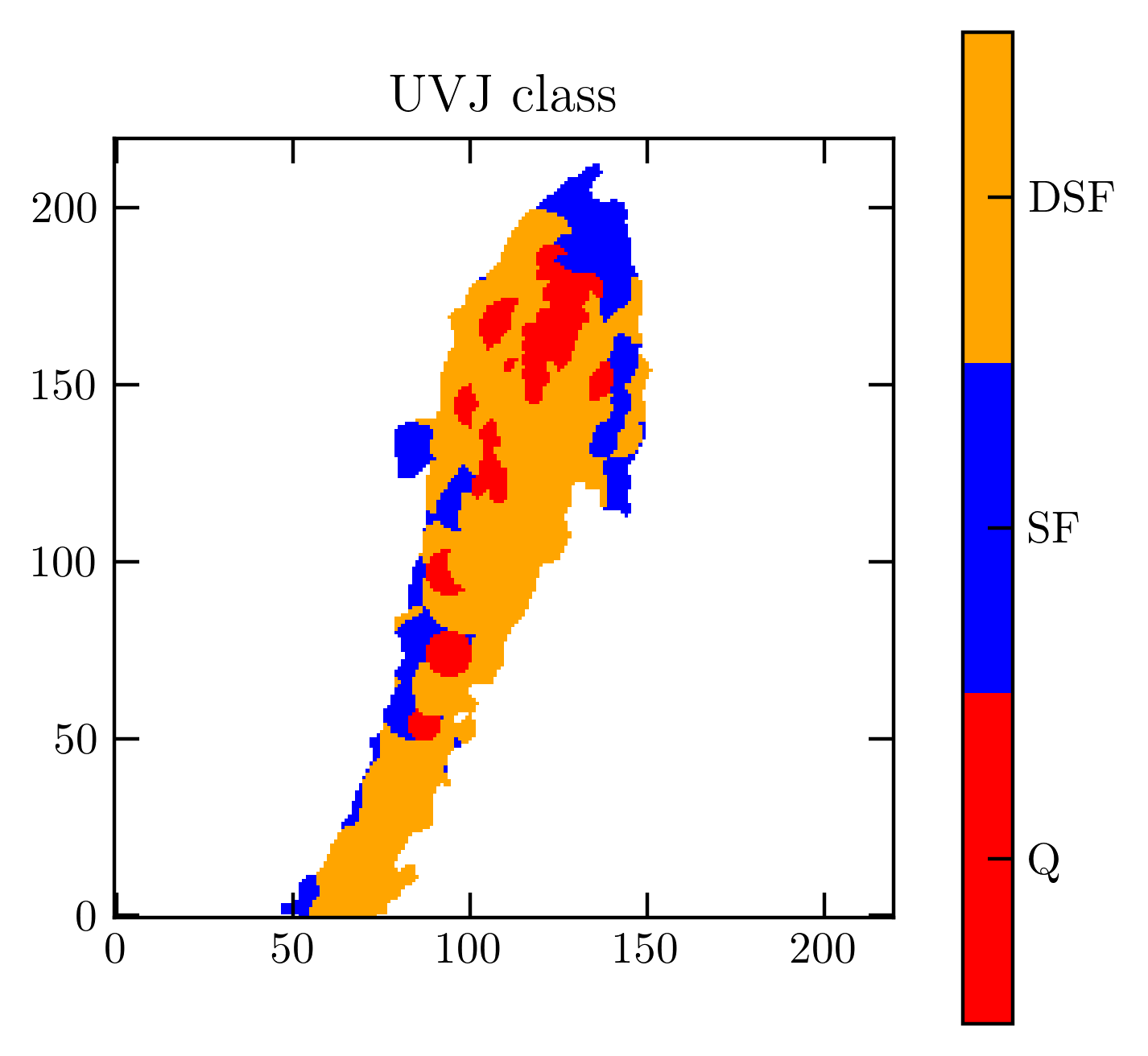}
\caption{
    Spatially-resolved image-plane $UVJ$ diagrams for DSFG-1 (Arc 1a), DSFG-3 (Arc 3c), and Arc 2a ({\it left panels}), and bin maps of $UVJ$ classifications ({\it right panels}); these are Quiescent (Q), Star-forming (SF), and Dusty Star-forming (DSF).
    The fields of view match those in Figs.~\ref{fig:pixbin_1a}, \ref{fig:pixbin_3c}, and \ref{fig:pixbin_2a}. The selection of quiescent galaxies is from \citet{Muzzin:2013aa}, while the distinction of SF vs. DSF is from \citet{Miller:2022ab}. Colors of points in the $UVJ$ diagrams show sSFRs derived from the full observed-frame near-IR SED fit. 
    A transparent gray square shows the location of the global $UVJ$ colors for each arc.
    The black vector in the bottom panel indicates the effect of increasing $A_V$ by 1 mag, per the \citet{Calzetti:2000aa} attenuation curve.
    \label{fig:UVJ}
}
\end{figure*}

\subsection{Resolved $UVJ$ diagrams}
\label{sec:UVJ}

The plane of $U-V$ and $V-J$ colors (or the well-known $UVJ$ diagram) is an efficient method to discern between stellar populations that are red due to age vs. those that are reddened by dust (e.g., \citealt{Labbe:2005aa,Wuyts:2007aa, Williams:2009aa, Brammer:2011aa, Whitaker:2011aa}).
For objects at $z\approx 2$, it was not practically feasible before JWST to reach the same resolution for rest-frame $J$-band (near-IR) imaging as for rest-frame (optical) $U$ and $V$.
With the advent of NIRCam, an essentially direct conversion from F115W, F200W, and F356W photometry to rest-frame $U$, $V$, and $J$ (respectively) at $z\approx 2$ is now possible.
For the range of $1.7 < z < 2.3$, \citet{Miller:2022ab} calibrated a linear relation for rest-frame $U-V$ and $V-J$ colors as a function of redshift and photometry in the NIRCam filters.
While the classification of stellar populations with the $UVJ$ diagram is imperfect, and made more uncertain by the assumptions of the conversion to rest-frame colors, it benefits from a reliance on very few pieces of information. No SED templates are imposed directly, and only 3 photometric measurements are employed (with which a robust SED fit would hardly be practical). 
Fortunately, the objects of interest in this work are all understood to lie within the redshift range for which the \citet{Miller:2022ab} calibration is valid, so we use this semi-independent method alongside the measurement of sSFR from the SED fits (\S \ref{sec:resolved_SED_maps} and \S \ref{sec:SN_host_maps}).

The left panels of Fig.~\ref{fig:UVJ} plot the rest-frame $U-V$ and $V-J$ colors of the bins in the DSFG-1 and DSFG-3 alongside arc 2a, all still shown in the image plane. As differential magnification is negligible within the small bin size (i.e., $U$, $V$, and $J$ are all magnified equally in each bin), there is no need to correct the $UVJ$ colors. 
The marker colors indicate sSFR as derived from the SED.
Before introducing the selection criteria, we first remark on the approximate similarity in the locus for Arcs 1a and 3c.
Arc 2a occupies a clearly different region, in particular towards redder $U-V$ colors.  
These reddest $U-V$ bins include substantially lower sSFRs, about 3 orders of magnitude lower than the lowest-sSFR bins in DSFG-1 and DSFG-3. 
Referring to Fig.~\ref{fig:pixbin_2a}, we find unsurprisingly that these are concentrated at the nucleus (central kpc) of the galaxy. 
Given the high stellar mass density here, $\Sigma_\star > 10^{10.5}~M_\odot~{\rm kpc}^{-2}$, it seems likely that this galaxy has an early formation redshift \citep{Estrada-Carpenter:2020aa} and has already built up a dense stellar bulge and started quenching inside-out (e.g., \citealt{Fang:2013aa,  Ji:2022aa, Ji:2022ab, Ji:2023aa}). This is consistent with (but not necessarily implicitly due to) the gravitational stabilization of gas by an increasing dominance of a spheroidal bulge component, known as morphological quenching \citep{Martig:2009aa}.

We use the higher-redshift selection box for quiescent galaxies defined by \citet{Muzzin:2013aa}. While the distinction between star-forming vs. dusty star-forming populations is more arbitrary and less pronounced, we also show the $V-J > 1.25$ cut from \citet{Miller:2022ab}. 
Differential magnification may indeed impact the global colors for each of the three arcs (shown as gray squares in Fig.~\ref{fig:UVJ}; see discussion of global properties in \S \ref{sec:integrated}). We find that DSFG-1 and DSFG-3 lie in very similar locations, safely in the star-forming/dusty star-forming region, but essentially close to the dividing line between the two.
Arc 2a, on the other hand, lies on the cusp of the division between quiescent and dusty star-forming. 
The right-column panels of Fig.~\ref{fig:UVJ} show the classification of bins from the $UVJ$ diagram as quiescent (Q), star-forming (SF), and dusty star-forming (DSF). 
Curiously, while there is a greater presence of quiescent regions in the massive galaxy seen in Arc 2a (relative to the DSFG arcs 1a and 3c), a clear majority of the galaxy's area consists of stellar populations consistent with the dusty star-forming region of the $UVJ$ diagram, with only its outskirts being classified as less-obscured star-forming.
Extending the quiescent selection box slightly to redder $V-J$ colors would change the classification of a large number of bins from dusty star-forming to quiescent. 
Yet, the overall classification of this object as quiescent would likely be erroneous, despite its low sSFR ($\sim 0.08 ~ {\rm Gyr}^{-1}$; \citetalias{Frye:2024aa}).
Less-obscured star formation is found in the outskirts of the galaxy, and the galaxy center (with peak $\Sigma_\star$) is coincident with the quiescent bins, perhaps signifying inside-out quenching.
In \S \ref{sec:SN_host_maps} and Fig.~\ref{fig:rSFMS_2a}, we found that 57\% of the galaxy's area was consistent with being above the center of the rSFMS evolved to $z=1.78$. 
In other words, $<43\%$ was quiescent based on sSFR, or at least on the lower end of the resolved main sequence;
now, using the independent $UVJ$ selection, we find that $\sim 20\%$ of the area is classified as quiescent.

For the DSFGs, Arcs 1a and 3c, there are remarkably few regions that fall into the quiescent selection. For Arc 1a, there is a clear separation with the southeast portion dominated by dusty star-forming populations and a northwest portion of regular star-forming populations. This could further support the orbiting clumps hypothesis (\S \ref{sec:resolved_SED_maps}), with both components actively forming stars, but one hosting greater dust content (or at least, somehow more obscured due to projection effects, although this region is actually at a lower redshift). 
As for Arc 3c, the inner regions of the galaxy are DSF with outskirts that are classified as SF, with isolated embedded quiescent regions. These seem consistent with the observation of multiple peaks in $\Sigma_\star$, perhaps as a result of a late-stage merger or the large-scale fragmentation of gas within the disk.
In both DSFG-1 and DSFG-3, the most active star formation (highest sSFR) is clearly concentrated in the DSF box, which is to be expected given the connection between starbursts and dust obscuration (e.g., \citealt{Whitaker:2017aa}).

\begin{figure*}[ht!]
\centering
\includegraphics[width=\textwidth]{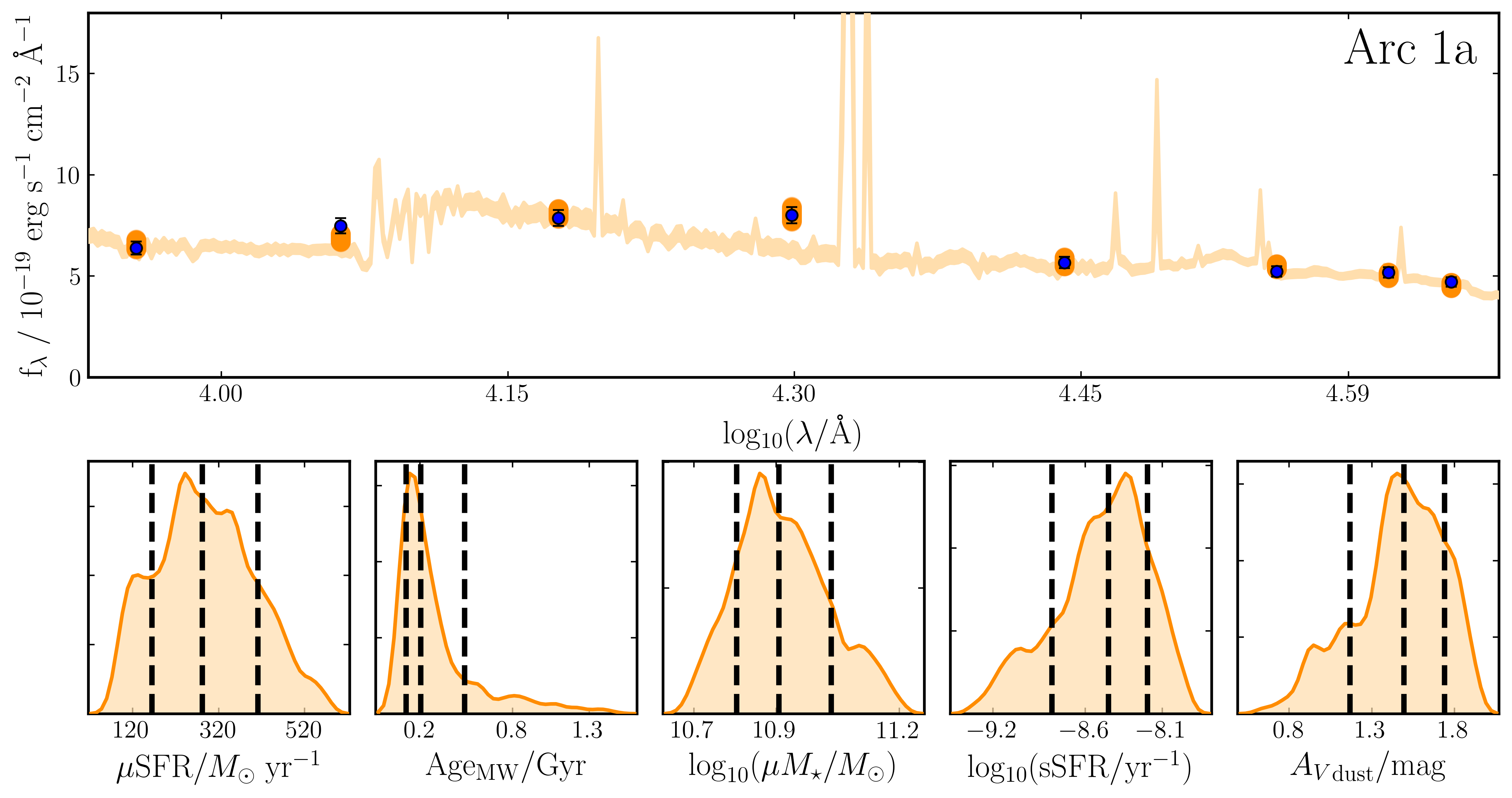}
\includegraphics[width=\textwidth]{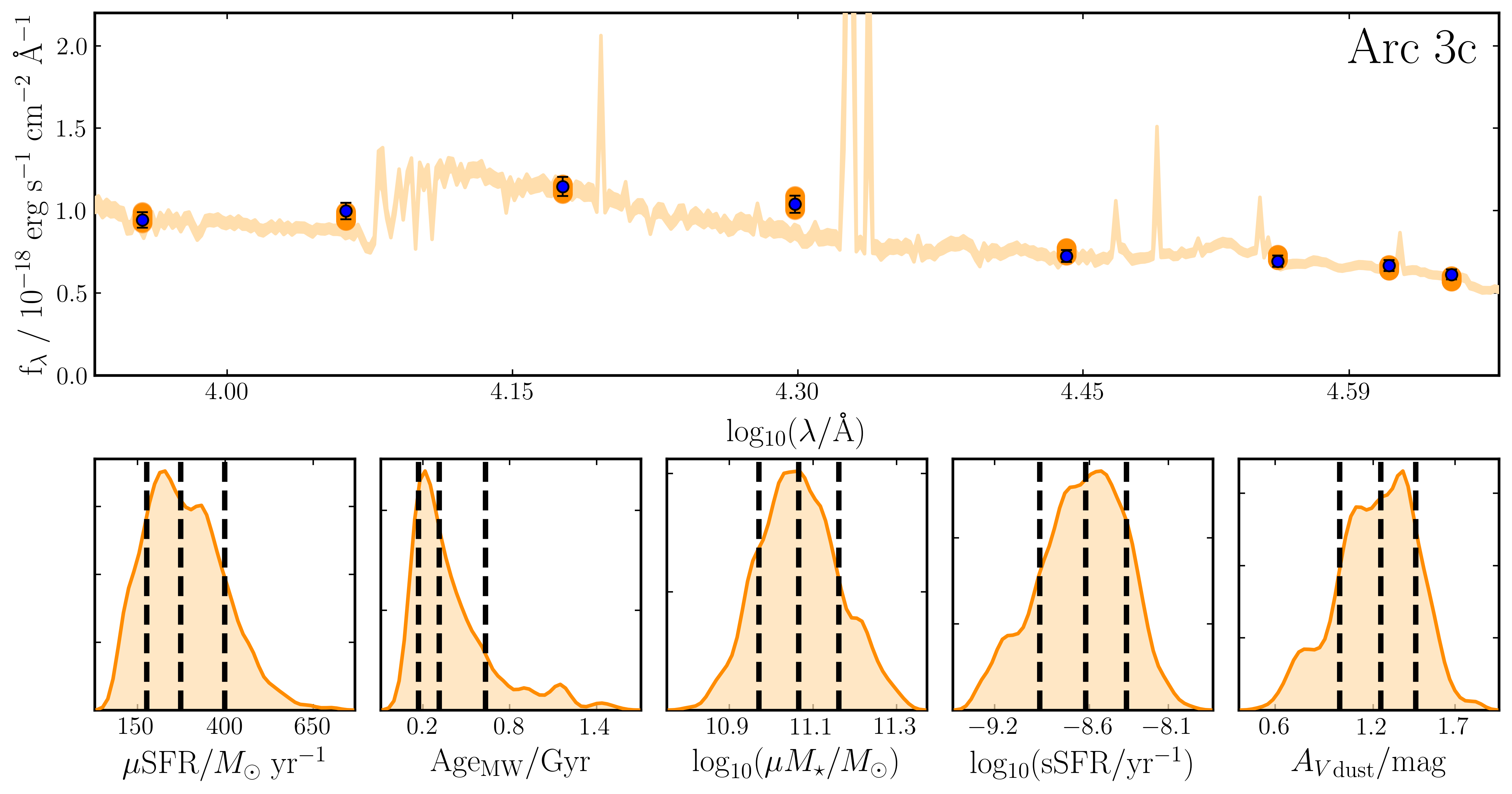}
\caption{Best-fit near-IR SED and properties from {\sc bagpipes} for Arc 1a ({\it top}) and Arc 3c ({\it bottom}). 
    The top subpanels show the measured photometry as blue data points, with the best-fit SED as an orange curve. The expected photometry from the SED is shown as orange points.
    The bottom subpanels of each show the posterior distribution for apparent SFR (before correction for magnification), the mass-weighted age, the apparent (pre-lensing correction) stellar mass, the sSFR, and the dust attenuation $A_V$. Dashed vertical lines show the 16th, 50th, and 84th percentiles.
    These are distinct objects (not multiple images) but are very similar in these given properties.
    \label{fig:SED_1a_3c}
}
\end{figure*}

\subsection{A comparison of integrated vs. resolved properties}
\label{sec:integrated}

In addition to the spatially-resolved SED fitting, we also use {\sc Bagpipes} to measure global properties of G165-DSFG-1 and G165-DSFG-3. For simplicity, we measure the photometry of the counterimage arcs 1a and 3c, as these contain the entire object (in contrast with the high-magnification arcs 1bc and 3ab, which arise from only a small fraction of their respective sources). Moreover, the magnification gradient for these counterimages, while not entirely negligible, is much smaller. After convolving all filters with a kernel to match the PSF of the coarsest-resolution filter (F444W), we use {\sc photutils} \citep{Bradley:2022aa} to capture Kron aperture fluxes \citep{Kron:1980aa}.

We fit the 8-filter photometry with {\sc bagpipes} in an identical manner as for the pixel-by-pixel fluxes, described in \S \ref{sec:pixelbypixel}, using the same set of priors.
While we measure photometry at 2 mm, 3 mm, and 6 GHz (Table \ref{tab:magnifications}), these are only included in a combined SED fit for the sum of Arc 1a, 1bc, 3ab, and 3c.
The resolution of these long-wavelength observations ($\sim 1 \farcs 1$, $\sim 1 \farcs 3$, and  $\sim 0 \farcs 5$, respectively) are much coarser than what is enabled by JWST (see Fig.~\ref{fig:SP_1a_3c}).
Moreover, the synthesized beam of the ALMA observations is highly elongated, owing to this being a high-declination ($+42\degrees$) target for ALMA's southern latitude. Unfortunately, this elongation is essentially aligned with the orientation of greatest magnification, so lensing does not offer an easy remedy. 
With these parameters of the currently-available data, it is therefore not worthwhile to attempt spatially-resolved SED fitting with both ALMA and JWST, as only a few resolving elements cover the extent of the arcs. 

\startlongtable
\begin{deluxetable*}{c|ccccccc}
\tablecaption{Properties of DSFG-1 and DSFG-3 from total IR dust SED, and corresponding SN rates.\label{tab:SFR_IR}}
\tablehead{
\colhead{Object} & \colhead{$\mu_{\rm 2 mm} {\rm SFR}_{\rm IR}^\dagger$} & \colhead{SFR$_{\rm IR}$} & \colhead{log[$L_{\rm IR} /L_\odot$]}  & \colhead{$R_{\rm CC}^\ddagger$} & \colhead{$R_{\rm CC} (1+z)^{-1}$} & \colhead{$R_{e,2{\rm mm}}$} & \colhead{$R_{e,2{\rm mm}}$}\\
\colhead{} & \colhead{[$M_\odot~{\rm yr}^{-1}$]} & \colhead{[$M_\odot~{\rm yr}^{-1}$]} & \colhead{} &  \colhead{[yr$^{-1}$]}  &  \colhead{[yr$^{-1}$]}  & \colhead{[arcsec]}  & \colhead{[kpc]}
}
\startdata
G165-DSFG-1  & $9100 \pm 1900$ & $ 320 \pm 70$  & $12.5 \pm 0.1$ &  $3.3 \pm 0.7$ &	$1.0 \pm 0.2$ & $0.20 \pm 0.02$ &	$1.6 \pm 0.2$ \\
\hline
G165-DSFG-3  & $2900 \pm 500$ & $400 \pm 80$ & $12.6 \pm 0.1$  &  $4.2 \pm 0.8$  & $1.3 \pm 0.2$	& $0.25 \pm 0.03$ & $2.1 \pm 0.2$\\
\enddata
\tablenotetext{^\dagger}{The total apparent (i.e. lensing-uncorrected) IR-derived SFR of $12000\pm 2000 ~M_\odot {\rm yr}^{-1}$ from \citetalias{Harrington:2016aa} is divided between Arcs 1abc and 3abc according to their ratio of millimeter flux, $S_{\rm 1abc} / S_{\rm 3abc} = 3.1 \pm 0.3$ (averaged for 2 mm and 3 mm).}
\tablenotetext{^\ddagger}{Expected rate of core-collapse supernovae tabulated as $R_{\rm CC} = k_{\rm CC} \cdot {\rm SFR}$; 
$k_{\rm CC}=0.0104 ~ M_\odot^{-1}$ 
(e.g., \citealt{Strolger:2015aa}). These are intrinsic SN rates; observed SN rates will be dilated by a factor of $(1+z)^{-1}$, but also increased for any SNe occurring in regions of the galaxy that are multiply imaged.} 
\tablecomments{
We derive the SFR of the two DSFGs independently using the far-IR photometry \citepalias{Harrington:2016aa} vs. the rest-frame UV-NIR photometry from NIRCam (which is shown in Table~\ref{tab:SFR_SED}). 
As the far-IR photometry largely does not resolve Arcs 1abc and 3abc, we make a preliminary estimate by apportioning the total ${\rm SFR}_{\rm IR}$ according to their share of flux at 2 mm and 3 mm (combining Arcs 1a and 1bc, and Arcs 3ab and 3c), as both arc systems are understood to be at similar enough redshifts. Then, these apportioned SFRs are corrected for the total 2 mm magnifications of Arc 1abc and Arc 3abc, respectively.
Based on these various SFRs, we estimate the rate of occurrence of core-collapse supernovae (for the rest-frame of each galaxy); the observable rate is likely bracketed by the SED- vs. IR-inferred SFRs, given that SNe in some very dusty environments may be missed (or, alternatively, the extreme star formation may lead to a porosity in the dust geometry that lends to a more favorable chance for discovery). 
}
\end{deluxetable*}

However, 
we can make a comparison of SED-fitting with vs. without the inclusion of far-IR rest-frame photometry for the unresolved, combined SED of both DSFG-1 and DSFG-3 together. 
Flux measurements that better constrain the peak of the dust SED at rest-frame 100 $\mu$m are available (\citealt{Canameras:2015aa}; \citetalias{Harrington:2016aa}), but these all have resolutions of $8 \farcs 5$\textemdash in the case of the LMT AzTEC\textemdash or much larger.
In Appendix~\ref{sec:appendix4}, we sum the NIR photometry of the four arcs to revisit the SED fit of \citet{Harrington:2016aa}.
As shown by \citet{Battisti:2019aa}, the stellar mass of DSFGs can be underestimated by 0.3 dex with only UV to near-IR coverage (a discrepancy that is greatest for objects with $M_\star > 10^{10}~M_\odot$). 
More broadly, for galaxies simulated with the Evolution and Assembly of GaLaxies and their Environments (EAGLE; \citealt{Schaye:2015aa})
project
and
the SKIRT dust radiative transfer code \citep{Baes:2011aa},
\citet{Dudzeviciute:2020aa} found that even with UV-to-radio coverage, SED-modeling resulted in
a systematic under-prediction of stellar mass by $0.5 \pm 0.1$ dex.
It also appears to be the case that inclusion of far-IR photometry in an SED fit tends to reduce the inferred SFR \citep{Pacifici:2023aa}, although \citet{Wise:2023aa} found that SFR can be under- or over-estimated (up to 1 dex) from UV to near-IR alone. For galaxies with lower SFR$_{\rm UV-NIR} \lesssim 300~M_\odot~{\rm yr}^{-1}$, they found that adding FIR data boosted the estimated SFR; for high SFR$_{\rm UV-NIR}$ galaxies, FIR data lowered the estimated SFR. 
Indeed, in our case, DSFG-1 and DSFG-3 have a combined SFR$_{\rm UV-NIR} \approx 90~M_\odot~{\rm yr}^{-1}$, but the unresolved UV-FIR SED yields SFR$_{\rm UV-FIR} = 350 \pm 70~M_\odot~{\rm yr}^{-1}$, a $\sim 0.6$ dex increase
(Table~\ref{tab:SFR_SED}).
As is also emphasized by \citet{Smail:2023aa}, spatially resolved ``pixel-by-pixel" SED modeling is made significantly more robust by the inclusion of long-wavelength photometry at similar angular resolution, highlighting the particular need for submillimeter interferometry in this analysis.

\startlongtable
\begin{deluxetable*}{c|ccccccc}
\tablecaption{Properties of DSFG-1 and DSFG-3 from rest-frame UV through near-IR SED fitting, and corresponding SN rates. \label{tab:SFR_SED}}
\tablehead{
\colhead{Object} &  \colhead{$\mu_{\rm 2\mu m} {\rm SFR}$} & \colhead{${\rm SFR}$} & \colhead{$R_{\rm CC}^\ddagger$} & \colhead{log[$\frac{\mu_{\rm 2 \mu m} M_\star}{M_\odot}$]} & \colhead{log[$\frac{M_\star}{M_\odot}$]} & \colhead{log[$\frac{\rm sSFR}{{\rm yr}^{-1}}$]} & \colhead{$A_V$} \\
\colhead{} & \colhead{[$M_\odot~{\rm yr}^{-1}$]} & \colhead{[$M_\odot~{\rm yr}^{-1}$]} & \colhead{[yr$^{-1}$]} & \colhead{} & \colhead{} & \colhead{} &\colhead{[mag]} 
}
\startdata
Arc 1a (SED)  &   $280 \pm 130$	& $50 \pm 20$	& $0.5 \pm 0.2$	& $10.9 \pm 0.1$ & $10.2 \pm 0.1$ & $-8.5 \pm 0.3$ &	$1.5 \pm 0.3$  	\\
Arc 1a (H$\alpha$)  & $320 \pm 140^\dagger$  	& $60 \pm 30^\dagger$	& $0.6 \pm 0.3$	& $10.5 \pm 0.1^\dagger$  & $9.7 \pm 0.1^\dagger$ &  $-8.4 \pm 0.3^\dagger$ &	$1.5 \pm 0.4^{\dagger\dagger}$   	\\
Arc 1a (bin-wise sum)  & $380 \pm 10$  	& $69 \pm 3$	& $0.72 \pm 0.03$	& $10.91 \pm 0.01$  & $10.17 \pm 0.02$ &  $-8.34\pm 0.02$ &	$\approx 1.0 \pm 0.8^{\dagger\dagger}$   	\\
\hline
Arc 3c (SED)  &   $260 \pm 110$	& $40 \pm 20$	& $0.4 \pm 0.2$	& $11.1 \pm 0.1$ & $10.3 \pm 0.1$ & $-8.7 \pm 0.3$  &	$1.2 \pm 0.2$  	\\
Arc 3c (bin-wise sum)  &   $440 \pm 10$	& $75 \pm 3$	& $0.78 \pm 0.03$	& $11.06 \pm 0.01$ & $10.29 \pm 0.02$ & $-8.42 \pm 0.02$  &	$1.2 \pm 0.7$  	\\
\hline\hline
Arcs 1abc+3abc (UV-FIR)$^\star$ & $5560 \pm 280$ & $350 \pm 70$	& $3.6 \pm 0.7$	& $11.68 \pm 0.03$ & $10.48 \pm 0.09$ & $-7.93 \pm 0.03$  &	$2.54 \pm 0.04$  	\\
\enddata
\tablenotetext{^\ddagger}{Expected rate of core-collapse supernovae tabulated as in Table~\ref{tab:SFR_IR}.} 
\tablenotetext{^\dagger}{The H$\alpha$-derived SFR and stellar masses from \citetalias{Frye:2024aa} are the summed values from NIRSpec apertures NS\_46 and NS\_969. The intrinsic (magnification-corrected) properties use the 2 $\mu$m magnifications from Table~\ref{tab:magnifications}. The sSFR is determined from this SFR$_{\rm H\alpha}$ and the stellar mass, which is determined from an SED model that incorporated H$\alpha$ flux as a constraint.} 
\tablenotetext{^{\dagger\dagger}}{As described in \S \ref{sec:integrated}, the dust attenuation $A_V$ is estimated from the Balmer-decrement derived $E(B-V)_{\rm gas} = 0.85 \pm 0.25$ for NS\_969 \citepalias{Frye:2024aa} and a \citet{Calzetti:2000aa} reddening law, $R_V = A_V / E(B-V) = 4.05$. We lastly impose a conversion to a stellar continuum color excess through $E(B-V)_{\rm stellar} = 0.44 E(B-V)_{\rm gas}$, per \citet{Calzetti:1997ab}. For the bin-wise dust attenuation, we provide the inner 68\% interval for the bins' values.} 
\tablenotetext{^\star}{Here, we use the total flux-weighted magnification for both DSFG-1 and DSFG-3 combined, $\mu = 16 \pm 3$, to derive intrinsic values.
}
\tablecomments{
We compute the UV-NIR SED-derived SFRs for Arcs 1a and 3c (see Fig.~\ref{fig:SED_1a_3c}), and correct these for respective magnifications at $2 \mu$m. These values are lower than the SFR$_{\rm IR}$ estimates, but this is expected given their large dust attenuations.
In Appendix~\ref{sec:appendix4}, we fit the summed SED of both DSFGs so that the unresolved far-IR photometry can be incorporated.
The intrinsic, combined SFR$_{\rm UV-FIR}$ is $\approx 4$ times the sum of SFR$_{\rm UV-NIR}$ for the two DSFGs, while the summed stellar masses are more consistent with the integrated value. 
We also compare with the properties derived through the H$\alpha$ measurement by \citetalias{Frye:2024aa}, incorporating the two NIRSpec apertures that covered this object (but applying our 2 $\mu$m magnifications, from Table \ref{tab:magnifications}). In all cases, we find quite close agreement with our independently-derived values.
}
\end{deluxetable*}

\citet{Zibetti:2009aa} first presented an important bias to consider when inferring the stellar mass of a galaxy: mass estimates based on global (galaxy-integrated) fluxes were 40\% (0.2 dex) lower than those determined by summing over the map of stellar mass derived from spatially-resolved SED fitting.
This effect is the direct result of the stellar mass contribution from dust-obscured regions being underestimated through the unresolved photometry (although \citealt{Sorba:2015aa, Sorba:2018ab} point out that an ``outshining" bias from the youngest stellar populations may be responsible, as well). 
\citet{Song:2023ab} found recently that without coverage of rest-frame near-IR in SED fitting, $M_\star$ can be overestimated by 0.2 dex, owing directly to an overestimate of both dust attenuation and stellar ages from only rest-frame UV and optical photometry. 
As shown in Table~\ref{tab:SFR_SED}, there is excellent agreement in the stellar masses determined through integrated photometry vs. a sum of stellar masses derived through resolved bin-wise SED fitting with {\sc piXedfit}, which may indeed be the result of photometric coverage out to rest-frame 1.4 $\mu$m.

Table~\ref{tab:SFR_SED} also compares with the H$\alpha$-inferred SFR and dust attenuation properties presented by \citetalias{Frye:2024aa} for G165-DSFG-1.
These are the result of combining measurements from NIRSpec slits NS\_46 and NS\_969, the latter of which is labeled ``Arc 1a" in their Table 4.
We find strong agreement for SFR, especially within uncertainties: global $\mu{\rm SFR}_{\rm NIR} = 280 \pm 130~M_\odot~{\rm yr}^{-1}$ vs. spatially-integrated  $\mu{\rm SFR}_{\rm NIR} = 380 \pm 10~M_\odot~{\rm yr}^{-1}$ vs. $\mu{\rm SFR}_{{\rm H}\alpha} = 320 \pm 140~M_\odot~{\rm yr}^{-1}$.
The stellar mass derived by \citetalias{Frye:2024aa} is not completely independent of our result, as they used a joint photometric and spectroscopic SED fit with FAST$++$ \citep{Kriek:2009ab, Schreiber:2018aa}.
Whether due to the different SED fitting algorithms or to the exclusion of spectroscopic information, the stellar mass we estimate\textemdash log$[\mu M_\star / M_\odot] = 10.9 \pm 0.1$\textemdash is 0.4 dex larger than the value quoted in \citetalias{Frye:2024aa}.
Our assumption of a ``single" SFH in the form of an exponentially declining ``tau" model may underestimate the stellar mass by $0.1 - 0.3$ dex relative to a multi-component SFH \citep{Michaowski:2014aa, Carnall:2019ab, Jain:2023aa}.
Given the typical uncertainty of 0.1 dex in stellar mass for SED modeling at $z>1$ (e.g., \citealt{Pacifici:2023aa}), 
or even slightly elevated uncertainty of $\sim 0.3$ dex for DSFGs \citep{Michaowski:2014aa},
it is perhaps most likely that the statistically significant discrepancy is explained by the information provided by the near-IR spectroscopy (or perhaps the extrapolation from limited slit coverage of the source).

Finally, we compare the estimate of dust attenuation that we find from modeling the UV-NIR SED against that derived from the Balmer decrement in H$\alpha$ vs. H$\beta$ from \citetalias{Frye:2024aa}. 
They measured $E(B-V) = 0.85 \pm 0.25$ for NS\_969 (with which we choose to compare here, given its larger uncertainty). For a \citet{Calzetti:2000aa} reddening law, or $R_V = A_V / E(B-V) = 4.05$, this yields $A_V = 3.0 \pm 1.0$.
However, this discrepancy with our smaller value of $A_V = 1.5 \pm 0.3$ can be 
explained by the typically larger color excess for ionized gas relative to stellar continuum. \citet{Calzetti:1997aa} quantify this as $E(B-V)_{\rm stellar} = 0.44 \times E(B-V)_{\rm gas}$ (cf. \citealt{Shivaei:2015aa}). Applying this correction to the value from \citetalias{Frye:2024aa}, we recover $A_V = 1.5 \pm 0.4$, in superb agreement with our finding.
For the diffuse ISM in the outskirts of galaxies, 
the ratio of 
$E(B-V)_{\rm stars} / E(B-V)_{{\rm H}\textsc{ii}}$ is larger than for stellar birth clouds in star-forming regions, and even approaches unity (e.g., \citealt{Robertson:2024aa}).
Our finding that a ratio of 0.44 is appropriate for DSFG-1 may actually be in line with our interpretation of widespread star-forming regions, but this claim is limited for now by the current spectroscopic coverage of just a narrow region of the arc.

\section{Discussion}
\label{sec:discussion}

\subsection{How $UVJ$ selection breaks down on resolved scales, and possible evidence for radial dust transport within the disk}
\label{sec:dust_transport}

There are at least a dozen bins for Arc 2a in Fig.~\ref{fig:UVJ} with very low sSFRs (log$[{\rm sSFR}/{\rm yr}^{-1}] < -11.5$) that are still safely outside of the classical quiescent selection region, and would instead be classified as dusty star-forming populations. 
The quiescent sequence has been found to evolve with age\textemdash \citealt{Belli:2019aa} found a best-fit relation for the median stellar age of ${\rm log}(t_{\rm 50}~ [{\rm yr}]) = 7.03 + 0.84(V-J) + 0.74(U-V)$.
These redder colors in resolved $U-V$ and $V-J$ can thus partly be the result of older stellar populations ($\sim 3$ Gyr),
but this model still severely underpredicts the finding of $V-J \approx 1.7$ and $U-V \approx 2.3$.
Ultimately, the most likely explanation seems to be that these quiescent regions are more dust-attenuated than typically expected. The reddening vector of $\Delta A_V$ shown in Fig.~\ref{fig:UVJ} implies that several of these low-sSFR bins have at least $A_V > 0.5$ mag.
This may also be borne out by the finding of $A_V \gtrsim 1$ for nearly the entirety of Arc 2a in Fig.~\ref{fig:pixbin_2a}.
As some added confirmation, SN H0pe is located nearby this dusty quiescent region, and estimated via light-curve fitting to be attenuated by $A_V = 1.21 \pm 0.05$ (\citealt{Pierel:2024aa}).
While dust in quiescent galaxies is still very much an area of active research, owing to the difficulty this presents for observations (e.g., \citealt{Gomez:2010aa, Rowlands:2012aa,  Magdis:2021aa, Donevski:2023aa}; and references therein), they are often assumed to be quite dust-poor. 
Revisions to the original $UVJ$ selection frequently exclude the $V-J < 1.5$ cut to accommodate older and dustier quiescent galaxies
\citep{van-der-Wel:2014aa, Whitaker:2015aa, Belli:2019aa}.
For Arc 2a, this correction would appropriately enclose the low-sSFR bins within the quiescent regime, but would also erroneously include regions with sSFRs that are over 2 dex larger (sSFR $\gtrsim 0.5$ Gyr$^{-1}$).

This might indicate a 
shortcoming of our interpretation of the classical $UVJ$ selection when applied to resolved scales.
In a similar way that the resolved star formation main sequence breaks down at $\sim 100$ pc and smaller, owing to stars migrating from their parent star-forming clouds, 
the $UVJ$ diagram presumably also breaks down when significant amounts of dust and metals formed in regions of active star formation are transported to adjacent, more quiescent regions.
This can occur over short distances within the plane of the disk through turbulent motions in the ISM
\citep{Cho:2003aa},
but 
\citet{Vorobyov:2006aa}
showed that spiral stellar density waves could help shepherd dust several kpc in the radial direction over the course of $\sim 1$ Gyr (see also \citealt{Mishurov:2014aa}).

The extra reddening seen may also be due to extraplanar dust that has been transported a few kpc into the circumgalactic medium by the likes of galactic-scale winds driven by supernovae and winds from massive stars \citep{MacLow:1989aa, Norman:1989aa, Heckman:1990aa, Heckman:2017aa, Richie:2024aa} or simply through momentum injection by radiation pressure (e.g., \citealt{Ferrara:1991aa, Murray:2005aa, Thompson:2015aa,  Barnes:2020aa, Kannan:2021aa}).
However, it is not clear if extraplanar dust (or even a thick disk of dust) can constitute a large enough fraction of the total dust mass, as it is 
estimated 
to be $<5\%$ \citep{Howk:2000aa, Popescu:2000aa, Bianchi:2011aa, Seon:2014aa}.
The vertical stability of the disk also plays a role, as diffuse distributions of dust appear to be more predominant in low-mass, slow-rotating disks \citep{Dalcanton:2004aa, Holwerda:2012aa}.
This extraplanar dust is also typically thought to be highly filamentary in nature, associated with chimney vents and extending a few kpc from the midplane
(\citealt{Wang:2001aa, Stein:2020ab, Mosenkov:2022aa}; see also review by \citealt{Veilleux:2020aa}).
It is 
unlikely to provide the necessary column densities to reproduce the observed $A_V$ over large scales \citep{Thompson:2004aa}, 
especially when considering the effect of grain destruction by sputtering (e.g., \citealt{Spitzer:1978aa}).
Admittedly, 
widespread large attenuations do not necessitate uniform, thick screens of dust; 
they merely need large quantities of dust to be spatially correlated with regions of visible starlight.
We thus consider it most likely that the bulk of the dust responsible for the attenuation is still largely contained (and well mixed) within the disk.
Over larger, galaxy-wide scales, this effect is likely to be less of a concern, as $UVJ$ colors will be biased in favor of the unobscured sightlines. 
For globally quiescent objects, this means that the more dust-free regions with older stellar populations will dominate. This appears to be the case for the global colors of Arc 2a, as the object just barely falls within the $UVJ$-quiescent regime. 
Relatively few studies so far have examined spatially-resolved $UVJ$ diagrams in the era of JWST (including \citealt{Miller:2022ab, Kokorev:2023aa, Suess:2023aa}), 
so our speculation on the effect of dust transport is necessarily limited.
Yet, we can expect a better understanding of these systematics (and in what other circumstances the selection criteria break down; e.g., \citealt{Leja:2019aa}) with time.

\subsection{Starbursts or not? The proximity of DSFG-1 and DSFG-3 to the star-forming main sequence}
\label{sec:deltaMS}

In \S \ref{sec:rSFMS} we considered the location of spatially-resolved regions relative to the resolved star formation main sequence, but now we examine the position of DSFG-1 and DSFG-3 as a whole.
For both DSFG-1 at $z=2.236 ~(t = 2.88~{\rm Gyr})$ and DSFG-3 at $z=2.1 ~(t = 3.07~{\rm Gyr})$, the \citet{Speagle:2014aa} model for the evolution of the main sequence\footnote{The main sequence follows ${\rm log}[{\rm SFR}(M_\star, t)] = (0.84 \pm 0.02 - 0.026\pm 0.003 \times t)~{\rm log}M_\star - (6.51 \pm 0.24 - 0.11 \pm 0.03 \times t)$, for the age of the Universe $t$ in Gyr. This model uses a \citet{Kroupa:2001aa} IMF, consistent with our SFRs.}
yields
${\rm log}[{\rm SFR}(M_\star, t)] \approx 0.8~{\rm log}M_\star - 6.2$.
Using our derived stellar mass of DSFG-1 of ${\rm log}M_\star = 10.2$, we find 
${\rm SFR}_{\rm MS} \approx 90~M_\odot~{\rm yr}^{-1}$.
The IR-inferred SFR$_{\rm IR} \approx 320~M_\odot~{\rm yr}^{-1}$ is therefore approximately 3.6 times that of the main sequence at this redshift.
As for DSFG-3, with stellar mass ${\rm log}M_\star = 10.3$, 
${\rm SFR}_{\rm MS} \approx 110 ~ M_\odot~{\rm yr}^{-1}$, and the inferred 
SFR$_{\rm IR} \approx 400~M_\odot~{\rm yr}^{-1}$ is likewise approximately 3.6 times that of the main sequence at this redshift.
With distance to the main sequence defined as $\Delta_{\rm MS} \equiv {\rm log}_{10}({\rm SFR} / {\rm SFR}_{\rm MS}$, both cases yield $\Delta_{\rm MS} \approx 0.6$ dex.
Adopting a $\Delta_{\rm MS} \ge 0.6$ dex threshold (e.g., \citealt{Rodighiero:2011aa}) for starbursts above the main sequence\textemdash chosen to be a factor of 2 above the typical scatter of $\approx0.3$ dex
\citep{Brinchmann:2004aa, Daddi:2007aa, Elbaz:2007aa, Noeske:2007aa, Salim:2007aa, Whitaker:2012aa, Speagle:2014aa, Tacchella:2016ab}\textemdash this means that DSFG-1 and DSFG-3 are only just marginally classified as starbursts.

While DSFGs with such high SFRs are often incorrectly presumed to lie universally within the starburst regime, \citet{Rodighiero:2011aa} found that only $20\% \pm 4\%$ of galaxies with ${\rm SFR} > 100~M_\odot~{\rm yr}^{-1}$ satisfied the $+0.6$ dex criterion. 
Similarly, 
\citet{Barrufet:2020aa} found that approximately 60\% of DSFGs (in a sample of 185) have SFRs consistent with the star-forming main sequence.
Indeed, many claims have been made that 
DSFGs simply occupy the
high-$M_\star$ end of the SFMS 
(e.g.,
\citealt{Finlator:2006aa, Dunlop:2011aa, Magnelli:2012aa, Michaowski:2012aa, Sargent:2012aa, Targett:2013aa, Koprowski:2016aa, Dunlop:2017aa, Elbaz:2018aa, Drew:2020aa, Pantoni:2021ab}; contrast with \citealt{Hainline:2011aa}, for example).
Furthermore, for a set of 15 other lensed DSFGs in the PASSAGES sample (not including G165), \citet{Kamieneski:2024aa} found that proxies for $\Delta_{\rm MS}$ (namely, molecular gas depletion time and $\Sigma_{\rm SFR}$) predicted broad consistency with the upper end of the star-forming main sequence, but not necessarily classification as starbursts.
In this work, we have also found that the radial profiles of sSFR (Figs.~\ref{fig:slit_1a} and \ref{fig:slit_3c}) are broadly consistent with those expected for starbursts from \citet{Nelson:2021aa}.

The finding that the DSFGs in the G165 field can be classified as starbursts may have some connection to the finding that they reside within a highly active star-forming environment (and possible galaxy overdensity) by \citetalias{Frye:2024aa}\textemdash although \citealt{Drew:2020aa} contend that distance to the SFMS may be a poor indicator of recent merger activity.
Regardless, the locus of high-$z$ DSFGs on the SFR$-M_\star$ plane helps to quantify the role of gas supply-dependent, secular evolution \citep{Dave:2012aa}. 
A more systematic examination of the stellar masses of DSFGs\textemdash especially now, with JWST helping to circumvent their prohibitive degree of dust obscuration\textemdash will ultimately shed more light on their connection to the star-forming main sequence.
For now, the inconsistency of DSFGs and SMGs lying on or above the main sequence \citep{Hodge:2020ab} may ultimately arise from the apparent heterogeneity of the overall population (e.g., \citealt{Hayward:2011aa, Hayward:2012aa, Magnelli:2012aa, Hayward:2013aa}).

\subsection{Prediction of observer-frame supernova rates from G165-DSFG-1 and G165-DSFG-3, and prospects for future discovery}
\label{sec:SN_rates}

The connection of core-collapse supernova rates to SFRs is straightforward, as CCSN progenitors are short-lived massive stars, believed to range in mass from $8 - 50~M_\odot$ \citep{Nomoto:1984aa, Tsujimoto:1997aa}. 
The supernova rate can thus be tabulated as $R_{\rm CC} = k_{\rm CC} \cdot {\rm SFR}$, where the scale factor is derived from the IMF $\phi(M)$ as
\begin{equation}
k_{\rm CC} = \frac{\int_{8~M_\odot}^{50~M_\odot} \phi(M) dM}{\int_{0.1~M_\odot}^{125~M_\odot} M \phi(M) dM}
\end{equation}
and essentially represents the fraction of the forming stellar populations that consists of CCSN progenitors.
These lower and upper mass limits and a 
\citet{Kroupa:2001aa}
IMF yield 
$k_{\rm CC} = 0.0104~M_\odot^{-1}$
(e.g. \citealt{Strolger:2015aa}), which we adopt for this work to remain consistent with the SED fitting.
A \citet{Salpeter:1955aa} IMF for the same $8-50~M_\odot$ progenitor range would yield $k_{\rm CC} = 0.0070~M_\odot^{-1}$ 
(e.g., \citealt{Mattila:2001aa, Young:2008aa}), but this difference is essentially canceled out by the corresponding effect on the SFR (e.g., \citealt{Madau:2014aa, Ziegler:2022aa}).

With this in mind, the predicted rest-frame supernova rate for G165-DSFG-1 is 
$3.3 \pm 0.7~{\rm yr}^{-1}$ 
using the FIR-derived SFR, 
or $0.5 \pm 0.2~{\rm yr}^{-1}$ using the lower, UV-NIR SED-derived SFR (see Tables~\ref{tab:SFR_IR} and \ref{tab:SFR_SED}). 
For G165-DSFG-3, these rates are comparable: 
$4.2 \pm 0.8~{\rm yr}^{-1}$
(FIR) and 
$0.4 \pm 0.2~{\rm yr}^{-1}$ (UV-NIR SED). 
Before taking into account the multiplexing provided by multiple images of any SNe, these rates translate to observed-frame values\footnote{Cosmological time dilation decreases the apparent rate of occurrence for SNe, $R_{\rm CC, obs} = R_{\rm CC} \cdot (1+z)^{-1}$, but it also stretches out the time for which a given supernova's light curve is detectable.} for G165-DSFG-1 of $R_{\rm CC, obs} \approx 0.2$ yr$^{-1}$ (SED) to $\approx 1.0$ yr$^{-1}$ (FIR), and for G165-DSFG-3 of $R_{\rm CC, obs} \approx 0.1$ yr$^{-1}$ (SED) to $\approx 1.3$ yr$^{-1}$ (FIR).

However, for the most reliable rate estimate, we use the SFR from the best-fit UV-through-FIR SED for DSFG-1 and DSFG-3 combined, ${\rm SFR}_{\rm UV-FIR} = 350 \pm 70~M_\odot~{\rm yr}^{-1}$. 
This results in a total expected rest-frame rate of $R_{\rm CC} = 3.6 \pm 0.7~{\rm yr}^{-1}$, or a remarkable observer-frame rate of 
$1.1 \pm 0.2~{\rm yr}^{-1}$.
This very high rate is derived only from the pair of DSFGs; 
however, as discussed by \citetalias{Frye:2024aa}, the H$\alpha$-derived SFR for the objects associated with Arcs 1abc and 2abc groups surpasses 500 $M_\odot~{\rm yr}^{-1}$, which comes from only the unobscured mode of star formation. 
Given that these groups appear to signal galaxy overdensities at $z\sim 2$, it may well be the case that $\gtrsim 1 - 2$ CCSNe yr$^{-1}$ may be observable for the G165 cluster. 
This makes the G165 field an extremely attractive target for follow-up monitoring campaigns with JWST, as even the six Frontier Field clusters \citep{Lotz:2017aa} all together would provide only 0.9 CCSNe per year, with a cadence of four 1-hour visits with F150W within a year \citep{Petrushevska:2018ab}. 
Given the best-fit $A_V \approx 2.5$ from the UV-FIR SED, it is entirely feasible to detect the descendent supernovae with only modest integration with JWST. 
An appropriate cadence would be designed to span a baseline of at least several months, so that enough time elapses for a clearly observable light curve evolution of any SNe (which is subject to a factor of $1+z$ in cosmological time dilation).

This last point is crucial, and helps to answer the inevitable question of why no CCSNe have yet been discovered in G165, given that two follow-up visits were undertaken with JWST on 22 April 2023 and 9 May 2023 after the discovery of the Type Ia SN H0pe in imaging on 30 March 2023. At only $+23$ d and $+40$ d from Earth's perspective, this is a mere $+8$ d and $+13$ d in the rest-frame at $z\sim 2$, which is hardly sufficient to guarantee a CCSN detection.
The cadence of observations for SN H0pe was designed for monitoring of a discovered event; a cadence for discovering new SNe will simply need to extend over a longer period of time.

\section{Summary and Conclusions} 
\label{sec:summary}

With this work, we have examined in detail the magnified, massive galaxies at $z\sim 2$ behind the G165 galaxy cluster, including two luminous dusty star-forming galaxies (log$(M_\star / M_\odot) \approx 10.2$; SFR$_{\rm IR}\approx 300-400~M_\odot~{\rm yr}^{-1}$) and a massive galaxy that appears to be exiting a phase of dusty star formation (log$(M_\star / M_\odot) \approx 10.8 - 11.0$ and SFR$_{{\rm H}\alpha}\approx 5-10~M_\odot~{\rm yr}^{-1}$; \citetalias{Frye:2024aa}).
This latter object has been the subject of intense interest as the host of a recently-discovered triply-lensed Type Ia supernova, SN H0pe.
Yet, the lensed DSFGs are responsible for the intense submillimeter flux by which the cluster was first discovered \citep{Canameras:2015aa, Harrington:2016aa, Frye:2019aa, Pascale:2022aa},
and we have sought to further elucidate their properties through recent imaging in the near-IR with JWST and in the millimeter-wave through ALMA.
We also present a new parametric lens model for the G165 cluster, developed with {\sc lenstool}. 
This model was constructed predominantly for the purpose of predicting lensing time delays for SN H0pe, in order to compare with values measured independently through the photometric and spectroscopic light curve, and thus infer the Hubble constant $H_0$.
However, this model is also invaluable for reconstructing the intrinsic (i.e. magnification-corrected) properties of the lensed DSFGs.

Coincidentally, what was initially thought to be a single DSFG strongly lensed by the foreground merging galaxy cluster has turned out to be two DSFGs with a peculiar (almost uncanny) similarity. 
The original known DSFG, with a detection of CO(3--2) at $z=2.236$ (\citetalias{Harrington:2016aa}; also \citealt{Harrington:2021aa}) 
was identified as a merging image pair Arc 1bc, and the counterimage Arc 1a by
\citet{Frye:2019aa} and \citet{Pascale:2022aa}.
With new NIRCam and ALMA continuum imaging, it has become clear that the object identified as Arc 3ab (merging pair) and Arc 3c (counterimage) on the same side of the lensing cluster also harbors considerable dust and star formation. 
In fact, while the Arc 1abc system (which we denote G165-DSFG-1) is considerably brighter in dust continuum (primarily due to the contribution from the high-magnification Arc 1bc) and likewise presumably CO emission, the intrinsic properties are very similar to the Arc 3abc system (G165-DSFG-3),
which may even have a larger SFR$_{\rm IR}$ ($400 \pm 80~M_\odot~{\rm yr}^{-1}$ vs. $320 \pm 70~M_\odot~{\rm yr}^{-1}$).

The redshift of DSFG-3 is still somewhat uncertain, as the current molecular line detections are all spatially unresolved and cannot distinguish between the two. It is most likely that it also lies at $z \approx 2.236$ with DSFG-1, so that their CO lines are blended at the 100 km s$^{-1}$ spectral resolution of LMT/RSR.
If they do in fact lie at the same redshift, then this proximity ($\approx 1\arcsec$, $<10$ kpc) would certainly imply that the objects are in the early stages of a spectacular gas-rich major merger (with a nearly 1:1 mass ratio, each with log$(M_\star / M_\odot) \approx 10.2$). 
The color gradients that are clearly apparent in Fig.~\ref{fig:SP_DSFG} may even be the result of tidal disruption of molecular gas after a first passage, driving an asymmetry of dust-obscured star formation towards the center of mass of the system (which is most readily observable in DSFG-1; Fig.~\ref{fig:slit_1a}). 
Moreover, the fainter Arc 3, 4, and 6 clumps that lie in between DSFG-1 and DSFG-3 (and predicted by the lens model to lie at the same redshift) could be additional evidence for such a significant gravitational interaction.
Future spectroscopic follow-up and kinematic information for the molecular gas is ultimately necessary to help resolve this open question.

Regardless of their distances along the line-of-sight, the presence of two strongly star-forming DSFGs only $1\arcsec$ apart in the sky is a remarkable finding. 
In fact, one could perhaps even make the case that this conjunction would be more impressive in the sky {\it without} the magnification and distortion contributed by the lensing galaxy cluster!
More quantitatively, they share similar IR-inferred SFRs, as well as UV-NIR SED-derived SFRs ($50 \pm 20~ M_\odot~{\rm yr}^{-1}$ for DSFG-1, $40 \pm 20~ M_\odot~{\rm yr}^{-1}$ for DSFG-3), stellar masses (${\rm log}(M_\star / M_\odot) = 10.2 \pm 0.1$ vs. $10.3 \pm 0.1$, respectively), and dust attenuation ($A_V = 1.5 \pm 0.3$ vs. $A_V = 1.2 \pm 0.2$). 
Their intrinsic (source-plane) dust continuum sizes are $1.6 \pm 0.2$ kpc and $2.1 \pm 0.2$ kpc.
Unsurprisingly, given the well-known correlation of radio with the far-IR (e.g., \citealt{van-der-Kruit:1971aa, de-Jong:1985aa, Helou:1985aa, Condon:1992aa, Lisenfeld:1996aa, Yun:2001aa, Bell:2003aa, Murphy:2006aa, Murphy:2006ab}), they also have similar radio continuum fluxes ($S_{\rm 6GHz}/\mu \approx 30 ~\mu{\rm Jy}$ for Arc 1a and $\approx 20 ~\mu{\rm Jy}$ for Arc 3c).
Their position in the $UVJ$ diagram is functionally identical ($U-V \approx 1.1$, $V-J \approx 1.2$), so their similarities cannot 
be attributed to limited flexibility in SED fitting.

Given the diversity and heterogeneity of DSFGs as a population\textemdash see \citet{Smail:2023aa} for another recent example of two DSFGs nearby to each other behind a lensing cluster, but with markedly different properties\textemdash this chance ultra-close alignment of two remarkably similar DSFGs is very peculiar.
While we do not wish to fully speculate on whether or not this is purely coincidence, especially given the uncertainty in exact redshift for one of the objects, it is maybe not entirely outside the realm of possibility that these objects could be associated with filamentary nodes of large-scale structure in the Universe.
As discussed in our introduction, DSFGs are widely believed to be associated with galaxy overdensities and protoclusters (see the theoretical work by \citealt{Chiang:2013aa}, and recent review by \citealt{Alberts:2022ab} and references therein).
This is similar to the association of protoclusters with high-redshift radio galaxies (HzRGs; e.g. \citealt{Venemans:2007aa, Miley:2008aa}) or Ly$\alpha$ emitters (e.g., \citealt{Keel:1999aa, Umehata:2015aa, Harikane:2019aa}) and extended Ly$\alpha$ nebulae, also known as Lyman Alpha Blobs (e.g. \citealt{Steidel:2000aa, Hennawi:2015aa, Arrigoni-Battaia:2018ab, Ramakrishnan:2023aa}; review by \citealt{Overzier:2016aa}).
These signposts, and their connections to each other, are still very much an active area of research.
And indeed, \citet{Frye:2024aa} found evidence of a substantial overdensity of $z\approx 2$ galaxies behind G165 through photometric redshifts (rivaling even the number of identified members of the foreground cluster at $z=0.35$). 
The authors also identified several groups of objects found serendipitously to lie at the same redshift (one association of six objects at $z=1.8$ and another of several galaxies at $z=2.2$) with the JWST NIRSpec Micro-Shutter Array (MSA).
Given the limitations of how MSA slits could be arranged, there are almost certainly other members of these associations that are not yet uncovered.
With this rich set of background objects, all strongly lensed by a massive merging galaxy cluster, the G165 field is sure to be the source of a wealth of future astrophysical discoveries.

We summarize the primary scientific findings of our work:

\begin{itemize}

\item Through resolved SED fitting of DSFG-1 and DSFG-3, we find complex distributions of $\Sigma_{\rm SFR}$, $\Sigma_\star$, and $A_V$ (which are of primary concern for this work). 
For DSFG-1, there is a noticeable asymmetry, with star formation and dust attenuation skewed higher towards the southeast, with the peak in $\Sigma_\star$ essentially occupying a highly elliptical central region rather than a single peak. These likely are related to the presence of multiple orbiting components at slightly different redshifts in this arc ($z=2.2355$ vs. $z=2.2401$). 

\item In the case of DSFG-3, there is not this clear asymmetry. Here, $\Sigma_{\rm SFR}$ is centrally peaked but again in a rather elliptical nuclear region. Stellar mass density $\Sigma_\star$, on the other hand, shows multiple distinct peaks throughout the disk plane, possibly signaling a history of (major or minor) merger activity. Dust attenuation $A_V$ is patchy throughout the object, with two distinct peaks, neither of which seem to be at the galaxy center.

\item We examine the position of the spatially-resolved regions (on scales of $150 - 1500$ pc) on the resolved star formation main sequence, which extends the global relation down to sub-galactic scales. There is a clear distinction between DSFG-1 and DSFG-3 vs. the more quiescent (or at least inside-out quenching) host of SN H0pe, examined through Arc 2a. For the DSFGs, we even observe a possible sequence at an elevated mode of star formation efficiency.

\item By extrapolating the model from \citet{Abdurrouf:2018aa} for the evolution of the rSFMS between $z\approx 1.8$ to 0 backwards to $z\approx 2$, we gain an initial sense of what areal fractions of the two DSFGs and Arc 2a are consistent with being above the rSFMS. 
For the DSFGs, these star-forming/starburst regions account for an overwhelming portion of the area ($80-100\%$), in contrast with only $\sim 60\%$ for Arc 2a. 
This supports the picture of many luminous DSFGs sustaining large-scale star-forming events, which are difficult to fully explain through major mergers.

\item The two DSFGs occupy similar locations in the $UVJ$ plane, as do the loci of their resolved, pixel-by-pixel colors. 
While not all regions are classified as dusty star-forming, only a small minority would be classified as $UVJ$-quiescent, further indicating the presence of large-scale galaxy-wide star formation events in tandem with the rSFMS results.

\item For the more quiescent galaxy seen in Arc 2a (according to its global $UVJ$ color), which is possibly undergoing inside-out quenching, we find a higher fraction of regions within the quiescent regime. 
Curiously, this object falls primarily within the dusty star-forming region of the $UVJ$-plane, with only a small area in the outskirts of the disk falling within the unobscured star-forming regime.

\item Comparing $UVJ$ colors and SED-derived sSFRs indicates a possible breakdown of the classical $UVJ$ selection on resolved scales. 
Moreover, we find evidence for a substantial degree of dust attenuation for the low-sSFR regions, including the region where SN H0pe is located. We suggest that this could be the result of efficient radial dust transport within the galaxy.

\item In comparing the properties we derive for DSFG-1 and DSFG-3 by using global galaxy-wide photometry vs. SED-fitting the spatially-resolved photometry and summing the relevant properties, we find approximate agreement for both SFRs and $M_\star$.
As Arc 1a also has an H$\alpha$ detection reported by \citetalias{Frye:2024aa}, we also compare SFR$_{{\rm H}\alpha}$ with ${\rm SFR}_{\rm SED}$, finding very close agreement. Comparing the attenuation $A_V$ found through SED fitting vs. through the Balmer decrement, we reach excellent agreement for $A_V = 1.5$, if we make the assumption that $E(B-V)_{\rm stellar} = 0.44\cdot E(B-V)_{\rm gas}$ \citep{Calzetti:1997aa, Calzetti:2000aa}.

\item Using the global properties of DSFG-1 and DSFG-3, we find that they lie directly on the $\Delta_{\rm MS} = +0.6$ dex threshold above the star-forming main sequence to be classified as starbursts.
The location of DSFGs on the SFMS is still subject to debate, but it is obvious that they cannot simply be uniformly considered starbursts or high-$M_\star$ end main sequence galaxies. 
In this case, we consider it perhaps notable that these DSFGs are starbursting despite being largely disk-like and without particularly obvious signs of merger activity, which is often responsible for a temporary elevation in sSFR.

\item Employing our lens model for the purpose of estimating the magnifications and time delays of the three images of SN H0pe, we find $\mu_a = 6.7 \pm 0.1$, $\mu_b = 9.8 \pm 0.3$, and $\mu_c = 8.9 \pm 0.3$, which are in reasonable agreement with the values estimated from the light curve photometry. 
For relative time delays, our model predicts arrival time after image SN2a to be $\Delta t_{c,a} = 53 \pm 3$ days and $\Delta t_{b,a} = 106 \pm 2$ days. 
These are used in the inference of $H_0$ by \citet{Pascale:2024aa}.

\item Using the combined SFR inferred from the rest-frame UV to FIR SED, 
along with the information obtained through our lens model, we estimate the combined rate of core-collapse supernovae in DSFG-1 and DSFG-3 to be $3.6 \pm 0.7~{\rm yr}^{-1}$.
With cosmological time dilation, this reduces to 
$1.1 \pm 0.2$ yr$^{-1}$.

\end{itemize}

In reality, the rate at which we can 
detect SNe depends also on the control time, or the amount of time that a survey is capable of detecting a given SN \citep{Zwicky:1938aa}.
The often patchy dust geometry of DSFGs also makes the prediction of 
how attenuated supernovae will be quite difficult.
For this reason, the near-IR sensitivity of JWST may provide the safest observing strategy, but detections with the Vera C. Rubin Observatory's Legacy Survey of Space and Time (LSST) or the Nancy Grace Roman Space Telescope
certainly cannot be ruled out
(e.g., \citealt{Hounsell:2018aa, Petrushevska:2020aa}).
Moreover, the large-scale nature of the star formation events in these DSFGs lends well to increasing the chances of multiplexing the observed SN rate through multiple imaging by lensing, as this makes it more likely that a CCSN will occur in a region of the galaxy inside the caustics.

To this last point, we re-emphasize that these estimates come only from the two galaxies DSFG-1 and DSFG-3 alone; based on the $\gtrsim 100~M_\odot~{\rm yr}^{-1}$ integrated (obscured and unobscured) SFR for the $z\sim 2$ group, \citetalias{Frye:2024aa} still estimate a rate of $\gtrsim 1$ SNe yr$^{-1}$ from the observer's perspective, making a compelling case for a continued monitoring campaign of G165.

\section*{Acknowledgments}

We wish to dedicate this paper to the memory of our colleague at the School of Earth and Space Exploration, Susan Selkirk. 

P.S.K. would like to thank Ian Smail, Evan Scannapieco, and Philip Mauskopf for helpful discussions to improve this paper.
This work was also improved meaningfully as a result of discussions P.S.K. had with colleagues while at the Aspen Center for Physics, which is supported by National Science Foundation grant PHY-2210452.

This work is based on observations made with the NASA/ESA/CSA James Webb Space Telescope. The data were obtained from the Mikulski Archive for Space Telescopes at the Space Telescope Science Institute, which is operated by the Association of Universities for Research in Astronomy, Inc., under NASA contract NAS 5-03127 for JWST. These observations are associated with JWST program \#~1176 (PEARLS; PI: R. Windhorst). The JWST data used can be found in MAST: \dataset[10.17909/225y-k062]{http://dx.doi.org/10.17909/225y-k062}.
R.A.W., S.H.C., and R.A.J. acknowledge support from NASA JWST Interdisciplinary Scientist grants NAG5-12460, NNX14AN10G and 80NSSC18K0200 from GSFC.
E.F.-J.A. acknowledges support from UNAM-PAPIIT project IA102023, and from CONAHCyT Ciencia de Frontera project ID:  CF-2023-I-506. 

This paper makes use of the following ALMA data: ADS/JAO.ALMA \#~2021.1.00607.S.
ALMA is a partnership of ESO (representing its member states), NSF (USA) and NINS (Japan), together with NRC (Canada), MOST and ASIAA (Taiwan), and KASI (Republic of Korea), in cooperation with the Republic of Chile. The Joint ALMA Observatory is operated by ESO, AUI/NRAO and NAOJ. The National Radio Astronomy Observatory is a facility of the National Science Foundation operated under cooperative agreement by Associated Universities, Inc.
This research has made extensive use of NASA's Astrophysics Data System.

We also acknowledge the indigenous peoples of Arizona, including the Akimel O'odham (Pima) and Pee Posh (Maricopa) Indian Communities, whose care and keeping of the land has enabled us to be at ASU's Tempe campus in the Salt River Valley, where much of our work was conducted.

%

\facilities{JWST(NIRCam), ALMA, VLA}


\software{{\sc APLpy} \citep{Robitaille:2012aa,Robitaille:2019aa},
          {\sc astropy} \citep{Astropy-Collaboration:2013aa},
          {\sc bagpipes} \citep{Carnall:2018aa},
          {\sc blobcat} \citep{Hales:2012aa},
          {\sc casa} \citep{McMullin:2007aa},
          {\sc glue} \citep{Beaumont:2015aa, Robitaille:2017aa},
          \lenstool\ \citep{Kneib:1993aa, Kneib:1996aa, Jullo:2007aa, Jullo:2009aa}, 
          Ned Wright's Cosmology Calculator \citep{Wright:2006aa},
          {\sc piXedfit}
          \citep{Abdurrouf:2021ab, Abdurrouf:2022aa, Abdurrouf:2022ab, Abdurrouf:2022ac},
          {\sc photutils} \citep{Bradley:2022aa}
          }

          %



\appendix

\section{Lens model parameters}
\label{sec:appendix}

In Table~\ref{tab:lens_bestfit}, we provide the full set of best-fit gravitational lens model parameters for our characterization of the G165 cluster with {\sc lenstool}, in accordance with our strategy described in \S \ref{sec:lens_modeling}. 
Additionally, we supply the range in priors for each parameter, which is distributed uniformly in all cases. 
For priors on the image system redshifts, these are chosen iteratively based on photometric redshift information, where we widen the prior range for parameters where previous model iterations led to their posterior distribution pushing up against the boundary conditions. 
For selecting a single best-fit version of our current model, we opt for the median value of the posterior for each parameter, with the uncertainties described by the inner 68\% confidence interval. This statistical measure is very likely an underestimate of the true systematic uncertainty in these parameters.
Lastly, we also provide the highest-likelihood (minimum $\chi^2$) solution. For virtually all parameters, these highest-likelihood values are contained within the 68\% confidence interval from the median solution, adding some assurance to the robustness of the model.
\citet{Pascale:2024aa} examine the impact of each parameter and unknown (i.e. image system redshifts) on the measurement of magnifications and time delays, in the context of inferring $H_0$ from SN H0pe.

\begin{figure}[ht!]
\centering
\includegraphics[width=0.65\textwidth]{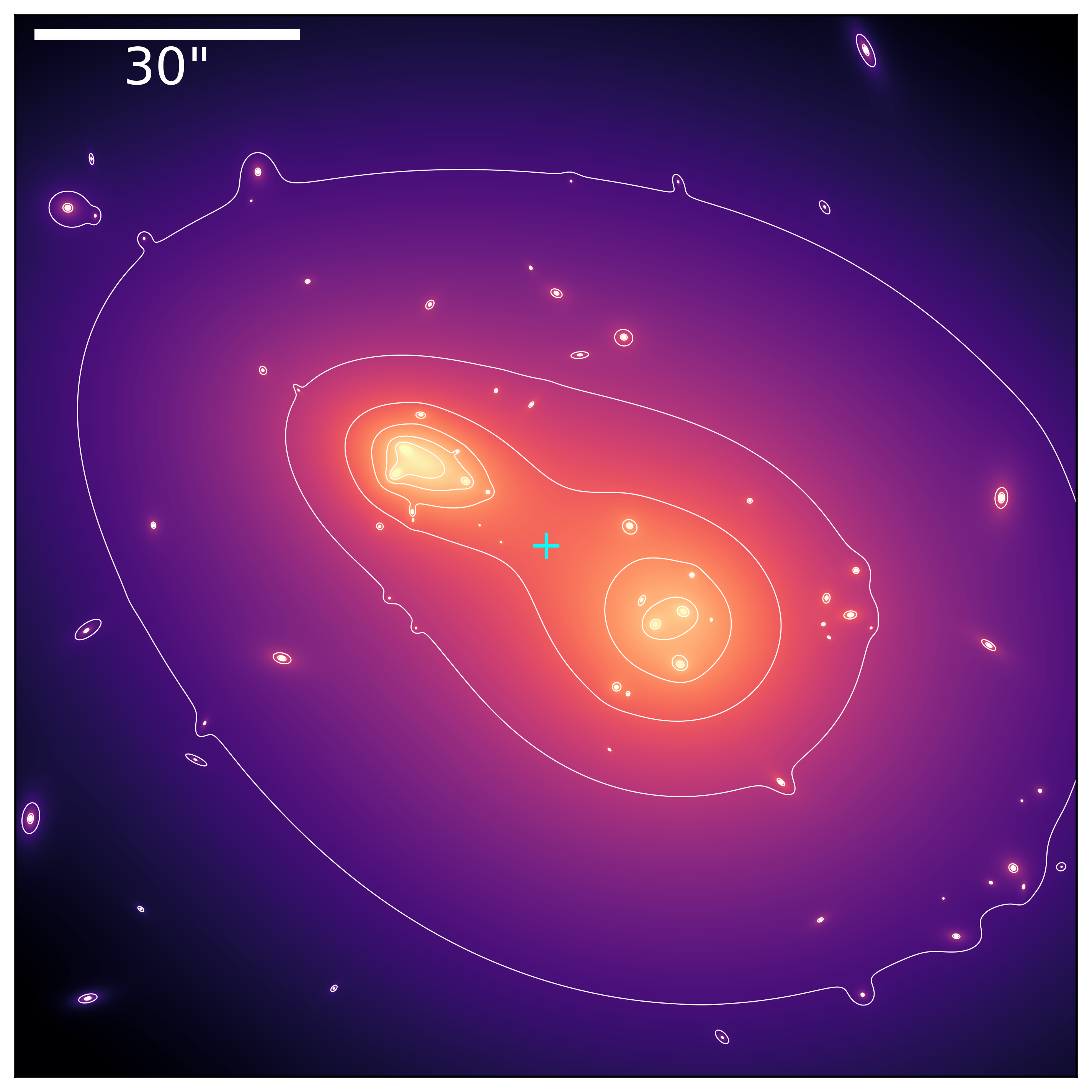}
\caption{Convergence ($\kappa$) map of the G165 field from the median best-fit lens model, which is the dimensionless surface mass density (i.e., relative to the critical surface density).
A cyan cross marks the center of this map, which is at
$(\alpha, \delta) =$ (11$^h$27$^m$15\fs 156, $+42^d$28$^m$30\fs 06).
White contours indicate $\kappa = [0.2, 0.4, 0.6, 0.8, 1.0, 1.2]$. 
    \label{fig:kappa}
}
\end{figure}

Finally, Fig.~\ref{fig:kappa} shows the convergence map $\kappa$ (or dimensionless surface mass density) of the full G165 field from our best-fit lens model.

\startlongtable
\begin{deluxetable*}{c|cccc}
\tablecaption{
Best-fit gravitational lens model parameters for the G165 cluster. \label{tab:lens_bestfit}
}
\tablehead{
\colhead{Parameter} &  \colhead{Units} & \colhead{Prior range} & \colhead{Median \& $1\sigma$ uncertainty} & \colhead{Highest-likelihood solution}
}
\startdata
         $x_1$  & [$\arcsec$]& [-16.8, -6.8] & $-13.4 \pm 0.1$ & -13.6 \\ 
         $y_1$  & [$\arcsec$]& [3.2, 13.2] & $+9.08 \pm 0.07$  & +9.18 \\ 
         $e_1$ & \textemdash& [0.0, 0.85] & $0.40 \pm 0.03$  & 0.44 \\ 
         $\theta_1$  & [$\degrees$]& [-90, 90] & $-24.0 \pm 0.2$ & -23.9 \\ 
         $r_{c1}$  & [kpc]& [0, 98.4] & $17.8 \pm 0.6$ & 16.9\\  
         $\sigma_1$  & [km s$^{-1}$]& [0, 800] & $476 \pm 3$ & 473 \\ \hline 
         $x_2$  & [$\arcsec$]& [11.3, 21.3] & $+14.39 \pm 0.03$ & +14.37 \\  
         $y_2$  &[$\arcsec$]& [-13.7, -3.7] & $-8.73 \pm 0.01$  & -8.75  \\  
         $e_2$ & \textemdash& [0, 0.85] & $0.11 \pm 0.00$ & 0.11 \\  
         $\theta_2$  & [$\degrees$]& [-90, 90] & $-18.8 \pm 0.3$  & -19.1 \\ 
         $r_{c2}$   & [kpc]& [0, 98.4] & $43.7 \pm 0.4$ & 43.5 \\  
         $\sigma_2$  & [km s$^{-1}$]& [0, 800] & $689 \pm 3$  & 688 \\ \hline 
         $\sigma_3$  & [km s$^{-1}$]& [0, 250] & $87.4 \pm 0.9$ & 86.3 \\ \hline 
 mag$_0$  & [mag]& [17.0]$\dagger$ & 17.0$\dagger$& \textemdash\\ 
 $\sigma_0$  & [km s$^{-1}$]& [120]$\dagger$ & 120$\dagger$& \textemdash \\ 
 $r_{\rm core,0}$  & [kpc]& [0.15]$\dagger$ & 0.15$\dagger$& \textemdash \\ 
 $r_{\rm cut,0}$  & [kpc]& [30]$\dagger$ & 30$\dagger$& \textemdash \\ \hline 
         $z_3$ &\textemdash & [2.0, 2.2] & $2.11^{+0.01}_{-0.01}$ & 2.11 \\ 
         $z_4$ &\textemdash & [2.0, 2.2]& $2.12^{+0.01}_{-0.01}$ & 2.11 \\ 
         $z_6$ &\textemdash & [2.0, 2.2] & $2.10^{+0.01}_{-0.01}$ & 2.11 \\ 
         $z_7$ \& $z_{18}$ &\textemdash & [1.4, 3.0]& $1.73^{+0.01}_{-0.01}$ & 1.74 \\ 
         $z_{10}$ &\textemdash & [1.5, 1.9] & $1.76^{+0.01}_{-0.01}$ & 1.73 \\  
         $z_{11}$ &\textemdash & [4.0, 4.6]& $4.60^{+0.00}_{-0.00}$  & 4.60 \\  
         $z_{12}$ &\textemdash & [2.7, 3.2]& $2.72^{+0.03}_{-0.02}$ & 2.71 \\  
         $z_{13}$ &\textemdash & [3.2, 3.8]& $3.60^{+0.02}_{-0.02}$ & 3.64 \\  
         $z_{14}$ &\textemdash & [2.0, 3.0]& $2.58^{+0.01}_{-0.01}$ & 2.57 \\  
         $z_{15}$ &\textemdash & [2.0, 2.5]& $2.11^{+0.02}_{-0.01}$ & 2.11 \\  
         $z_{16}$ &\textemdash & [1.4, 1.8]& $1.79^{+0.01}_{-0.01}$ & 1.80 \\  
         $z_{17}$ &\textemdash & [5.5, 6.5]& $5.65^{+0.05}_{-0.06}$ & 5.62 \\  
         $z_{19}$ &\textemdash & [3.0, 4.0]& $3.91^{+0.06}_{-0.08}$ & 4.00 \\  
         $z_{20}$ &\textemdash & [3.0, 4.0]& $3.96^{+0.03}_{-0.05}$ & 3.90 \\  
         $z_{21}$ &\textemdash & [1.0, 4.0]& $1.74^{+0.04}_{-0.03}$ & 1.74 \\ 
\enddata
\tablenotetext{^\dagger}{Parameter held fixed.}
\tablecomments{ 
Priors for all parameters are taken to be distributed uniformly within the provided range.
}
\end{deluxetable*}

\section{Testing the redshift solution for DSFG-3}
\label{sec:appendix1}

\begin{figure}[ht!]
\centering
\includegraphics[width=0.7\columnwidth]{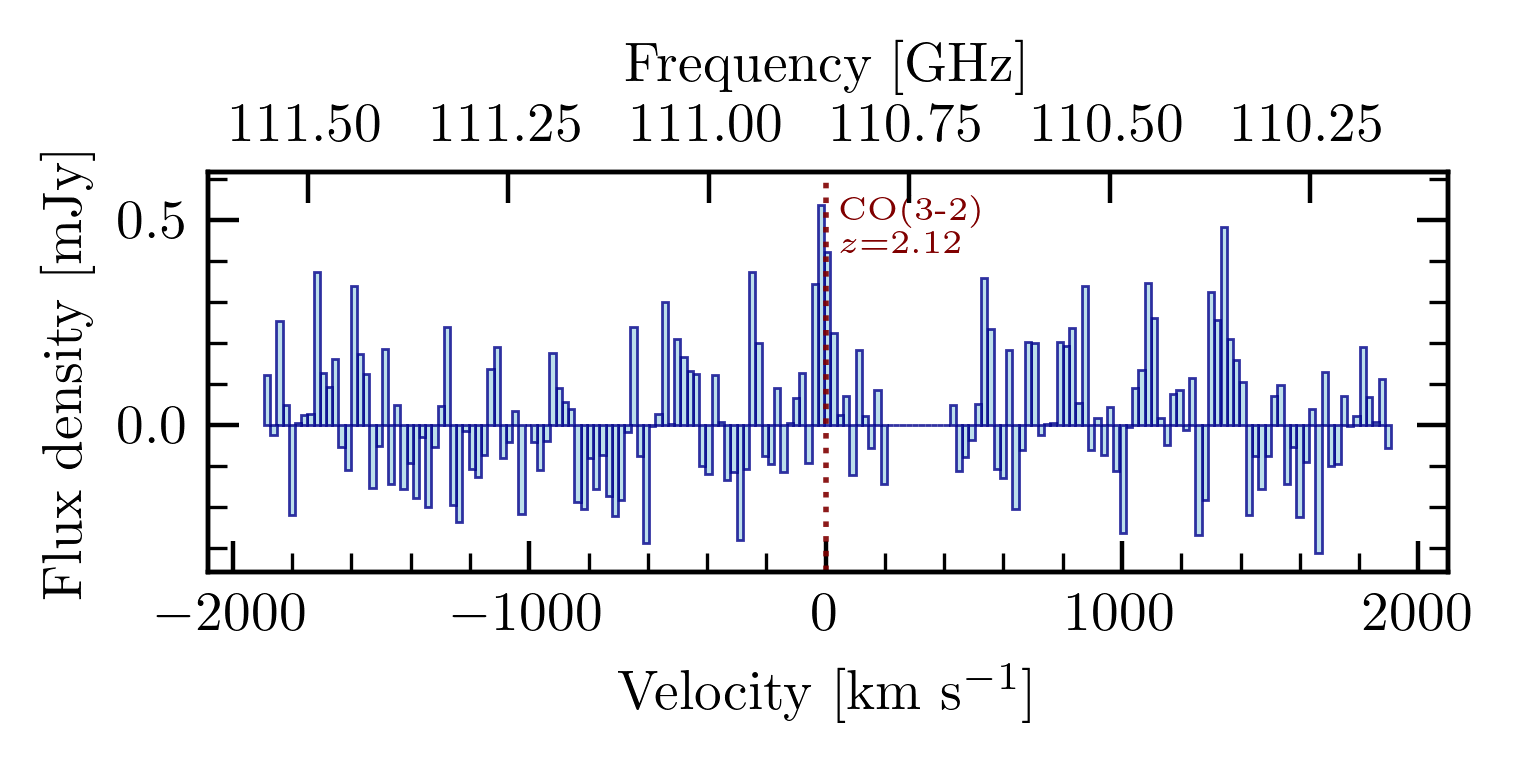}
\caption{
ALMA band 3 continuum-subtracted and Hanning-smoothed spectrum extracted from an aperture approximately equal in size to the synthesized beam, centered on the peak of continuum emission for Arc 3c. 
The spectral axis is converted to velocity at the rest frequency ($\approx 345.796$ GHz) of CO(3--2) at a redshift of $z=2.120$.
Two spectral windows are imaged together, with the small frequency coverage gap at $\approx +300~{\rm km}~{\rm s}^{-1}$.
This detection is only marginal (and there is a nearly equally believable detection at $\sim1300~{\rm km}~{\rm s}^{-1}$, which corresponds to $z\approx 2.13$), but is in support of the other evidence in favor of this redshift, so we acknowledge it as a possibility.
    \label{fig:zspec}
}
\end{figure}

Given the narrow range in the geometric redshift for DSFG-3, and approximate concordance with the photometric redshift, we searched for molecular line detections in the frequency ranges covered by the ALMA observations in Band 3 and Band 4 for possible line detections; with this spectral setup, these are $\nu_{\rm B3}/{\rm GHz} \in [109.08,110.70] \cup [110.77,112.52]$ and $\nu_{\rm B4}/{\rm GHz} \in [136.46,138.25] \cup [138.59,140.39]$, respectively. 
Unfortunately, the line detected by RSR is not contained within this spectral tuning.
We find no strong lines within the available spectral windows, which is consistent with our assumption of $z=2.23$.
Fig.~\ref{fig:zspec} shows
the highest S/N feature, which is an unlikely but
possible identification of a narrow CO(3--2) line at (observed) 110.849 GHz, corresponding to a redshift of $z_{\rm spec}=2.12 \pm 0.01$. 
Given the very marginal detection (S/N $\approx 3$), we certainly do not consider this a secure, robust spectroscopic redshift on its own. 
If the source is actually at $z=2.12$, there might be reason for a non-detection with the 1-hour integration with LMT, as CO(2--1) for $z=2.12$ is at 73.89 GHz, near to the low-frequency edge of the RSR bandwidth ($73 - 111$ GHz), whereas CO(3--2) would be very near the upper-frequency edge, at 110.83 GHz.

Fortunately, whether Arc 3 is at the assumed redshift of $z=2.23$ or at the secondary solution of $z=2.12$ is of relatively little import for many of the results of this work pertaining to the distribution of properties. 
Perhaps the greatest impact is $\sim 30\%$ lower derived SFR$_{\rm UV-NIR}$ if $z=2.12$ is accurate, which similarly would reduce the predicted CCSN rates.

\section{Calculating time delays for SN H0pe} 
\label{sec:td}

SN H0pe and its host galaxy appear consistent with a ``naked cusp" lensing morphology, with three images on the same side of the lens. This results from the source lying interior to the tangential caustic but exterior to the radial caustic in the source plane (e.g., \citealt{Maller:1997aa, Bartelmann:1998aa, Keeton:1998aa, Lewis:2002aa}).
If this interpretation is correct, then it is lensed into only the three images of SN H0pe and its host galaxy (Arcs 2a, 2b, 2c) that have already been identified.

\begin{figure}[ht!]
\centering
\includegraphics[width=0.7\columnwidth]{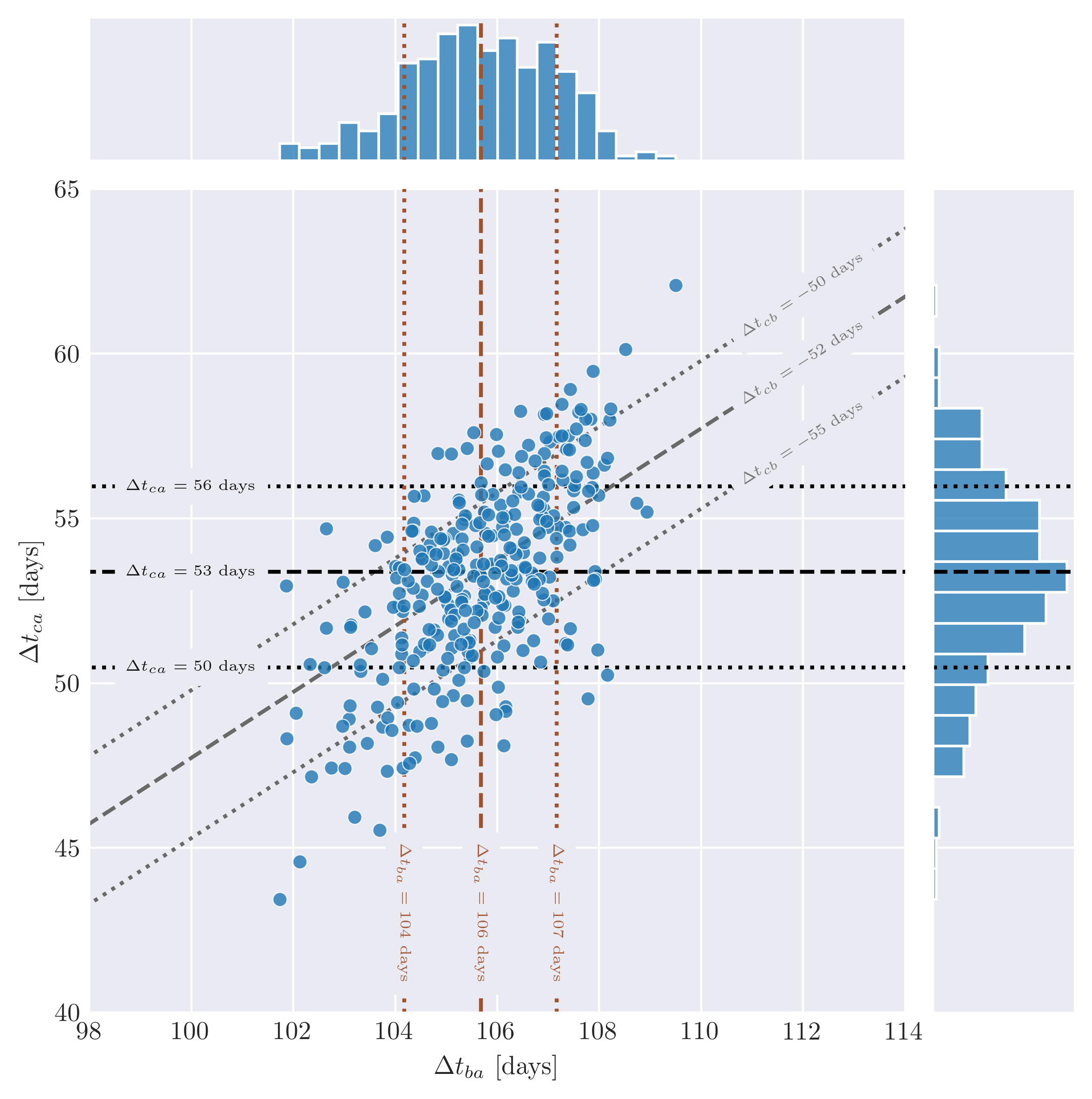}
\caption{ 
    The time delay covariance plot for SN H0pe, showing 300 resampled iterations of the lens model posterior distribution, each yielding a different value of $\Delta t_{ca}$ and $\Delta t_{ba}$ (i.e. the delay in arrival time for images 2c and 2b following image 2a, respectively). Maroon vertical lines mark the median and 68\% confidence interval for $\Delta t_{ba}$, black horizontal lines mark the same for $\Delta t_{ca}$, and gray diagonal lines for $\Delta t_{cb}$. 
    Histograms on the top and right of the covariance plot show the one-dimensional distributions of $\Delta t_{ca}$ and $\Delta t_{ba}$.
    \label{fig:time_delays}
}
\end{figure}

With our best-fit lens model for G165, we computed the light travel-time surface as 
\begin{equation}
\tau(\boldsymbol{\theta}) = \frac{D_{\Delta t}}{c}  \cdot \phi(\boldsymbol{\theta}, \boldsymbol{\beta}),
\end{equation}
where 
\begin{equation}
\phi(\boldsymbol{\theta}, \boldsymbol{\beta}) = \frac{1}{2} (\boldsymbol{\theta} - \boldsymbol{\beta})^2 - \psi(\boldsymbol{\theta})^2
\end{equation}
following the notation of \citet{Treu:2016aa} and \citet{Treu:2022ac}.
Here, $\boldsymbol{\theta}$ is the set of apparent image-plane sky positions, $\boldsymbol{\beta}$ is the set of corresponding true (unlensed) source-plane positions, $\phi(\boldsymbol{\theta}, \boldsymbol{\beta})$ is the (scalar) Fermat potential, and $\psi(\boldsymbol{\theta})$ is the scaled plane-projected gravitational potential. $D_{\Delta t}$ is the time-delay distance defined as
\begin{equation}
D_{\Delta t} \equiv (1+z_d)\frac{D_d D_s}{D_{ds}}
\end{equation}
for lens plane redshift $z_d$ and the angular diameter distances $D_d$, $D_s$, $D_{ds}$ between observer and lens plane, between observer and source plane, and between lens and source planes, respectively. 
The Hubble constant $H_0$ is connected through $D_{\Delta t} \propto H_0^{-1}$ \citep{Suyu:2010aa}.
For this reason, with a model for the lensing potential $\psi(\boldsymbol{\theta})$, and an observation of actual time delay between images through detection of a transient or variable source, one may infer $H_0$.

To propagate the uncertainty in the lens model forward to the time delay predictions, we again randomly selected 300 realizations from the MCMC sampling, and calculated the light travel times at the position of SN H0pe in images 2a, 2b, and 2c. We then calculated the delays after the first image to arrive (2a), $\Delta t_{ac}$ and $\Delta t_{ab}$. 
The results are shown in Fig.~\ref{fig:time_delays}.
As a caveat, the uncertainties on time delays may be closely dependent on the positional uncertainty assumed for the model (here, taken to be $0 \farcs 03$). 
These values, along with the convergence $\kappa$ and shear $\gamma$, are reported in Table~\ref{tab:timedelays}. We also include our model-derived magnifications for the locations of SN H0pe, alongside the respective values derived from the photometric light curve information from \citet{Pierel:2024aa}. These absolute measurements of magnification are possible only because SN H0pe is a standard candle Type Ia supernova. We find general agreement within uncertainties. As discussed by \citet{Pascale:2024aa}, the fidelity of a lens model in reproducing the SN Ia-derived magnifications can be used to weight the various lens models involved in the inference of $H_0$.
We also find consistency with the predictions from many of the other lens models compiled by \citeauthor{Pascale:2024aa} (see their Fig. 3), where the model from this work is referred to as Model 3.

\startlongtable
\begin{deluxetable*}{c|cccccc}
\tablecaption{Model-derived time delay properties for SN H0pe. \label{tab:timedelays}}
\tablehead{
\colhead{Image} & \colhead{Model time delay $\Delta t_{x,a}^\ddagger$} & \colhead{SN Ia $\Delta t_{x,a}^\dagger$} & \colhead{$\mu_{\rm model}^\ddagger$} & \colhead{$\mu_{\rm SN~Ia}^\dagger$} & \colhead{$\kappa^\ddagger$} & \colhead{$\gamma^\ddagger$}    \\
\colhead{} & \colhead{[days]} & \colhead{[days]} & \colhead{} & \colhead{} & \colhead{} & \colhead{}
}
\startdata
SN 2a  & \textemdash & \textemdash &$6.7 \pm 0.1$ &  $5.1^{+1.5}_{-1.7}$	& $0.571 \pm 0.004$ 	& $0.186 \pm 0.002$	  	\\
SN 2b  & $106 \pm 2$  & $117 \pm 10$ & $9.8 \pm 0.3$ & $9.2^{+3.5}_{-2.5}$ 	& $0.705 \pm 0.003$	 &	$0.434 \pm 0.004$   	\\
SN 2c  & $53 \pm 3$ & $68 \pm 11$ & $8.9 \pm 0.3$ & $7.4^{+1.4}_{-1.4}$	& $0.588 \pm 0.005$	 & $0.243 \pm 0.002$ 	
\enddata
\tablenotetext{^\dagger}{Magnifications and time delays derived from the SN Ia lightcurve determined by \citet{Pierel:2024aa}}
\tablenotetext{^\ddagger}{The uncertainties in these values are purely statistical and likely underestimate the true systematic uncertainties. All are based on a fiducial $H_0 = 70~{\rm km}~{\rm s}^{-1}~{\rm Mpc}^{-1}$. A more in-depth discussion, along with an investigation into the error budget of these values, is provided by \citet{Pascale:2024aa}}
%
\end{deluxetable*}

\section{Source-plane reconstruction and resolved SED fitting of high-magnification arcs as a consistency check}
\label{sec:appendix2}

\begin{figure}[htb]
\centering
\raisebox{0.15\height}{\includegraphics[width=0.55\columnwidth]{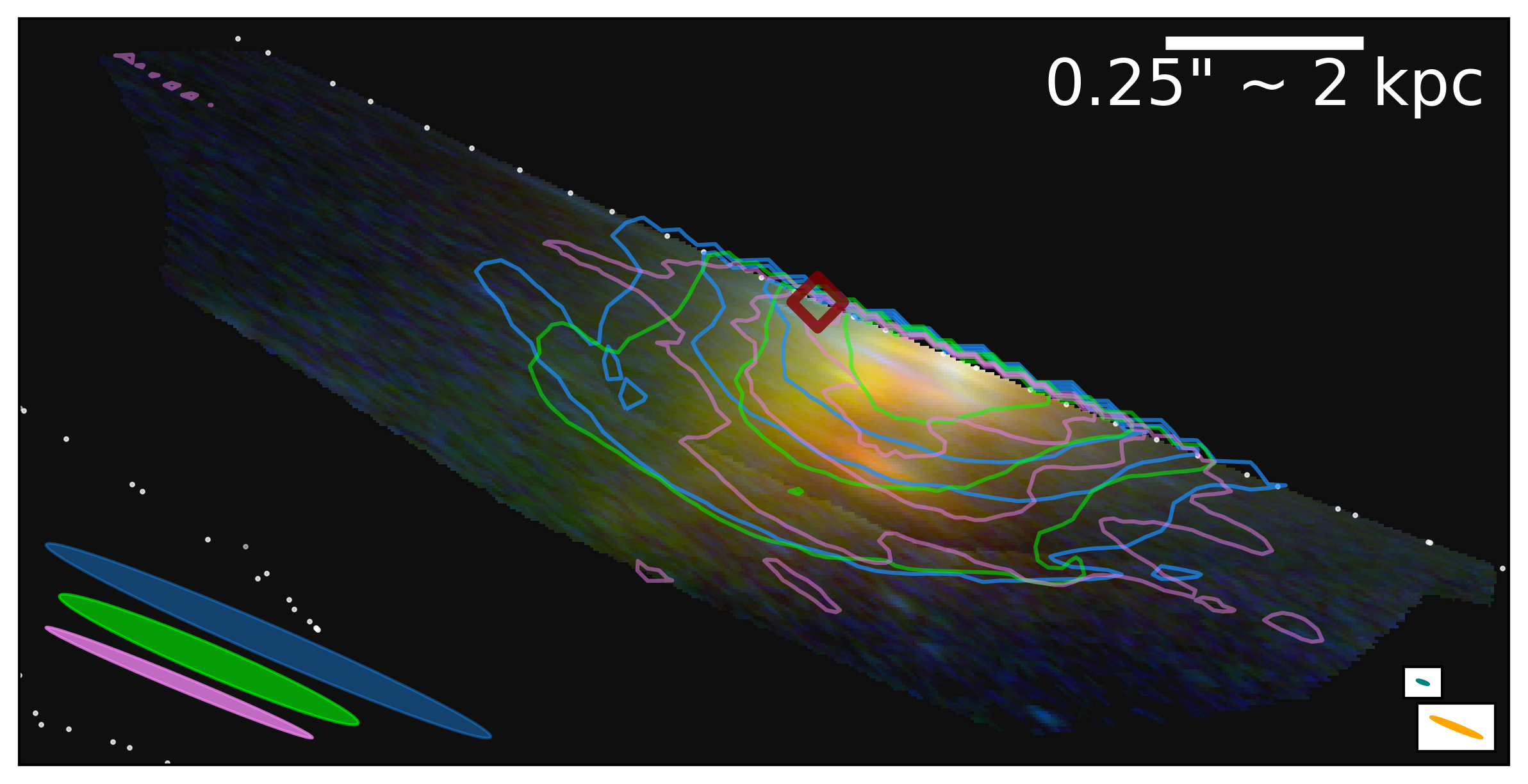}}
\includegraphics[width=0.35\columnwidth]{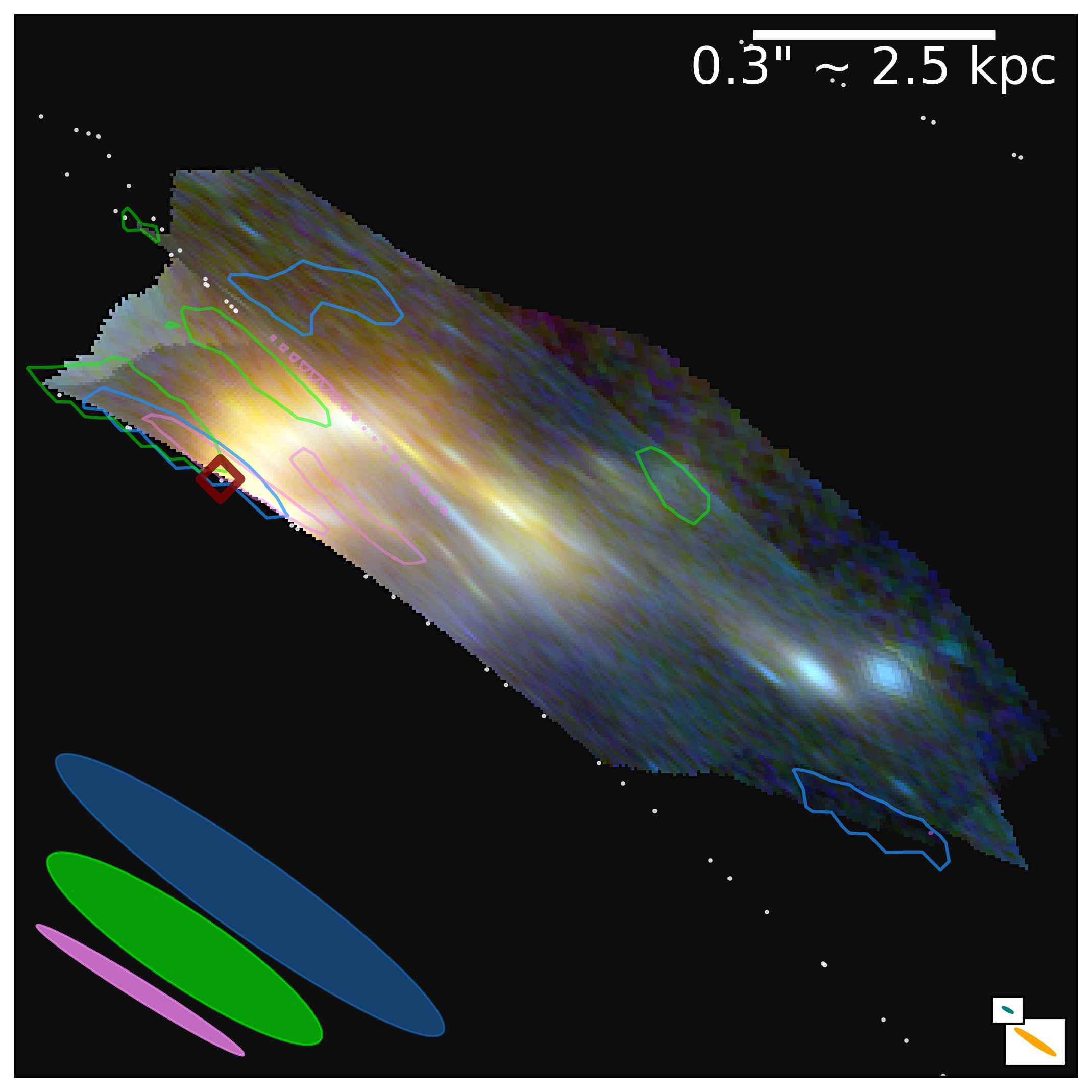}
\caption{
    Source-plane reconstruction of only the highly-distorted images 1bc ({\it top}) and 3ab ({\it bottom}), as with Fig.~\ref{fig:SP_1a_3c}, zoomed-in to show greater detail. 
    As the PSF varies much more severely across the extent of the arc, owing to the large magnification gradient, the beams shown are purely representative, and correspond to the position marked with a maroon diamond. This position near to the line of symmetry marks a bluer clump within the galaxy, which is denoted in the image plane as Arcs 1.2b and 1.2c by \citetalias{Frye:2024aa}.
    \label{fig:SP_high_mag}
}
\end{figure}

Fig.~\ref{fig:SP_high_mag} shows the source-plane reconstructions of Arcs 1bc and 3ab, which are the highly-magnified fold-configuration images. 
While these reconstructions are subject to very different effective source-plane PSFs compared to the lower-magnification counterparts in Fig.~\ref{fig:SP_1a_3c}, they reveal broad agreement and instill some confidence in the lens model as applied to these objects.

\begin{figure*}[ht!]
\centering
\includegraphics[width=0.97\textwidth]{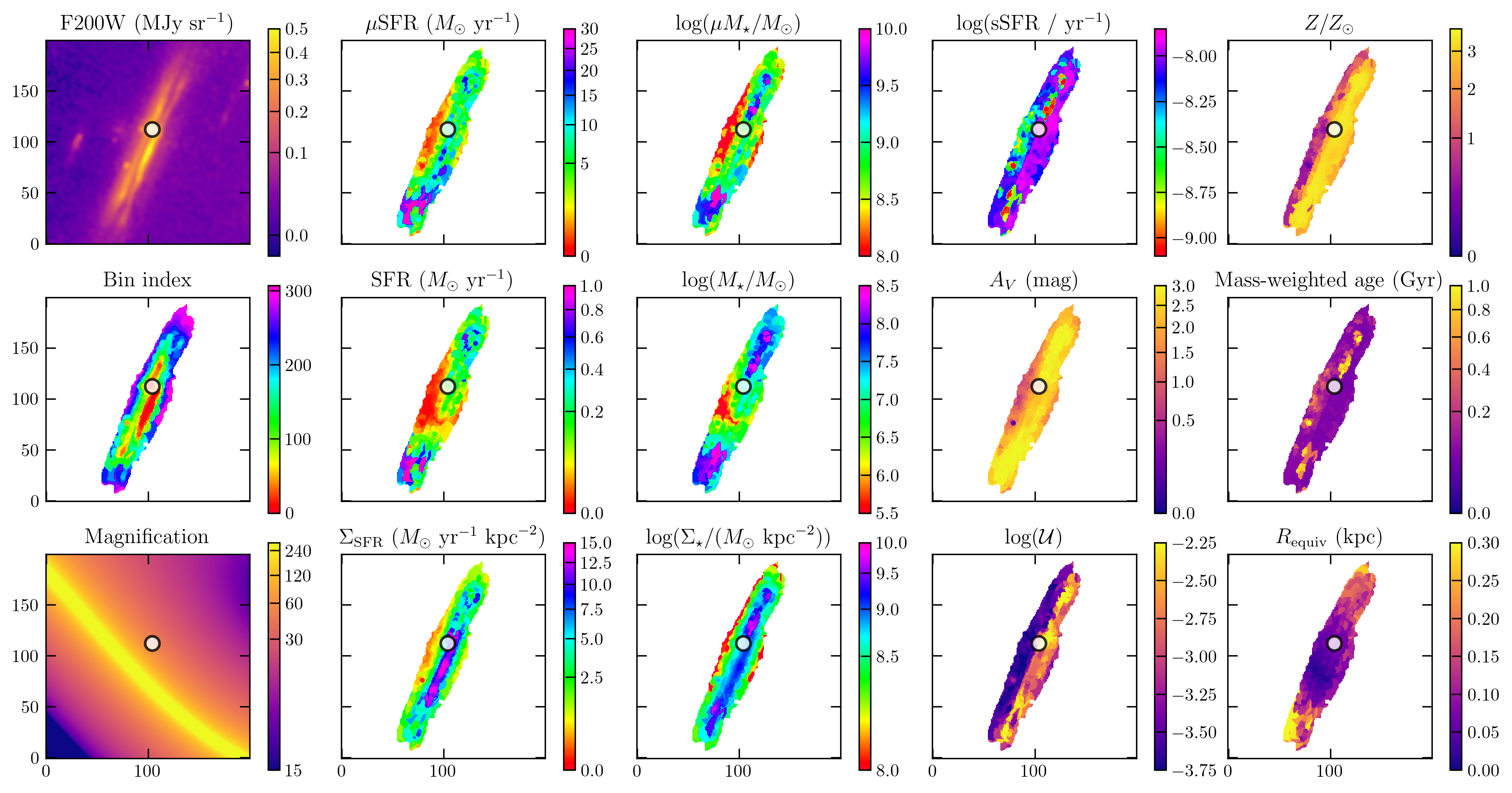}
\caption{
    As with Fig.~\ref{fig:pixbin_1a}, but for the highly-magnified Arc 1bc fold image system, with a fixed redshift of $z=2.236$. A black circle with white fill covers a foreground perturber (possibly within the galaxy cluster), for which the derived properties are not accurate (as the background redshift is held fixed).
    The field of view of each panel is $200\times 200$ pix, or $6 \times 6\arcsec$. 
    \label{fig:pixbin_1bc}
}
\end{figure*}

\begin{figure*}[ht!]
\centering
\includegraphics[width=0.97\textwidth]{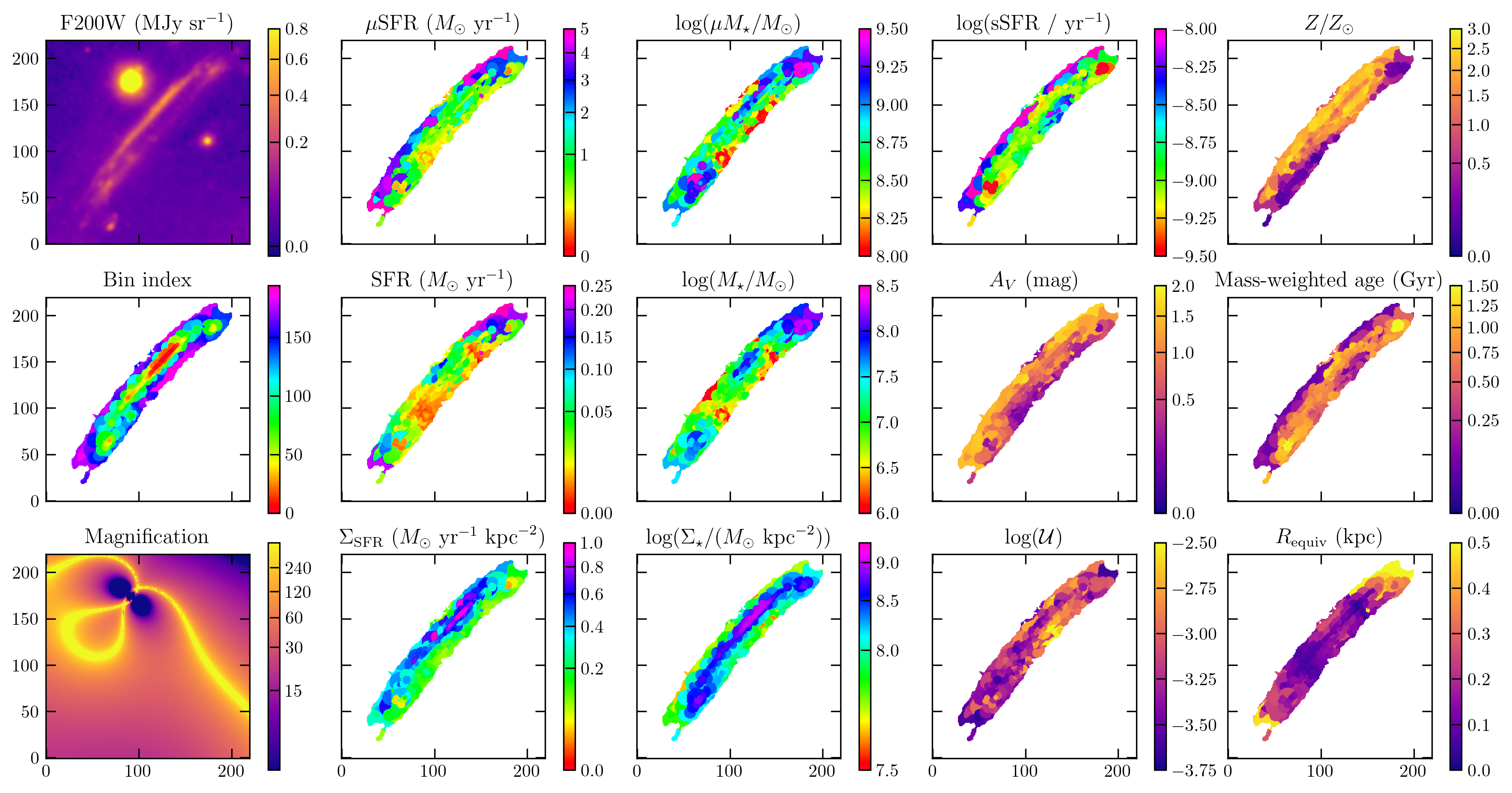}
\caption{
    As with Fig.~\ref{fig:pixbin_1bc}, but for Arc 3ab at an assumed fixed redshift of $z=2.1$.
    The field of view of each panel is $220\times 220$ pix, or $6 \farcs 6 \times 6 \farcs 6$. 
    \label{fig:pixbin_3ab}
}
\end{figure*}

Figs.~\ref{fig:pixbin_1bc} and \ref{fig:pixbin_3ab} show the same properties as Figs.~\ref{fig:pixbin_1a} and \ref{fig:pixbin_3c}, but for their high-magnification counterparts, Arcs 1bc and 3ab. 
As both DSFGs barely cross over the caustic curves at their respective redshifts (Fig.~\ref{fig:SP_DSFG}), only the southeasternmost portion of DSFG-1 and northwesternmost portion of DSFG-3 are triply-imaged into 1bc and 3ab, and the remaining structure is singly-imaged into the lower-magnification arcs 1a and 3c. 
In the image plane, we indicate the portion of each object that gets triply-imaged using red dividing lines in Figs.~\ref{fig:slit_1a} and \ref{fig:slit_3c}. 
This includes roughly one-third of the area of DSFG-1 studied with resolved SED fitting, and closer to one-fourth of DSFG-3.

At such small intrinsic scales (below 100 pc), it is not unreasonable to consider that the utility of SED fitting may begin to break down,
especially when criteria of energy balance between UV/optical and far-IR are imposed \citep{Smith:2018ab}.
Even without this assumption, the lower average S/N of individual bins resulting in greater uncertainties and degeneracies in derived physical properties.
It is also rather difficult to interpret 
spatial variations
in the image plane, as the arcs are the result of a $\sim$kpc-scale source-plane region being magnified and highly distorted after lensing.
Nonetheless, there is rough agreement in the key properties of $\Sigma_{\rm SFR}$, $\Sigma_\star$, $A_V$ as we find in the corresponding regions of Arcs 1a and 3c. 
Based on what we recover from this SED fitting, we identify a patchy distribution in dust attenuation $A_V$, but there is not 
significant
variation here (although this is not a particularly robust characterization of the sub-kpc dust geometry). 
Moreover, it is difficult to determine if the clumpy structure of $\Sigma_{\rm SFR}$ and $\Sigma_\star$ reveals the actual distribution of small-scale star-forming complexes, or if the variation is just the result of stochasticity in the SED fitting or inhomogeneous dust geometries (even below 100 pc scales).

\section{Integrated properties of SN H0pe host, Arc 2a}
\label{sec:appendix3}

For ease of comparison, we also use the same choice of priors to fit the SED of Arc 2a, the host galaxy of SN H0pe. 
We find that several parameters have wide posterior distributions or bimodalities, which limit our interpretations. 
This may be due to different stellar populations contributing comparably to the integrated galaxy light (as revealed through the resolved SED fitting), or to degeneracies (e.g. between age and dust) that are poorly broken by the UV-NIR photometry given the shape of the SED. 
Intriguingly, \citetalias{Frye:2024aa} found $A_V = 0.9 \pm 0.2$ for Arc 2a from the Balmer decrement, which is in line with the lower-$A_V$ peak of our posterior.
Given these bimodalities, for integrated properties, we instead assume the quantities determined by \citetalias{Frye:2024aa}, who incorporate the NIRSpec spectrum into the SED fit. This approach has a better chance at breaking these degeneracies and is thus more reliable.

\begin{figure*}[ht!]
\centering
\includegraphics[width=\textwidth]{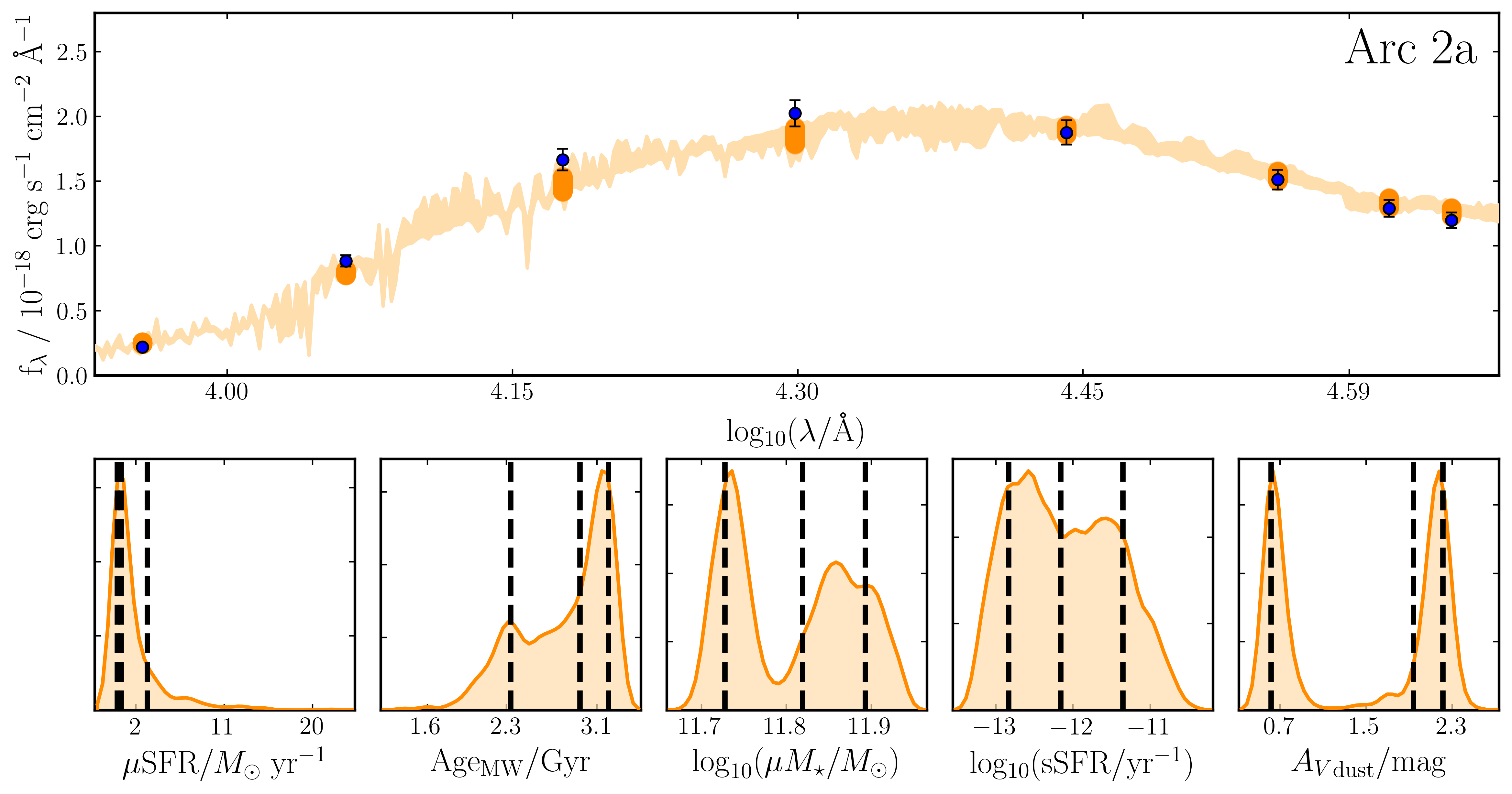}
\caption{Best-fit near-IR SED and properties from {\sc bagpipes} for Arc 2a at $z=1.78$, as with Fig.~\ref{fig:SED_1a_3c}.
    \label{fig:SED_2a}
}
\end{figure*}

\section{Simultaneous UV-FIR fitting of the spatially-integrated SED}
\label{sec:appendix4}

Inclusion of the 2mm and 3mm photometry presented in this work is desirable for constraining the dust SED, but their location at the tail end of the Rayleigh-Jeans regime renders them largely inconsequential for the fit.
Instead, we fold them into a combined SED fit for Arcs 1a, 1bc, 3ab, and 3c, so that we can take advantage of unresolved photometry measured by \citet{Harrington:2016aa} from the Wide-field Infrared Survey Explorer ({\it WISE}) at $12/22 \mu$m \citep{Wright:2010aa}, {\it Herschel} Spectral Photometric Imaging Receiver (SPIRE) at $250/350/500 \mu$m \citep{Griffin:2010aa}, and LMT/AzTEC at 1.1mm \citep{Wilson:2008aa}.
{\it Planck} photometry is excluded from the fit given its high confusion noise.
Then, NIRCam photometry is summed together for all four arcs to match.

\begin{figure*}[ht!]
\centering
\includegraphics[width=\textwidth]{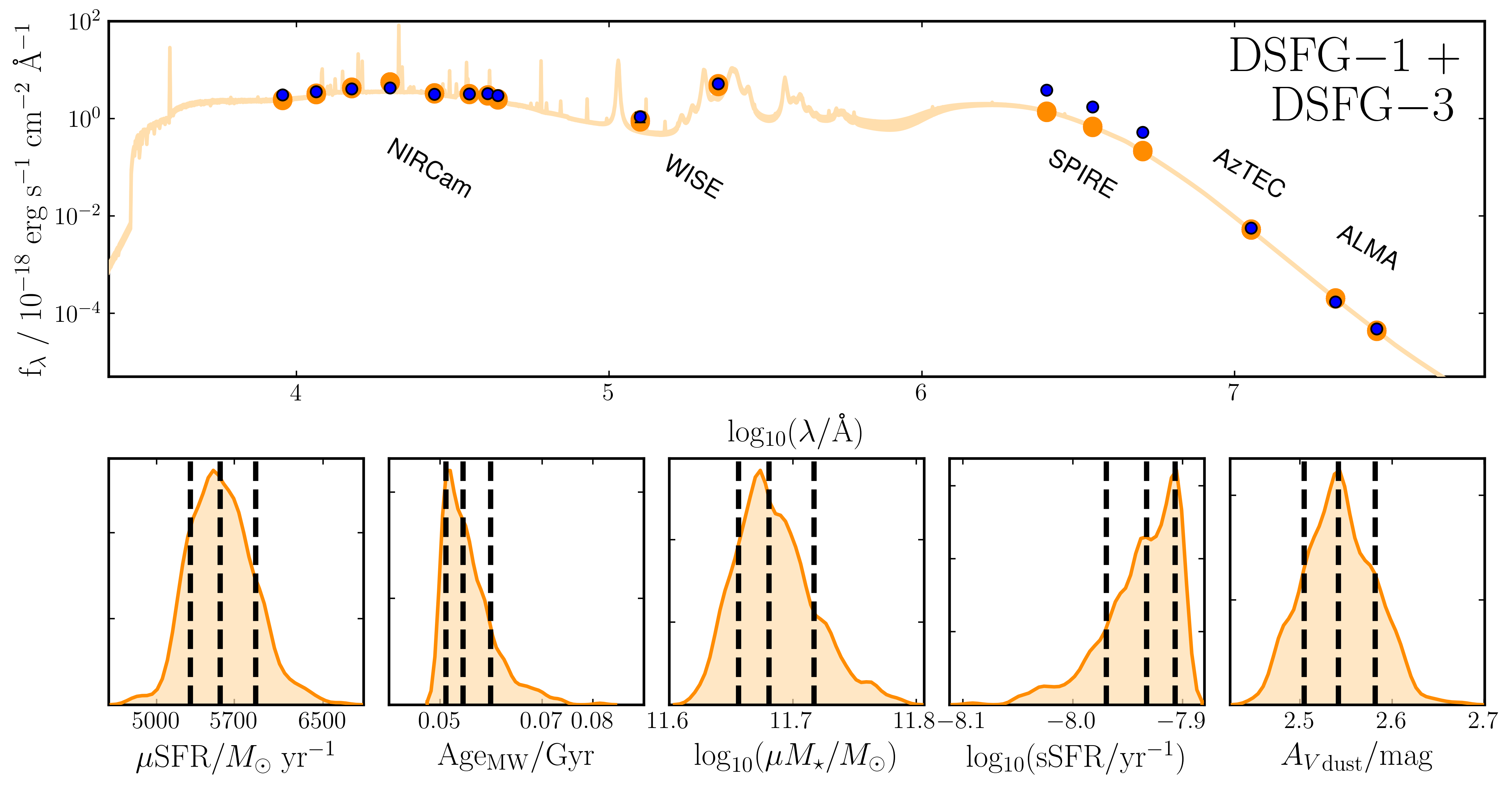}
\caption{{\it Top panel:} Combined SED fit for DSFG-1 and DSFG-3 at $z=2.236$, incorporating unresolved photometry to better constrain the dust SED. 
{\it Bottom panels:}
Posterior distributions of SFR, mass-weighted age, stellar mass, sSFR, and $A_V$. SFR and $M_\star$ are amplified by lensing magnification; corrected intrinsic values are given in Table~\ref{tab:SFR_SED}.
    \label{fig:SED_UV_FIR}
}
\end{figure*}

Given the increased number of constraints on the SED, we include additional parameters for our {\sc bagpipes} optimization, which pertain to the dust emission model (e.g., \citealt{Draine:2007aa}.)
These include the PAH mass fraction ($q_{\rm PAH} \in [0.5, 4.0]$), the lower cutoff of the incident starlight intensity on the dust ($U_{\rm min} \in [1, 25]$, effectively describing the dust temperature distribution), and the fraction of dust heated by $U > U_{\rm min}$ ($\gamma \in [0.01,0.99]$); these priors are all uniformly distributed, and consistent with those used by \citet{Williams:2019ab}.
As the far-IR coverage captures regions of higher attenuation than just UV-NIR alone, we also expand the upper limit on the prior on dust attenuation to $A_V < 8$.
Redshift is held fixed at $z=2.236$.
The best-fit dust emission parameters are 
$q_{\rm PAH} = 3.9 \pm 0.2$,
$U_{\rm min} = 22 \pm 2$,
and 
$\gamma = 0.09 \pm 0.06$.
The derived magnification-corrected SFR is SFR$_{\rm UV-FIR}= 350 \pm 70~M_\odot~{\rm yr}^{-1}$.
The results are shown in 
Fig.~\ref{fig:SED_UV_FIR} and Table~\ref{tab:SFR_SED}.

The failure to fully reproduce the {\it Herschel}/SPIRE photometry may be due to the upper bound of $U_{\rm min}$ (and by extension, dust temperature) being insufficiently large, as {\sc bagpipes} can accommodate only $U_{\rm min} \leq 25$.
Another possibility is that the 3mm photometry is contaminated non-trivially by synchrotron emission, as the radio SED is not included in the model. 
We also test different choices in SFH (a single burst model and an exponential$+$burst model), but find no significant improvement over the single exponential model. 
With the exponential$+$burst model, SFR reduces by $\lesssim 5$\% (within statistical uncertainties).

\end{document}